%% file: TOPQ_2012_21_MASTER.tex
\documentclass[reprint,aps,prd,a4paper,showpacs]{revtex4-1}

\usepackage{atlasphysics}
\usepackage{graphicx}
\usepackage{subfigure}
\usepackage{mathrsfs}
\usepackage{multirow}
\usepackage{url}
\usepackage{verbatim}
\usepackage{slashed}
\usepackage[hyperindex,breaklinks,colorlinks]{hyperref} 
\usepackage{mathptmx} 
\usepackage{lineno}

\usepackage{preprintcover}  
\PreprintCoverPaperTitle{Comprehensive measurements of {\bfseries{\emph t}}-channel single top-quark production cross sections
       at $\mathbf{\sqrt{\mathrm{{\bf{\emph s}}}} = 7}$~TeV with the ATLAS detector}  
\PreprintIdNumber{CERN-PH-EP-2014-133}  
\PreprintCoverAbstract{
This article presents measurements of the $t$-channel single top-quark ($t$) and 
top-antiquark ($\bar{t}$) total production cross sections $\sigma(tq)$ and $\sigma(\bar{t}q)$, 
their ratio $R_{t}=\sigma(tq)/\sigma(\bar{t}q)$, and a measurement of the inclusive production 
cross section $\sigma(tq + \bar{t}q)$ in proton--proton collisions at $\sqrt{s} = 7$~TeV at the LHC.
Differential cross sections for the $tq$ and $\bar{t}q$ processes are measured as a function of the 
transverse momentum and the absolute value of the rapidity of $t$ and $\bar{t}$, respectively.
The analyzed data set was recorded with the ATLAS detector and corresponds to an 
integrated luminosity of 
4.59~fb$^{-1}$. Selected events contain one charged lepton, large missing transverse momentum, and two or 
three jets. 
The cross sections are measured by performing a binned maximum-likelihood fit to the 
output distributions of neural networks.
The resulting measurements are 
$\sigma(tq)= 46 \pm 1\, (\mathrm{stat.}) \pm 6\, (\mathrm{syst.})\, \mathrm{pb}$, 
$\sigma(\bar{t}q)= 23 \pm 1\, (\mathrm{stat.}) \pm 3\, (\mathrm{syst.})\, \mathrm{pb}$,
$R_{t}=2.04\pm 0.13\, (\mathrm{stat.})\, \pm 0.12\, (\mathrm{syst.})$, 
and $\sigma(tq + \bar{t}q)= 68 \pm 2\,(\mathrm{stat.})\;\pm 8\,(\mathrm{syst.})\,\mathrm{pb}$, 
consistent with the Standard Model expectation. 
The uncertainty on the measured cross sections is dominated by systematic uncertainties,
while the uncertainty on $R_{t}$ is mainly statistical.
Using the ratio of $\sigma(tq + \bar{t}q)$ to its theoretical prediction, and assuming
that the top-quark-related CKM matrix elements obey the relation
$|V_{tb}|\gg |V_{ts}|, |V_{td}|$, we determine $|V_{tb}|=1.02 \pm 0.07$.}  
\PreprintJournalName{Physical Review D}  
\begin{document}

\title{Comprehensive measurements of {\bfseries{\emph t}}-channel single top-quark production cross sections
       at $\mathbf{\sqrt{\mathrm{{\bf{\emph s}}}} = 7}$~TeV with the ATLAS detector}
\author{The ATLAS collaboration}

\date{\today}

\begin{abstract}
This article presents measurements of the $t$-channel single top-quark ($t$) and 
top-antiquark ($\bar{t}$) total production cross sections $\sigma(tq)$ and $\sigma(\bar{t}q)$, 
their ratio $R_{t}=\sigma(tq)/\sigma(\bar{t}q)$, and a measurement of the inclusive production 
cross section $\sigma(tq + \bar{t}q)$ in proton--proton collisions at $\sqrt{s} = 7$~TeV at the LHC.
Differential cross sections for the $tq$ and $\bar{t}q$ processes are measured as a function of the 
transverse momentum and the absolute value of the rapidity of $t$ and $\bar{t}$, respectively.
The analyzed data set was recorded with the ATLAS detector and corresponds to an 
integrated luminosity of 
4.59~fb$^{-1}$. Selected events contain one charged lepton, large missing transverse momentum, and two or 
three jets. 
The cross sections are measured by performing a binned maximum-likelihood fit to the 
output distributions of neural networks.
The resulting measurements are 
$\sigma(tq)= 46\pm 1\, (\mathrm{stat.}) \pm 6\, (\mathrm{syst.})\, \mathrm{pb}$, 
$\sigma(\bar{t}q)= 23 \pm 1\, (\mathrm{stat.}) \pm 3\, (\mathrm{syst.})\, \mathrm{pb}$,
$R_{t}=2.04\pm 0.13\, (\mathrm{stat.})\, \pm 0.12\, (\mathrm{syst.})$, 
and $\sigma(tq + \bar{t}q)= 68 \pm 2\,(\mathrm{stat.})\;\pm 8\,(\mathrm{syst.})\;\mathrm{pb}$, 
consistent with the Standard Model expectation. 
The uncertainty on the measured cross sections is dominated by systematic uncertainties,
while the uncertainty on $R_{t}$ is mainly statistical.
Using the ratio of $\sigma(tq + \bar{t}q)$ to its theoretical prediction, and assuming
that the top-quark-related CKM matrix elements obey the relation
$|V_{tb}|\gg |V_{ts}|, |V_{td}|$, we determine $|V_{tb}|=1.02 \pm 0.07$.
\end{abstract}

\maketitle

\section{Introduction}
\label{sec:intro}
\begin{figure}[!h!tpb]
  \centering
\subfigure[]{
  \includegraphics[width=0.21\textwidth]{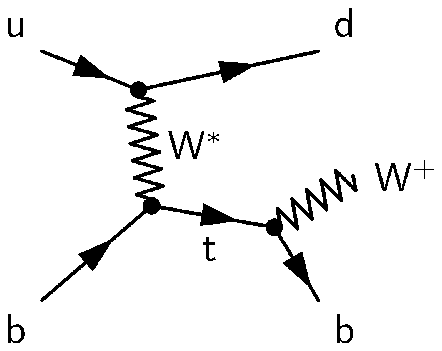}
  \label{subfig:top_quark}
}
\hspace*{0.03\textwidth}
\subfigure[]{
  \includegraphics[width=0.21\textwidth]{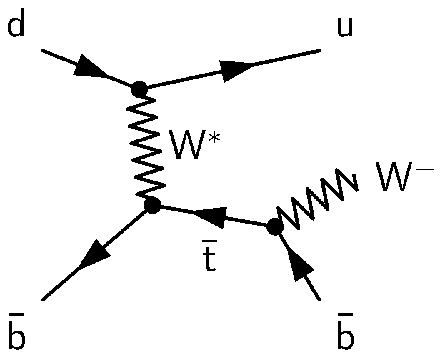}
  \label{subfig:top_antiquark}
}
\caption{\label{fig:Feynman_tchan}Representative leading-order Feynman diagrams of \subref{subfig:top_quark} 
   single top-quark production and \subref{subfig:top_antiquark} single top-antiquark 
   production via the $t$-channel exchange of a virtual $W^*$ boson, including the decay of the top-quark 
   and top-antiquark, respectively.}
\end{figure}

In proton--proton ($pp$) collisions at the LHC, top quarks are produced at unprecedented rates, 
allowing studies that were intractable before. 
The production of single top quarks via weak charged-current interactions is among the top-quark 
phenomena becoming accessible to precise investigations. 
In leading-order (LO) perturbation theory, single top-quark production is described by three 
subprocesses that are distinguished by the virtuality of the exchanged $W$ boson. 
The dominant process is the $t$-channel exchange depicted in 
Fig.~\ref{fig:Feynman_tchan}, which is the focus of the measurements presented in this article.
A light quark from one of the colliding protons interacts with a $b$-quark from 
another proton by exchanging a virtual $W$ boson ($W^*$). Since the $u$-quark density of the proton 
is about twice as high as the $d$-quark density,
the production cross section of single top quarks $\sigma(tq)$, shown in  
Fig.~\ref{fig:Feynman_tchan}\subref{subfig:top_quark}, is expected to be about twice the 
cross section of top-antiquark production $\sigma(\bar{t}q)$, 
shown in Fig.~\ref{fig:Feynman_tchan}\subref{subfig:top_antiquark}.
At LO, subleading single top-quark processes are the associated production of a $W$ boson 
and a top quark~($Wt$) and the $s$-channel production of $t\bar{b}$, analogous to the Drell--Yan process.

In general, measurements of single top-quark production provide insights into the properties 
of the $Wtb$ vertex. The cross sections are proportional to the square of the coupling at the 
production vertex. In the
Standard Model (SM), the coupling is given by the Cabibbo--Kobayashi--Maskawa (CKM) matrix element $V_{tb}$~\cite{CKM1,CKM2} 
multiplied by the universal electroweak coupling constant. Angular 
distributions of top-quark decay products give access to the Lorentz structure
of the $Wtb$ vertex, which has a vector--axial vector structure in the SM.
As illustrated in Fig.~\ref{fig:Feynman_tchan}, the $t$-channel process features a $b$-quark in 
the initial state if described in LO Quantum Chromodynamics (QCD), and therefore the cross section 
depends strongly on the $b$-quark parton distribution function (PDF), which is derived from the 
gluon PDF by means of the DGLAP evolution~\cite{Altarelli:1977zs,Dokshitzer:1977sg,Gribov:1972ri}.
A measurement of the combined top-quark and top-antiquark cross section
$\sigma(tq+\bar{t}q) = \sigma(tq) + \sigma(\bar{t}q)$
is well suited to constrain $V_{tb}$ or the $b$-quark PDF.
In addition, the measurement of $\sigma(tq+\bar{t}q)$ is sensitive to various models
of new physics phenomena~\cite{Tait:2000sh}, such as extra heavy quarks, gauge bosons, or
scalar bosons.

Separate measurements of $\sigma(tq)$ and $\sigma(\bar{t}q)$ extend the sensitivity to 
the PDFs of the $u$-quark and the $d$-quark, exploiting the different initial states of the
two processes, shown in Fig.~\ref{fig:Feynman_tchan}. At a center-of-mass energy of 
$\sqrt{s} = 7 \tev$, the typical momentum
fraction $x$ of the initial-state light quarks is in the range of $0.02 \lesssim x \lesssim 0.5$,
with a median of 0.17 for $u$-quarks and a median of 0.13 for $d$-quarks.
The additional measurement of the cross-section ratio 
$R_t \equiv  \sigma(tq)/\sigma(\bar{t}q)$ is sensitive to the ratio of the two PDFs in the 
$x$-range specified above and
features smaller systematic uncertainties because of partial cancelations of common uncertainties.
The measurements of $\sigma(tq)$, $\sigma(\bar{t}q)$, and $R_t$ provide complementary inputs 
in constraining PDFs to data currently used in QCD fits.
Investigating $R_t$ also provides a way of searching for
new-physics contributions in single top-quark (top-antiquark) 
production~\cite{AguilarSaavedra:2008gt} and of elucidating the nature of physics 
beyond the SM if it were to be observed~\cite{Gao:2011fx}.

In this article we present measurements of $\sigma(tq+\bar{t}q)$, $\sigma(tq)$, 
$\sigma(\bar{t}q)$, and the cross-section ratio $R_t$ at a center-of-mass energy of 
$\sqrt{s} = 7 \tev$, using the full data set corresponding to an integrated luminosity of 
4.59~fb$^{-1}$. Final calibrations for the $7\tev$ data set are used, resulting in reduced systematic
uncertainties.
The measurement of $\sigma(tq+\bar{t}q)$ is used to determine the value of the
CKM matrix element $|V_{tb}|$. Additionally, for the first time, differential cross sections are measured as
a function of the transverse momentum of the top quark, $p_\mathrm{T}(t)$, and the top antiquark, 
$p_\mathrm{T}(\bar{t})$, and as a function of the absolute value of the rapidities $|y(t)|$ and 
$|y(\bar{t})|$, respectively.
 
In $pp$ collisions at $\sqrt{s}=7\,\tev$,
the total inclusive cross sections of top-quark and top-antiquark production in the $t$-channel 
are predicted to be
\begin{eqnarray*}
  \sigma(tq)       & = & 41.9 ^{+1.8}_{-0.9}\ \mathrm{pb}, \\
  \sigma(\bar{t}q) & = & 22.7 ^{+0.9}_{-1.0}\ \mathrm{pb}, \ \ \mathrm{and} \\
  \sigma(tq+\bar{t}q) & = & 64.6 ^{+2.7}_{-2.0}\ \mathrm{pb} ,
\end{eqnarray*}
with approximate next-to-next-to-leading-order (NNLO) precision~\cite{Kidonakis:2011wy},
assuming a top-quark mass of $m_t=172.5\,\gev$ and using the 
{\textsc MSTW2008 NNLO}~\cite{Martin:2009iq} PDF set.
The quoted uncertainty contains the scale uncertainty and the correlated PDF--$\alpha_{\mathrm{s}}$ uncertainty.
The contributions due to the resummation of soft-gluon bremsstrahlung included in the approximate 
NNLO result are relatively small and the cross-section predictions are therefore very close to the 
plain next-to-leading-order (NLO) calculation~\cite{Campbell:2009ss}.  
All predictions used in this article are based on the 
``five-flavor scheme'', involving a $b$-quark in the initial state 
(see Fig.~\ref{fig:Feynman_tchan}). 
An alternative approach is to consider the Born process $qg\rightarrow tqb$, 
where the $b$-quark does not enter in the QCD evolution of the PDFs 
and the strong coupling constant, referred to as ``four-flavor scheme''.
Recently, computations of differential cross sections have become available at approximate NNLO 
precision~\cite{Kidonakis:2013yoa}, complementing the predictions at NLO~\cite{Campbell:2009ss}. 
Measurements of these differential quantities will allow more stringent tests of the 
calculations. In addition, a thorough study of
differential cross sections can give hints about the potential presence of 
flavor-changing neutral currents or four-fermion operators in the single top-quark production 
process~\cite{Coimbra:2012ys}.

Single top-quark production in the $t$-channel was first established 
in $p\bar{p}$ collisions at $\sqrt{s}=1.96\tev$ at the Tevatron~\cite{Abazov:2009pa}.
Measurements of $t$-channel single top-quark and $Wt$ production at the LHC at $\sqrt{s}=7\,\tev$ 
were performed by the ATLAS collaboration~\cite{Aad:2012ux,Aad:2012xca} and the CMS 
collaboration~\cite{Chatrchyan:2012ep,Chatrchyan:2012zca}.
The ATLAS measurements used only a fraction of the recorded data, 
corresponding to 1.04~fb$^{-1}$ in the $t$-channel analysis. 
At $\sqrt{s}=8\,\tev$ the CMS collaboration measured the $t$-channel cross sections and 
the cross-section ratio $R_t$~\cite{Khachatryan:2014iya}.

The measurements presented in this article are based on events in the lepton+jets channel, 
in which the lepton 
can be either an electron or a muon originating from a $W$-boson decay. 
The analysis has acceptance for signal events involving $W\rightarrow \tau\nu$ decays if the 
$\tau$ lepton decays subsequently to either $e\nu_e\nu_\tau$ or $\mu\nu_\mu\nu_\tau$. 
The experimental signature of candidate events is thus given by one charged lepton (electron or 
muon), large values of the magnitude of the missing transverse momentum \MET, and two or three 
hadronic jets with high transverse momentum. 
The acceptance for $t$-channel events is dominated by the 2-jet signature, where one jet is a 
$b$-quark jet, while the second jet is a light-quark jet. A significant fraction of single 
top-quark events are also present in the 3-jet channel, whereas the $t\bar{t}$ background is
dominant in the 4-jet channel.
For this reason, the analysis is restricted to events with two or three jets.

Several other processes feature the same signature as single top-quark events, the main backgrounds
being $W+$jets production and top-quark-antiquark ($t\bar{t}$) pair production. Since a typical
signature-based event selection yields only a relatively low signal purity, a dedicated analysis 
strategy is developed to separate signal and background events.
In both the $2$-jet and $3$-jet channels, several observables discriminating between signal and 
background events are combined by a neural network (NN) to one discriminant (NN output). 
The cross-section measurements are based on a simultaneous fit to these multivariate discriminants. 
In the 2-jet channel, a cut on the NN discriminant 
is applied to obtain a sample of events enriched in $t$-channel single top-quark events,
facilitating the measurement of differential cross sections.

\section{Data samples and samples of simulated events}
~
The analysis described in this article uses $pp$ collision data collected at a
center-of-mass energy of $7\tev$ with the ATLAS detector~\cite{ATL-2008-001_sgtop} 
at the LHC between March and November 2011.
In this data-taking period, the average number of $pp$ collisions per bunch crossing
was nine.
The selected events were recorded based on single-electron or single-muon triggers.
Stringent detector and data quality requirements are applied, resulting in a data set 
corresponding to an integrated luminosity of $4.59\pm0.08$~fb$^{-1}$~\cite{LumiPaper2012}.

\subsection{The ATLAS detector}
The ATLAS detector~\cite{ATL-2008-001_sgtop} is built from a set of 
cylindrical subdetectors, which cover almost the full solid angle around the 
interaction point~\footnote{ATLAS uses a right-handed coordinate system with its origin at the nominal interaction point in the center
of the detector and the $z$-axis along the beam direction. The $z$-axis is parallel to the anti-clockwise beam 
viewed from above. The pseudorapidity 
$\eta$ is defined as $\eta=-\ln[\tan(\theta/2)]$, where the polar angle 
$\theta$ is measured with respect to the $z$-axis. The azimuthal angle 
$\phi$ is measured with respect to the $x$-axis, which points toward the 
center of the LHC ring.  Transverse momentum and energy are defined as $\pT = 
p\sin\theta$ and $\ET = E\sin\theta$, respectively. The $\Delta R$ distance 
in ($\eta$,$\phi$) space is defined as $\Delta R=\sqrt{(\Delta\eta)^2+(\Delta\phi)^2}$.}. 
ATLAS is composed of an inner tracking detector (ID) close to the 
interaction point, surrounded by a superconducting solenoid 
providing a $2\,$T axial magnetic field, electromagnetic 
and hadronic calorimeters, and a muon spectrometer (MS).
The ID consists of a silicon pixel detector, a silicon microstrip detector (SCT), and a straw-tube 
transition radiation tracker (TRT).
The electromagnetic calorimeter is a lead and liquid-argon (LAr) sampling calorimeter with high granularity.
An iron/scintillator tile calorimeter
provides hadronic energy measurements in the central pseudorapidity range.
The endcap and forward regions are instrumented with LAr calorimeters for
both the electromagnetic and hadronic energy measurements.
The MS consists of three large superconducting toroids with eight coils each, a system of trigger chambers,
and precision tracking chambers.

\subsection{Trigger requirements}
\label{sec:trigger}
ATLAS employs a three-level trigger system. The first level (L1) is built from custom-made hardware, 
while the second and third levels are software based and collectively referred to as 
the High Level Trigger (HLT). The data sets used in this analysis are defined by high-$p_\mathrm{T}$ 
single electron or single muon triggers~\cite{Gabaldon:2012zz}. During the data-taking period slightly different trigger 
conditions were used to cope with the increasing number of multiple $pp$ collisions per bunch crossing (pile-up).

At L1, electron candidate events are required to have an electromagnetic energy deposit of 
$E_{\rm T} > 14 \gev$; in the second part of the data-taking period the requirement was 
$E_{\rm T} > 16 \gev$. At the HLT level, the full granularity of the calorimeter and tracking information is 
available. The calorimeter cluster is matched to a track and the trigger electron object has 
to have $E_{\rm T} > 20 \gev$ or $E_{\rm T} > 22 \gev$, exceeding the corresponding L1 requirements by $6 \gev$. 

The single muon trigger is based on muon candidates reconstructed in the muon spectrometer.
At L1, a threshold of $p_\mathrm{T}=10 \gev$ is applied. At the HLT level, the requirement is
tightened to $p_\mathrm{T}>18\,\rm{GeV}$.

\subsection{Simulated events}
Samples of simulated $t$-channel single top-quark events are produced with the NLO matrix-element generator 
\textsc{POWHEG-BOX}~\cite{SAMPLES-POWHEG} interfaced to 
\textsc{PYTHIA}~\cite{SAMPLES-PYTHIA} (version 6.4.27) for 
showering and hadronization. 
In \textsc{POWHEG-BOX} the four-flavor scheme calculation is used to simulate $t$-channel single top-quark
production.
The events are generated using the fixed four-flavor NLO PDF set {\sc CT10}4f~\cite{Lai:2010vv} 
and the renormalization and factorization scales are calculated event-by-event~\cite{Frederix:2012dh}
with $\mu_R = \mu_F = 4 \cdot \sqrt{m_{b}^2 + p_{{\rm T},b}^2}$,
where $m_{b}$ and $ p_{{\rm T},b}$ are the mass and $\pT$ of the $b$-quark
from the initial gluon splitting.

Samples of $t\bar{t}$ events, $Wt$ events, and $s$-channel single top-quark events are generated with 
\textsc{POWHEG-BOX} 
interfaced to {\sc PYTHIA} using the {\sc CT10} NLO PDF set~\cite{Lai:2010vv}.
All processes involving top quarks are produced assuming $m_t = 172.5\gev$, and the parameters
of the \textsc{PYTHIA} generator controlling the modeling of the parton shower and the 
underlying event are set to the values of the Perugia 2011 tune~\cite{Skands:2010ak}.  

Vector-boson production in association with jets ($W/Z$+jets) is simulated using the multileg LO 
generator \textsc{ALPGEN}~\cite{SAMPLES-ALPGEN} (version 2.13) using the CTEQ6L1 PDF set~\cite{cteq6l}. 
The partonic events are showered with \textsc{HERWIG}~\cite{SAMPLES-HERWIG} (version 6.5.20), and 
the underlying event is simulated with the \textsc{JIMMY}~\cite{Butterworth:1996zw} model (version 4.31) 
using values of the ATLAS Underlying Event Tune~2~\cite{ATL-PHYS-PUB-2011-008}. 
$W+$jets and $Z+$jets events with up to five additional partons are generated.
The MLM matching scheme~\cite{alwall-2008-53} is used to remove overlap between 
partonic configurations generated by the matrix element and by parton shower evolution.
The double counting between the inclusive $W+n$-parton samples and
samples with associated heavy-quark pair-production is removed utilizing
an overlap removal based on a $\Delta R$ matching.
The diboson processes $WW$, $WZ$ and $ZZ$ are generated using \textsc{HERWIG} and \textsc{JIMMY}.

After the event generation step, all samples are passed through the full simulation of the ATLAS 
detector~\cite{ATL-2010-005_sgtop} based on \textsc{GEANT4}~\cite{SAMPLES-G4} and are then 
reconstructed using the same procedure as for collision data. The simulation includes the effect of 
multiple $pp$ collisions per bunch crossing. The events are weighted such that 
the distribution of the number of collisions per bunch crossing is the same as in collision data.

\section{Physics object definitions}
\label{sec:objects}
In this section the definition of the physics objects is given, namely reconstructed 
electrons, muons, and jets, as well as \MET. The definition of these objects involves
the reconstructed position of the hard interaction.
Primary interaction vertices are computed from reconstructed tracks that are compatible with
coming from the luminous interaction region. The hard-scatter primary vertex is chosen
as the vertex featuring the highest $\sum \pT^2$, the sum running over all tracks with $\pT > 0.4\gev$ associated
with the vertex.

\subsection{Electrons}
\label{sec:electron_def}
Electron candidates are selected from energy deposits (clusters) in the LAr electromagnetic calorimeter matched to 
tracks~\cite{Aad:2014fxa} 
and are required to have $\ET >25\gev$ and $|\eta_\mathrm{cl}|<2.47$, where $\eta_\mathrm{cl}$ denotes 
the pseudorapidity of the cluster.  Clusters falling in the calorimeter barrel/endcap 
transition region, corresponding to $1.37<|\eta_\mathrm{cl}|<1.52$, are ignored.
The energy of an electron candidate is taken from the cluster, while its
$\eta$ and $\phi$ are taken from the track.
The $z$-position of the track has to be compatible
with the hard-scatter primary vertex. Electron candidates are further required to
fulfil stringent criteria regarding calorimeter shower shape, track quality, 
track--cluster matching, and fraction of high-threshold hits in the TRT to ensure high identification 
quality.

Hadronic jets mimicking the signature of an electron, electrons from $b$-hadron or $c$-hadron decays,
and photon conversions constitute the major backgrounds for high-\pT~electrons originating from the decay of
a $W$ boson. Since signal electrons from $W$-boson decay are typically isolated from jet activity, 
these backgrounds can be suppressed via isolation criteria that require minimal calorimeter activity 
(calorimeter isolation) and only few tracks (track isolation) in an ($\eta$,$\phi$) region around the electron.
Electron candidates are isolated by imposing thresholds on 
the scalar sum of the transverse momenta of calorimeter energy deposits $\Sigma p_{\rm T}^{\rm calo}$
within a surrounding cone of radius $\Delta R = 0.2$, excluding the energy deposit associated with the
candidate, and on the scalar sum of the transverse momenta of tracks $\Sigma p_{\rm T}^{\rm track}$ 
in a cone of radius $\Delta R = 0.3$ around the candidate excluding the track associated with the 
electron candidate. 
The $\Sigma p_{\rm T}^{\rm calo}$ variable is corrected for pile-up effects as a function of the 
number of reconstructed vertices.
The thresholds applied to $\Sigma p_{\rm T}^{\rm calo}$ and $\Sigma p_{\rm T}^{\rm track}$ vary as 
a function of the electron $p_{\rm T}$, the electron $\eta$, and the number of 
reconstructed primary vertices and are chosen such that the efficiency for electrons from $W$-boson or 
$Z$-boson decays to pass this isolation requirement is 90\%. 

\subsection{Muons}
Muon candidates are reconstructed by combining track segments found in the ID and in the MS~\cite{Aad:2014zya}.
The momentum as measured using the ID is required to agree with the
momentum measured using the MS after correcting for the predicted muon energy loss in the calorimeter.
Only candidates that have $p_\mathrm{T}>25\;\gev$ and $|\eta|<2.5$ are considered.
Selected muons must additionally satisfy a series of requirements on the number of track hits present 
in the various tracking subdetectors.
Muon tracks are required to have at least two hits in the pixel detector, and six or more hits in 
the SCT. Tracks are rejected if they have more than two missing hits in the SCT and pixel
detectors, or tracks with an excessive number of outlier hits in the TRT.
Isolated muon candidates are selected by requiring $\Sigma p_{\rm T}^{\rm calo} <4\,\gev$
within a surrounding cone of radius $\Delta R=0.2$, and $\Sigma p_{\rm T}^{\rm track} <2.5\,\gev$ within a surrounding cone of 
radius $\Delta R = 0.3$. The efficiency of this combined isolation requirement varies between 95\% 
and 97\%, depending on the data-taking period.

The reconstruction, identification and trigger efficiencies of electrons and muons
are measured using tag-and-probe methods on samples enriched with
$\Zboson~\to\ell\ell$, $J/\psi ~\to
\ell\ell$, or $\Wboson^{\pm}~\to\ell\nu$ ($\ell=e,\mu$) events~\cite{Aad:2014fxa,Aad:2014zya}.

\subsection{Jets and missing transverse momentum}
Jets are reconstructed using the anti-$k_{t}$ algorithm~\cite{CAC-0801} 
with a radius parameter of 0.4, using topological clusters~\cite{Lampl:1099735} identified in the calorimeter
as inputs to the jet clustering. 
The jet energy is corrected for the effect of multiple $pp$ interactions, both in collision 
data and in simulated events.
Further energy corrections apply factors depending on the jet energy and the jet $\eta$
to achieve a calibration that matches the energy of stable particle jets 
in simulated events~\cite{Aad:2011he}.
Differences between data and Monte Carlo simulation are evaluated using in situ techniques 
and are corrected for in an additional step~\cite{Aad:2014bia}.
The in situ calibration exploits the $\pt$ balance in $Z$+jet, $\gamma$+jet, and dijet events. 
$Z$+jet and $\gamma$+jet data are used to set the jet energy scale (JES) in the central detector region, 
while $\pt$ balancing in dijet events is used to achieve an $\eta$ intercalibration 
of jets in the forward region with respect to central jets. 

Jets with separation $\Delta R<0.2$ from selected electron candidates are 
removed, as in these cases the jet and the electron are very likely to correspond 
to the same physics object. 
In order to reject jets from pile-up events, a quantity called the jet-vertex fraction 
$\epsilon_\mathrm{jvf}$ is defined as the ratio of $\sum \pT$ for all
tracks within the jet that originate from the hard-scatter primary vertex to the $\sum \pT$ of all tracks matched to the jet.
It is required that $\epsilon_\mathrm{jvf}>0.75$ for those jets that have associated tracks.
The $\epsilon_\mathrm{jvf}$ criterion is omitted for jets without matched tracks.
An overlap removal between jets and muons is applied, removing any muon 
with separation $\Delta R<0.4$ from a jet with $p_{\rm{T}}>25\,\rm{GeV}$ and $\epsilon_\mathrm{jvf} > 0.75$.
In the same way an overlap removal is applied between jets and electrons, removing any electron 
separated from a jet by $0.2<\Delta R<0.4$.

Only jets having $\pT >30$~$\gev$ and $|\eta|<4.5$ are 
considered. Jets in the endcap/forward-calorimeter transition region, corresponding to 
$2.75<|\eta|<3.5$, must have $\pT >35$~$\gev$.

The $\MET$~is a measure of the momentum of the escaping neutrinos, but is also affected by
energy losses due to detector inefficiencies.
The $\MET$ is calculated based on the vector sum of energy deposits in the calorimeter projected
onto the transverse plane
and is corrected for the presence of electrons, muons, and jets~\cite{EtmissPerformance2012}.

\subsection{Identification of {\bfseries{\emph b}}-quark jets}
\label{sec:bjets}
The identification of jets originating from the fragmentation of $b$-quarks is one of the most 
important techniques for selecting top-quark events. Several properties can be used to distinguish $b$-quark 
jets from other jets: 
the long lifetime of $b$-hadrons, the large $b$-hadron mass, and the large branching ratio to leptons.  
The relatively long lifetime of $b$-flavored hadrons results in a significant flight path length,
leading to reconstructable secondary vertices and tracks with large impact parameters relative to the 
primary vertex.

Jets containing $b$-hadrons are identified in the region $|\eta|<2.5$ by 
reconstructing secondary and tertiary vertices from the tracks associated with 
each jet and combining lifetime-related information in a neural network~\cite{ATLAS-CONF-2011-102}.
Three different neural networks are trained corresponding to an optimal 
separation of $b$-quark jets, $c$-quark jets, and light-quark jets. 
The output of the networks is given in terms of probabilities $p_b$, $p_c$, and $p_{\mathrm{l}}$, which
are then combined to form a final discriminant. 
In order to achieve excellent rejection of $c$-quark jets the ratio $p_b / p_c$ is calculated. 
The chosen working point corresponds to a  
$b$-tagging efficiency of about 54\% for $b$-quark jets in $t\bar{t}$ events.
The misidentification efficiency is 4.8\% for $c$-quark jets and 0.48\% for light-quark jets, 
as derived from simulated \ttbar\ events. 
Jets passing the requirement on the identification discriminant are called $b$-tagged jets. 
Scale factors, determined from collision data, 
are applied to correct the $b$-tagging efficiency in simulated events to match the data.

\section{Event selection}
\label{sec:selection}
The event selection requires exactly one charged lepton, $e$ or $\mu$, 
exactly two or three jets, and $\MET>30\gev$.
At least one of the jets must be $b$-tagged. 
A trigger matching requirement is applied according to which the lepton must lie within 
$\Delta R=0.15$ cone around its trigger-level object. 
Candidate events are selected if they contain at least one good primary vertex 
candidate with at least five associated tracks. Events containing jets with
transverse momentum $\pt > 20\gev$ failing to satisfy quality criteria against 
misreconstruction~\cite{Aad:2011he} are rejected. 

Since the multijet background is difficult to model precisely, its contribution is reduced by 
requiring the transverse mass of the lepton-$\MET$ system, 
\begin{equation}
  m_\mathrm{T}\left(\ell\MET\right) = \sqrt{2 p_\mathrm{T}(\ell) \cdot \MET 
  \left[1-\cos \left( \Delta \phi \left(\ell, \MET \right) \right) \right]}\ , 
\label{eq:mTW}
\end{equation} 
to be larger than $30\gev$.
Further reduction of the multijet background is achieved by 
placing an additional requirement on events with a charged lepton 
that is back-to-back with the leading jet in $p_\mathrm{T}$. 
This is realized by the following condition between the lepton $p_\mathrm{T}$ and 
the $\Delta \phi \left(j_1, \ell \right)$:
\begin{equation}
\pT\left(\ell\right) > 40\gev \cdot \left(1 -  \frac{\pi - |\Delta \phi\left(j_1, \ell \right)|}{\pi -1} \right)
\end{equation}
where $j_1$ denotes the leading jet.

In the subsequent analysis, signal events are divided into different analysis channels according to the 
sign of the lepton charge and the number of jets. In the 2-jet channels, exactly one
jet is required to be $b$-tagged. To further reduce the $W$+jets background in these channels, 
the absolute value of the difference in pseudorapidity $|\Delta\eta|$ of the 
lepton and the $b$-tagged jet is required to be smaller than~2.4.
In the 3-jet channels, events with exactly one 
and exactly two $b$-tagged jets are considered and separated accordingly. 
In the 3-jet-2-tag category no distinction is made between events with positive and negative lepton charge, since this
channel is dominated by $\ttbar$ background and can be used to further constrain the uncertainty on the $b$-tagging efficiency.
Finally, the resulting channels are referred to as:
2-jet-$\ell^+$, 2-jet-$\ell^-$, 3-jet-$\ell^+$-1-tag, 3-jet-$\ell^-$-1-tag,
and 3-jet-2-tag.

A control region is defined to be orthogonal to the signal region in the same kinematic phase space 
to validate the modeling of the backgrounds by simulated events. Events in these control regions 
feature exactly one $b$-tagged jet, which was identified with a less stringent $b$-tagging algorithm
than used to define the signal region.
The signal region is excluded from the control region by applying a veto.

\begin{figure*}[!tb]
  \centering
  \subfigure[]{
     \includegraphics[width=0.45\textwidth]{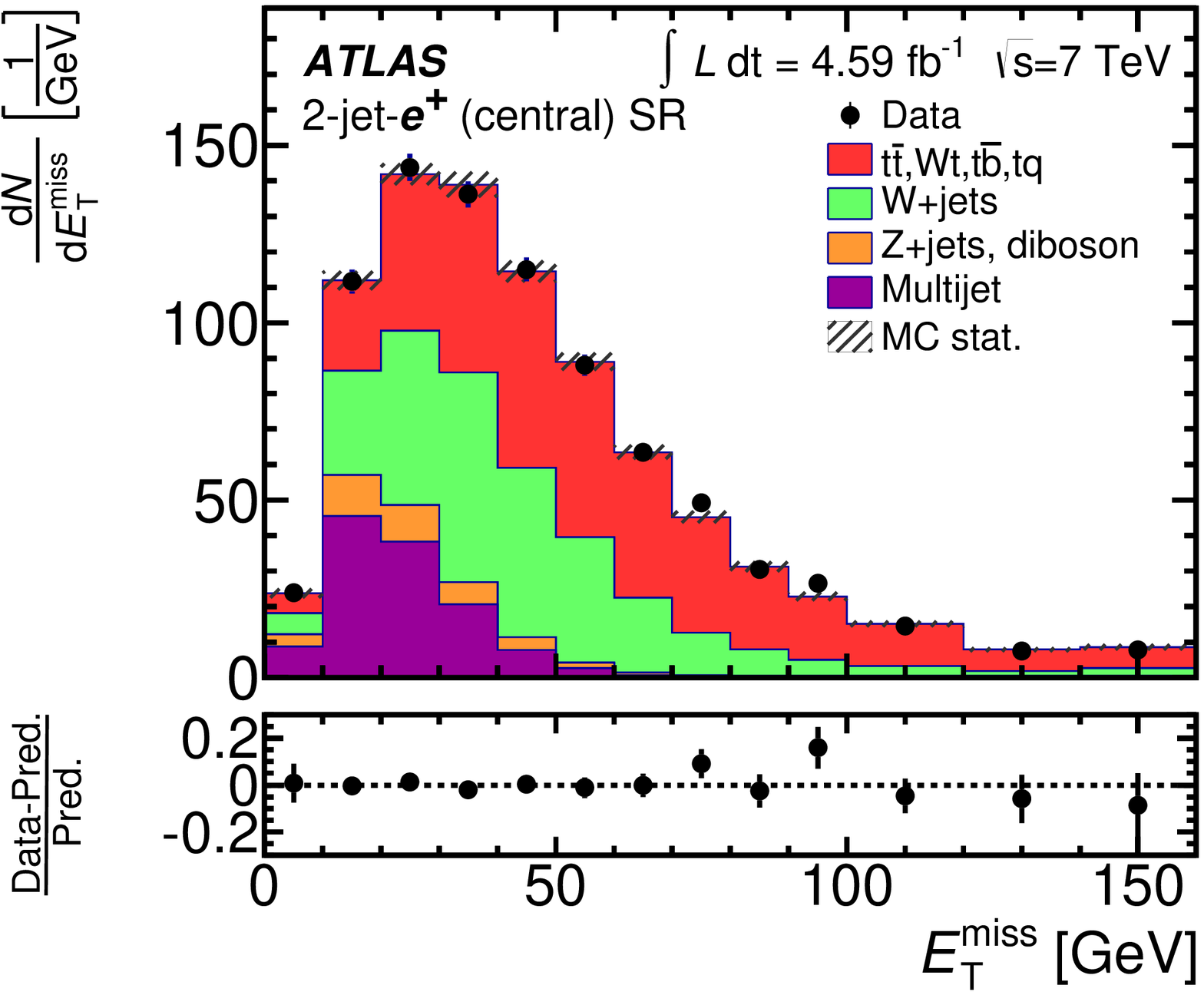}
  }
  \subfigure[]{
     \includegraphics[width=0.45\textwidth]{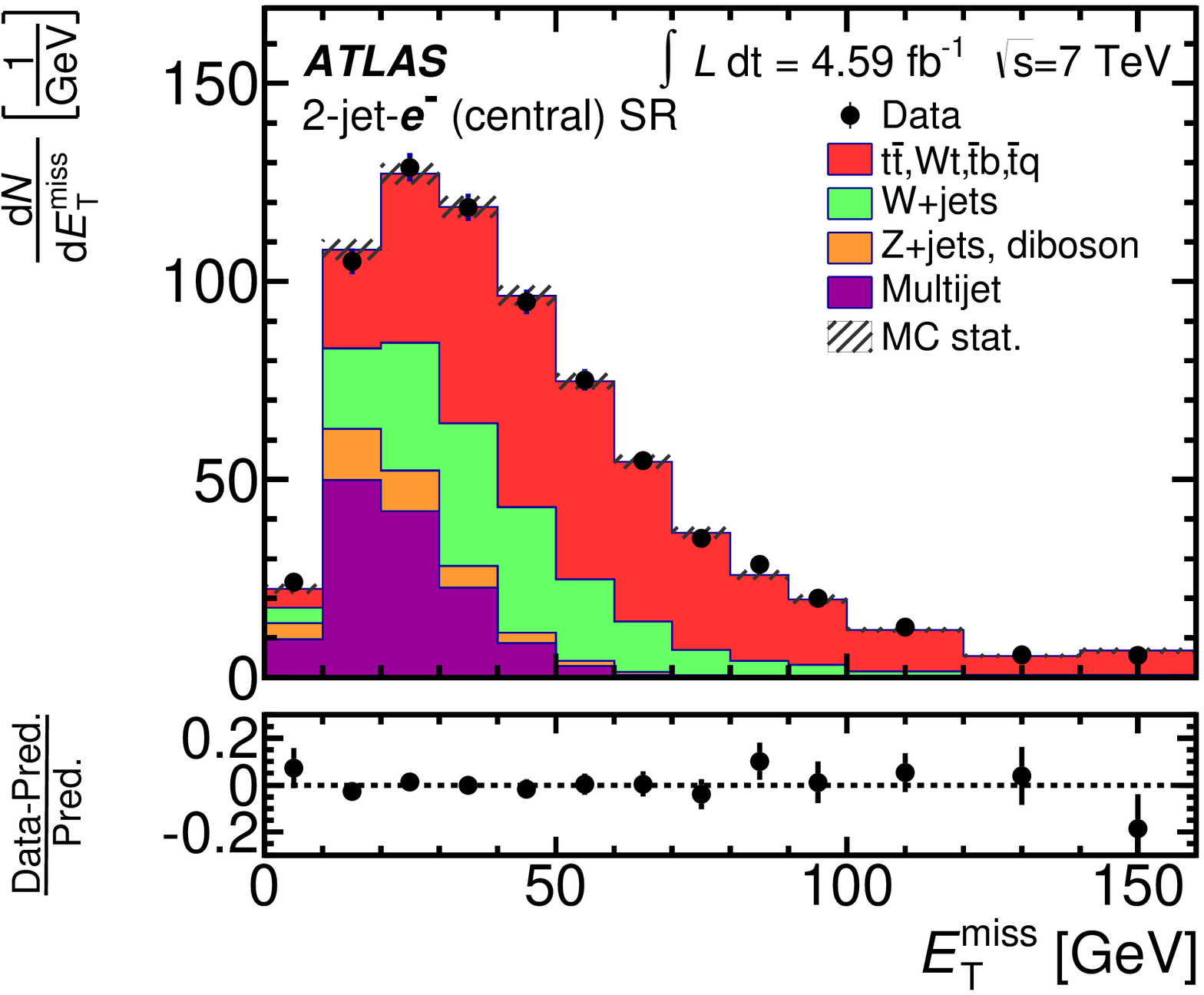}
  }
  \caption{\label{fig:missetfit_2Jets} \MET~distributions in the signal region~(SR) for the (a) 2-jet-$e^+$ 
     and (b) 2-jet-$e^-$ channels for central electrons.
     The distributions are normalized to the result of a binned maximum-likelihood fit described in Sec.~\ref{sec:QCDestimate}.  
     The relative difference between the observed and expected number of events 
     in each bin is shown in the lower panels.
  }
\end{figure*}
\section{Background estimation}
\label{sec:bgestimation}
One of the largest backgrounds to single top-quark processes in the lepton+jets channel is 
$W+$jets production. If one of the jets contains $b$-hadrons or $c$-hadrons, these events
have the same signature as signal events. Due to possible misidentification of a light-quark jet as a 
$b$-quark jet, $W+$light-jets production also contributes to the background. 
An equally important background comes from top-quark-antiquark ($t\bar{t}$) pair production events, 
which are difficult to separate from single top-quark events, since they contain top quarks as well. 
Another background is due to multijet production via the strong interaction. In these events a hadronic 
jet is misidentified as 
a lepton, usually an electron, or a real high-$\pT$ lepton is produced within a jet due to 
the semileptonic decay of a heavy-flavor ($b$ or $c$) hadron and satisfies the lepton isolation criteria.
Other smaller backgrounds come from diboson ($WW$, $WZ$, and $ZZ$) and $Z+$jets production.

\subsection{{\bfseries{\emph W}}/{\bfseries{\emph Z}}+jets background}
\label{sec:WjetsEstimate}
The $W$+jets background is initially normalized to the theoretical prediction and then subsequently
determined simultaneously both in the context of the multijet background estimation and as part of the extraction 
of the signal cross section. The estimated number of events of the much smaller 
$Z$+jets background is calculated using the theoretical prediction.

The cross sections for inclusive $W$-boson production and $Z$-boson production are predicted with NNLO precision 
using the {\sc FEWZ} program~\cite{Anastasiou:2003ds}, 
resulting in a LO-to-NNLO scale factor of 1.2 and an uncertainty of 4\%. The uncertainty includes the uncertainty
on the PDF and scale variations.
The scale factor is applied to the prediction based on the LO {\sc ALPGEN} calculation
for the $W$+$b\bar{b}$, $W$+$c\bar{c}$, and $W$+light-jets samples. 
An uncertainty for associated jet production is estimated using
variations of the factorization and renormalization scale and the \textsc{ALPGEN}
matching parameter. These variations yield an uncertainty of 5\% for the production of two additional 
light-quark jets and 15\% for two additional heavy-quark jets.
An additional relative uncertainty of 50\% is assigned to the $W$+$b\bar{b}$ and $W$+$c\bar{c}$ 
production rates to take uncertainties on heavy-flavor production into account. 
This uncertainty is estimated using a tag-counting method in control regions~\cite{Aad:2012ux}.

The {\sc ALPGEN} prediction for the $W$+$c$ process 
is scaled by a factor of $1.52$ that is obtained from a study
based on NLO calculations using {\sc MCFM}~\cite{Campbell:2010ff}.
Normalization uncertainties on the factorization and renormalization scale and PDF uncertainties are 24\%.

The processes $W$+$b\bar{b}$, $W$+$c\bar{c}$, and $W$+light-jets, 
being asymmetric in lepton charge, are combined and are used as a single process in the binned 
maximum-likelihood fit to determine the signal yield. 

\subsection{Multijet background}
\label{sec:QCDestimate}
Multijet background events pass the signal selection if a jet is misidentified as an isolated lepton or if the 
event has a non-prompt lepton that appears to be isolated. Since it is neither possible to simulate
a sufficient number of those events nor possible to calculate the rate precisely, different 
techniques are developed to model multijet events and to estimate the production rate.
These techniques employ both, collision data as well as simulated events.  

In the electron channel, misidentified jets are the main source of multijet background events. 
This motivates the jet-lepton method in which an electron-like jet is selected with special requirements 
and redefined as a lepton.
This jet has to fulfil the same $p_{\rm T}$ and $\eta$ requirements as a signal electron, and  
contain at least four tracks to reduce the 
contribution from converted photons. In addition, the jet must deposit 80--95\% of its energy in the electromagnetic calorimeter.
Events are selected using the same criteria as for the signal selection except for the selection 
of the electron.
The event is accepted if exactly one such `jet lepton' is found.
The jet-lepton selection is applied to a {\sc Pythia} dijet sample and the resulting set of events
is used to model the multijet background in the electron channel.

To determine the normalization of the multijet background in the electron channel,
 a binned maximum-likelihood fit to observed data in the \MET~distribution is performed after
 applying all selection criteria except for the \MET~requirement. In each channel two fits are 
performed separately; one for electrons in the central ($|\eta| < 1.5$) region and one for the
 endcap ($|\eta| > 1.5$) region of the electromagnetic calorimeter.
The multijet template is fitted together with templates derived from Monte Carlo simulation for all other
background processes whose rate uncertainties 
are accounted for in the fitting process in the form of additional constrained nuisance parameters.
For the purpose of these fits the contributions from $W$+light-jets
and $W$+$b\bar{b}$, $W$+$c\bar{c}$, $W$+$c$,
the contributions from $t\bar{t}$ and single top-quark production, 
and the contributions from $Z$+jets and diboson 
production, are each combined into one template. Distributions normalized to the fit results in the
2-jet-$e^+$ and 2-jet-$e^-$ signal regions for central electrons are shown in Fig.~\ref{fig:missetfit_2Jets}.

In the muon channel, the matrix method~\cite{Aad:2012qf} is used to obtain both the normalization and shape of the multijet background.
The method estimates the number of multijet background events in the signal region based on loose and tight lepton isolation
definitions, the latter selection being a subset of the former. Hence, the loose selection is defined to contain
leptons of similar kinematics, but results in much higher event yields and is, except for the muon isolation requirement,
identical to the signal selection. 
The number of multijet events $N^\mathrm{tight}_\mathrm{fake}$ passing the tight (signal) isolation requirements can be expressed as,
\begin{equation}
N^\mathrm{tight}_\mathrm{fake} = \frac{\epsilon_\mathrm{fake}}{\epsilon_\mathrm{real} - \epsilon_\mathrm{fake}} \cdot (N^\mathrm{loose} \epsilon_\mathrm{real} - N^\mathrm{tight}),
\label{eqn:intro-mm-tight_fake}
\end{equation}
where $\epsilon_\mathrm{real}$ and $\epsilon_\mathrm{fake}$ are the efficiencies for real and fake loose leptons being
selected as tight leptons, $N^\mathrm{loose}$ is the number of selected events in the loose sample, and $N^\mathrm{tight}$ is the number of selected events in the signal sample. 
The fake efficiencies are determined from collision data in a sample of selected muon candidates with high
impact parameter significance which is defined by the impact parameter divided by its uncertainty.
The real efficiencies are also estimated from collision data using a ``tag-and-probe'' method, which 
is based on the identification of a tight lepton and a loose lepton in events originating 
from a leptonically decaying $Z$ boson.

An uncertainty of 50\% is applied to the estimated yield of multijet background events based on comparisons of the rates 
obtained by using alternative methods, i.e. the matrix method in the electron channel
and the jet-lepton method in the muon channel, and using an alternative variable, i.e. $m_{\mathrm{T}}(\ell\MET)$ instead of
$\MET$ for the binned maximum-likelihood fit.

\subsection{$\mathrm{{\bf{\emph t}}}\bar{\mathrm{{\bf{\emph t}}}}$ production and other backgrounds}
\label{sec:ttbar}
The $t\bar{t}$ cross section is calculated at NNLO in QCD including resummation of
next-to-next-to-leading logarithmic (NNLL) soft gluon terms~\cite{Cacciari:2011hy,Baernreuther:2012ws,Czakon:2012zr,Czakon:2012pz,Czakon:2013goa}
with Top++2.0~\cite{Czakon:2011xx}. 
The PDF and $\alpha_{\mathrm{s}}$ uncertainties are calculated using the PDF4LHC prescription~\cite{Botje:2011sn} with the MSTW2008 
NNLO~\cite{Martin:2009iq,Martin:2009bu} at 68\% confidence level (CL), 
the CT10 NNLO~\cite{Lai:2010vv,Gao:2013xoa}, and the
NNPDF2.3~\cite{Ball:2012cx} PDF sets, and are added in quadrature to the scale uncertainty, 
yielding a final uncertainty of 6\%.

Since $Wt$ production is charge symmetric with respect to top-quark and top-antiquark production, 
the combined cross section of 
$\sigma(Wt) = 15.7 \pm 1.1\;$pb~\cite{Kidonakis:2010ux} is used in the analysis.
The predicted cross sections for $s$-channel production are $\sigma(t\bar{b}) = 3.1 \pm 0.1 \;$pb
and $\sigma(\bar{t}b) = 1.4 \pm 0.1\;$pb~\cite{Kidonakis:2010tc}.
The predictions of $\sigma(Wt)$, $\sigma(t\bar{b})$, and $\sigma(\bar{t}b)$ are given at 
approximate NNLO precision, applying soft-gluon resummation.
Theoretical uncertainties including PDF and scale uncertainties are 4.4\%~\cite{Kidonakis:2010tc} 
for $s$-channel single top-quark production and 7.0\%~\cite{Kidonakis:2010ux} for $Wt$ production.
The PDF uncertainties are evaluated using the 40 associated eigenvector PDF sets of MSTW 2008 
at 90\% CL.
The cross sections given above are used to compute the number of expected single top-quark  
events by normalizing the samples of simulated events.

All top-quark background processes are shown combined in the figures and used as a single process in the analysis. The charge asymmetry in $s$-channel 
production is taken from the approximate NNLO prediction. 

Diboson events ($WW$, $WZ$ and $ZZ$) are normalized to the NLO cross-section prediction calculated with {\sc MCFM}~\cite{Campbell:2010ff}. 
The cross-section uncertainty for these processes is 5\%. 

\subsection{Event yields}
Table~\ref{tab:evtyield} provides the event yields after event selection.
The yields are presented for the tagged channels, where exactly one $b$-tagged jet is required,
separated according to the lepton charge and for the 3-jet-2-tag channel.
Small contributions from the $tq$ process in the $\ell^-$ regions and the 
$\bar{t}q$ process in the $\ell^+$ regions originate from lepton charge misidentification.
\begin{table*}[!thbp]
  \begin{center}
    \begin{ruledtabular}
  \caption{Predicted and observed events yields for the 2-jet and 3-jet channels considered in this measurement. 
  The multijet background is estimated using data-driven techniques (see Sec.~\ref{sec:QCDestimate});
  an uncertainty of 50\% is applied. 
  All the other expectations are derived using theoretical cross sections 
  and their uncertainties (see Sec.~\ref{sec:WjetsEstimate} 
  and Sec.~\ref{sec:ttbar}).
\label{tab:evtyield}}
    \begin{tabular}{lcr@{\extracolsep{0pt}$\:\pm\:$}l@{\hspace{0.4cm}}@{\extracolsep{\fill}}
                     r@{\extracolsep{0pt}$\:\pm\:$}l@{\hspace{0.4cm}}@{\extracolsep{2cm}}
                     r@{\extracolsep{0pt}$\:\pm\:$}l@{\hspace{0.4cm}}@{\extracolsep{\fill}}
                     r@{\extracolsep{0pt}$\:\pm\:$}l@{\hspace{0.4cm}}@{\extracolsep{\fill}}
                     r@{\extracolsep{0pt}$\:\pm\:$}l@{\hspace{0.4cm}}@{\extracolsep{\fill}}}

    &&\multicolumn{4}{c}{2-jet channels} & \multicolumn{6}{c}{3-jet channels} \\
                     && \multicolumn{2}{c}{$\ell^+$} & \multicolumn{2}{c}{$\ell^-$} & 
                       \multicolumn{2}{c}{$\ell^+$} & \multicolumn{2}{c}{$\ell^-$} & \multicolumn{2}{c}{2-tag}\\
    \hline
    $tq$ & & $2550$ & $220$ & $3.6$ & $0.3$ & $845$ & $74$ & $1.2$ & $0.1$ & $309$ & $26$ \\
    $\bar{t}q$ & & $1.5$ & $0.1$ & $1390$ & $120$ & $0.52$ & $0.05$ & $435$ & $38$ & $162$ & $14$ \\
    $\ttbar,Wt,t\bar{b},\bar{t}b$ & & $5250$ & $530$ & $5130$ & $510$ & $8200$ & $820$ & $8180$ & $820$ & $5850$ & $580$ \\
    $W^+$+$b\bbar$,$c\cbar$,light jets && $5700$ & $2500$ & $16.3$ & $8.2$ & $2400$ & $1200$ & $11.5$ & $5.7$ & $200$ & $100$ \\
    $W^-$+$b\bbar$,$c\cbar$,light jets && $9.2$ & $4.6$ & $3400$ & $1700$ & $4.1$ & $2.0$ & $1470$ & $740$ & $137$ & $68$ \\
    $W$+$c$ & & $1460$ & $350$ & $1620$ & $390$ & $388$ & $93$ & $430$ & $100$ & $6.5$ & $1.6$ \\
    $Z$+jets, diboson && $370$ & $220$ & $310$ & $180$ & $190$ & $120$ & $180$ & $110$ & $22$ & $13$ \\
    Multijet & & $750$ & $340$ & $740$ & $370$ & $320$ & $160$ & $440$ & $220$ & $21$ & $11$ \\
    \multicolumn{11}{c}{ } \\
    Total expectation && $16100$ & $2600$ & $12600$ & $2000$ & $12400$ & $1500$ & $11100$ & $1100$  & $6710$ & $610$  \\

    Data &~~~~~& \multicolumn{2}{c}{$16198$$\;\;\,$} & \multicolumn{2}{c}{$12837$$\;\;\;$} & \multicolumn{2}{c}{$12460$$\;\;\;\;\;$} & \multicolumn{2}{c}{$10819$$\;\;\;$} & \multicolumn{2}{c}{$6403$$\;\;\;$} \\

  \end{tabular}
 \end{ruledtabular}
 \end{center}
\end{table*}

\section{Signal and background discrimination}
\label{sec:signal_discrim}
To separate $t$-channel single top-quark signal events from 
background events, several kinematic variables are combined to form powerful discriminants by
employing neural networks. A large number of potential input variables were 
studied, including not only kinematic variables of the identified physics objects, but also
variables obtained from the reconstruction of the $W$ boson and the top quark.

\subsection{Top-quark reconstruction}
When reconstructing the $W$ boson, the transverse momentum of the neutrino is given by
the $x$- and $y$-components of the $\MET$, while the unmeasured 
$z$-component of the neutrino momentum $p_z(\nu)$ is inferred by 
imposing a $W$-boson mass constraint on the lepton--neutrino system. 
Since the constraint leads to a quadratic equation for $p_z(\nu)$, 
a two-fold ambiguity arises. In the case of two real solutions, 
the one with the lower $|p_z(\nu)|$ is chosen.
In case of complex solutions, which can occur due to the low $\MET$ resolution,
a kinematic fit is performed that rescales the neutrino $p_x$ and $p_y$ such that 
the imaginary part vanishes and at the same time the transverse components
of the neutrino momentum are kept as close as possible to the $\MET$.
As a result of this algorithm, the four-momentum of the neutrino is
reconstructed. 

The top quark is reconstructed by adding the four-momenta of the reconstructed
$W$ boson and the $b$-tagged jet. Several angular variables, invariant masses and
differences in $\pT$ are defined using the reconstructed physics objects.

\subsection{Selection of discriminating variables}
\label{subsec:prepro}
The NeuroBayes~\cite{Feindt:2006pm} tool is used for preprocessing the input variables 
and for the training of the NNs.
The ranking of the variables in terms of their discrimination power is automatically determined as part of the
preprocessing step and is independent of the training procedure~\cite{Aad:2012ux}.
Only the highest-ranking variables are chosen for the training of the NNs. 
Separate NNs are trained in the $2$-jet channel and $3$-jet channel.
In the training, no separation is made according to lepton charge or lepton flavour. 
Dedicated studies show that training in the channels separated by lepton charge does 
not lead to an improvement in sensitivity.

\begin{figure*}[!tb]
  \centering
  \subfigure[]{
     \includegraphics[width=0.45\textwidth]{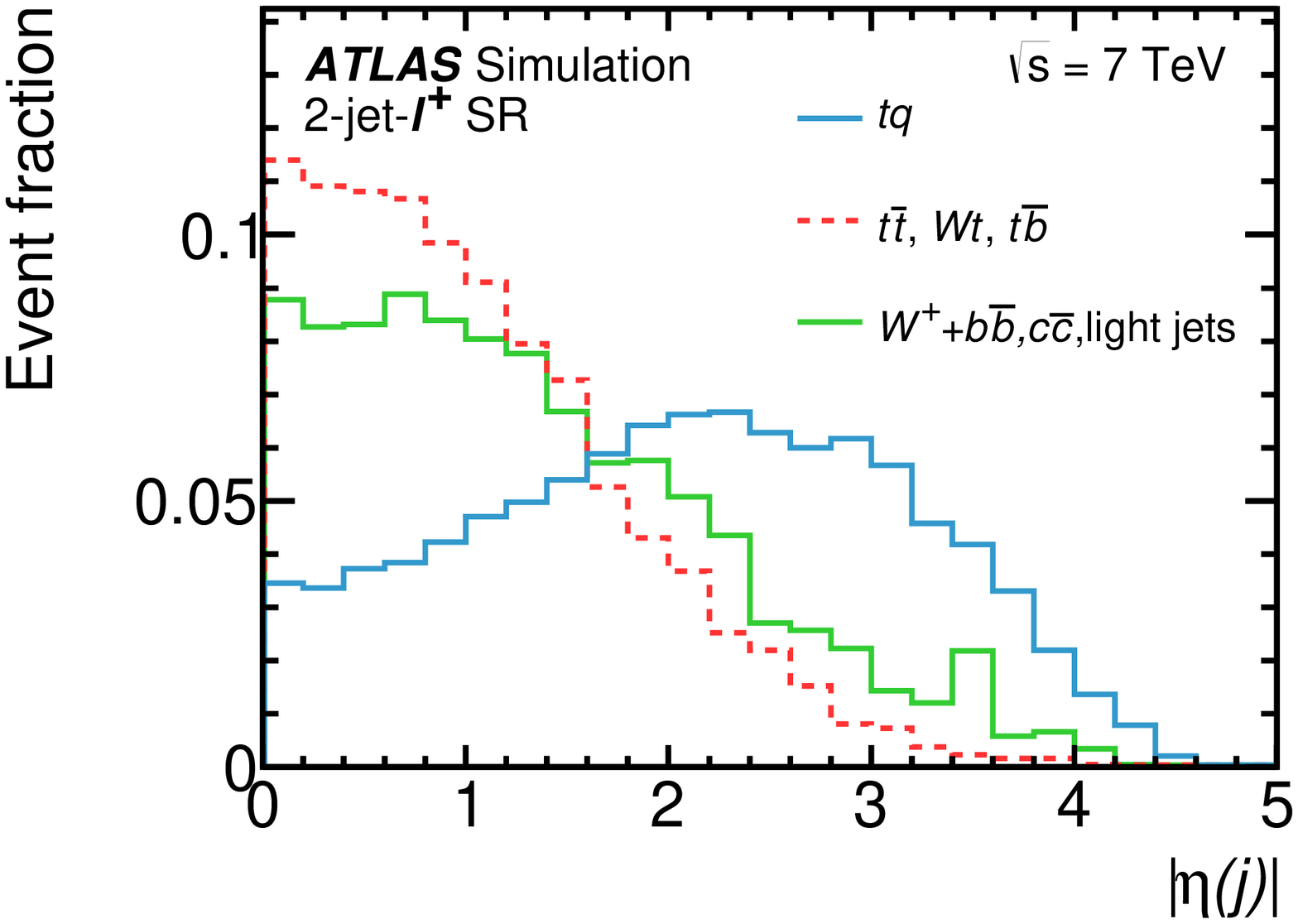}
     \label{subfig:2j_plus_lightEtashape}
  }
  \subfigure[]{
     \includegraphics[width=0.45\textwidth]{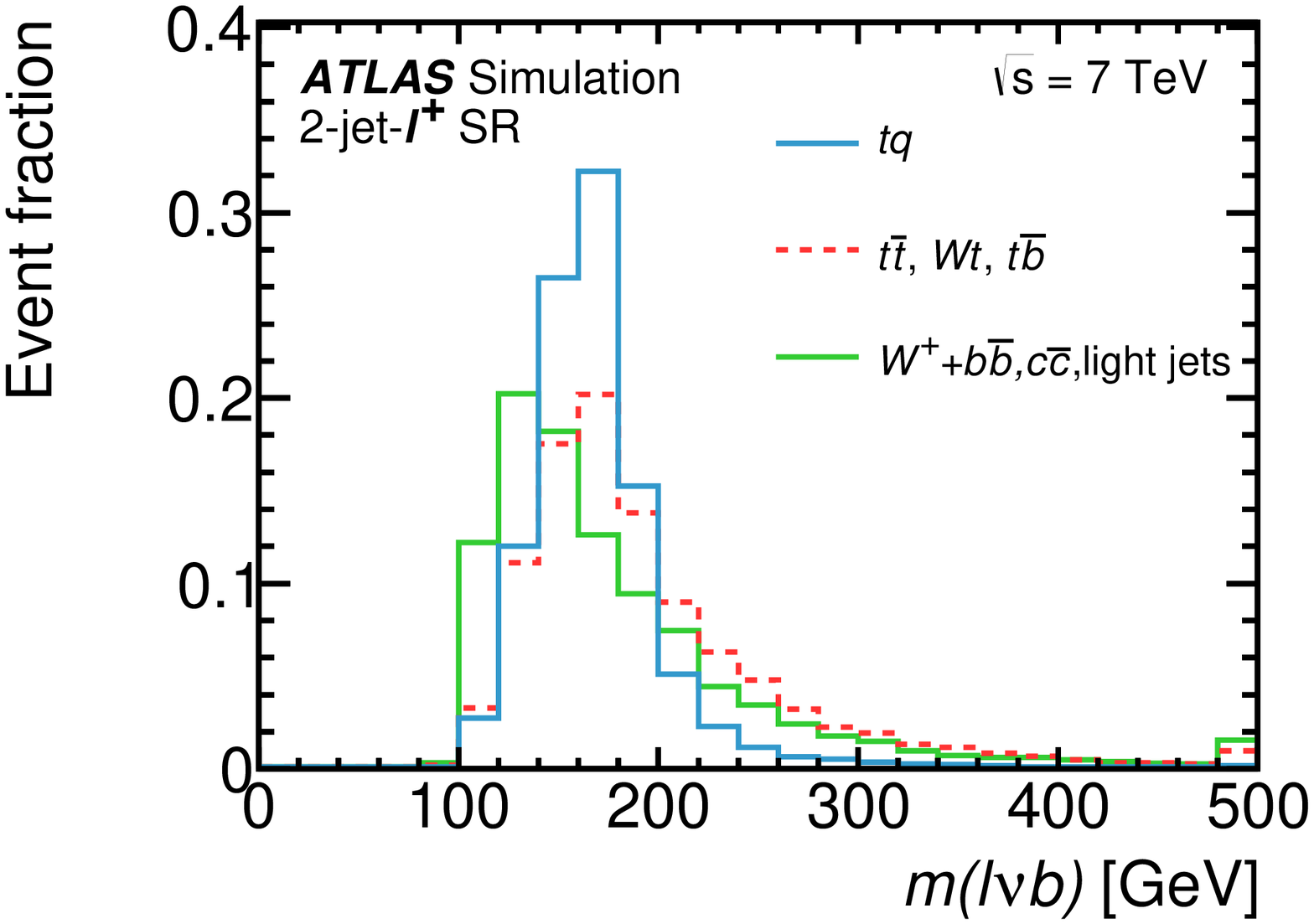}
     \label{subfig:2j_plus_mlnubshape}
  }
  \caption{\label{fig:sep_input2j} Probability densities of the two most important discriminating variables in the $2$-jet channels,
     shown in the 2-jet-$\ell^+$ channel in the signal region (SR). 
     The distributions are normalized to unit area.
     The absolute value of the pseudorapidity of the untagged jet $|\eta(j)|$ is shown in~\subref{subfig:2j_plus_lightEtashape}, and
     the invariant mass of the reconstructed top quark $m(\ell\nu b)$ is shown 
     in~\subref{subfig:2j_plus_mlnubshape}.
  }
\end{figure*}

As a result of the optimization procedure in the $2$-jet channel, 13 kinematic variables 
are identified as inputs to the NN. In the $3$-jet channel, 11 variables are used. 
It was found that reducing the number of variables further would result in a considerable loss of sensitivity.
The input variables to the NNs are listed in Table~\ref{tab:inputVariables}. The separation between 
signal and the two most important backgrounds, the top-quark background and the combined 
$W$+light-jets, $W$+$c\bar{c}$, and $W$+$b\bar{b}$ background, is shown in Fig.~\ref{fig:sep_input2j} for the two most important discriminating variables in the 
$2$-jet channel.

\begin{table*}[p]
  \centering
  \caption{\label{tab:inputVariables}Input variables of the NNs in the $2$-jet channels and in the 
    $3$-jet channels.
    The definitions of the variables use the term {\it leading jet} and
    {\it 2$^{nd}$ leading jet}, defined as the jet with the highest or 
    2$^\mathrm{nd}$ highest $p_{\rm T}$, respectively. In the 2-jet channels, exactly one jet is required
    to be $b$-tagged. The jet that is not $b$-tagged is denoted {\it untagged} jet.
    }
  \begin{ruledtabular}
  \begin{tabular}{ll}
  \multicolumn{2}{c}{Variables used in the $2$-jet channels and the $3$-jet channels} \\ \hline
  $m(\ell\nu b)$ & The invariant mass of the reconstructed top quark. \\
  $m_{\mathrm{T}}(\ell\MET)$ & The transverse mass of the lepton--\MET system, as defined in 
  Eq.~(\ref{eq:mTW}). \\
  $\eta(\ell\nu)$ & The pseudorapidity of the system of the lepton and the reconstructed neutrino. \\
  $m(\ell b)$ & The invariant mass of the charged lepton and the $b$-tagged jet. \\
  $H_{\mathrm{T}}$ & The scalar sum of the transverse momenta of the jets, the charged lepton,
  and the \MET. \\
                & \\ 
  \multicolumn{2}{c}{Variables used in the $2$-jet channels only} \\ \hline
  $m(jb)$ & The invariant mass of the untagged jet and the $b$-tagged jet. \\
  $|\eta(j)|$ &  The absolute value of the pseudorapidity of the untagged jet. \\
  $\Delta R\left(\ell,j\right)$ & $\Delta R$ between the charged lepton and the untagged jet. \\
  $\Delta R\left(\ell\nu b,j\right)$ & $\Delta R$ between the reconstructed top quark and the 
  untagged jet. \\
  $|\eta\left(b\right)|$ & The absolute value of the pseudorapidity of the $b$-tagged jet. \\
  $|\Delta p_{\mathrm{T}}\left(\ell,j\right)|$ & The absolute value of the difference between 
  the transverse momentum of the charged lepton and the untagged jet. \\
  $|\Delta p_{\mathrm{T}}\left(\ell\nu b,j\right)|$ & The absolute value of the difference between
  the transverse momentum of the reconstructed top quark and \\
   & the untagged jet. \\ 
  \MET & The missing transverse momentum. \\
                & \\ 
  \multicolumn{2}{c}{Variables used in the $3$-jet channels only} \\ \hline
  $|\Delta y\left(j_{1},j_{2}\right)|$ & The absolute
    value of the rapidity difference of the leading and 2$^\mathrm{nd}$ leading jets. \\
  $m\left(j_{2}j_{3}\right)$ & The invariant mass of the 2$^\mathrm{nd}$ leading jet and 
  the 3$^\mathrm{rd}$ leading jet. \\
  $\cos\theta\left(\ell,j\right)_{\ell\nu b \; \mathrm{ r. f.}}$ & The cosine of the angle $\theta$ 
  between the charged lepton and the leading untagged jet in the rest frame \\
     & of the reconstructed top quark. \\
  $\Sigma\eta\left(j_{i}\right)$ & The sum of the pseudorapidities of all jets in the event. \\
  $m\left(j_{1}j_{2}\right)$ & The invariant mass of the two leading jets. \\ 
  $p_{\mathrm{T}}\left(\ell\nu b\right)$ & The transverse momentum of the 
  reconstructed top quark. \\ 
  \end{tabular}
  \end{ruledtabular}
\end{table*}   

\begin{figure*}[p]
  \centering
  \subfigure[]{
     \includegraphics[width=0.45\textwidth]{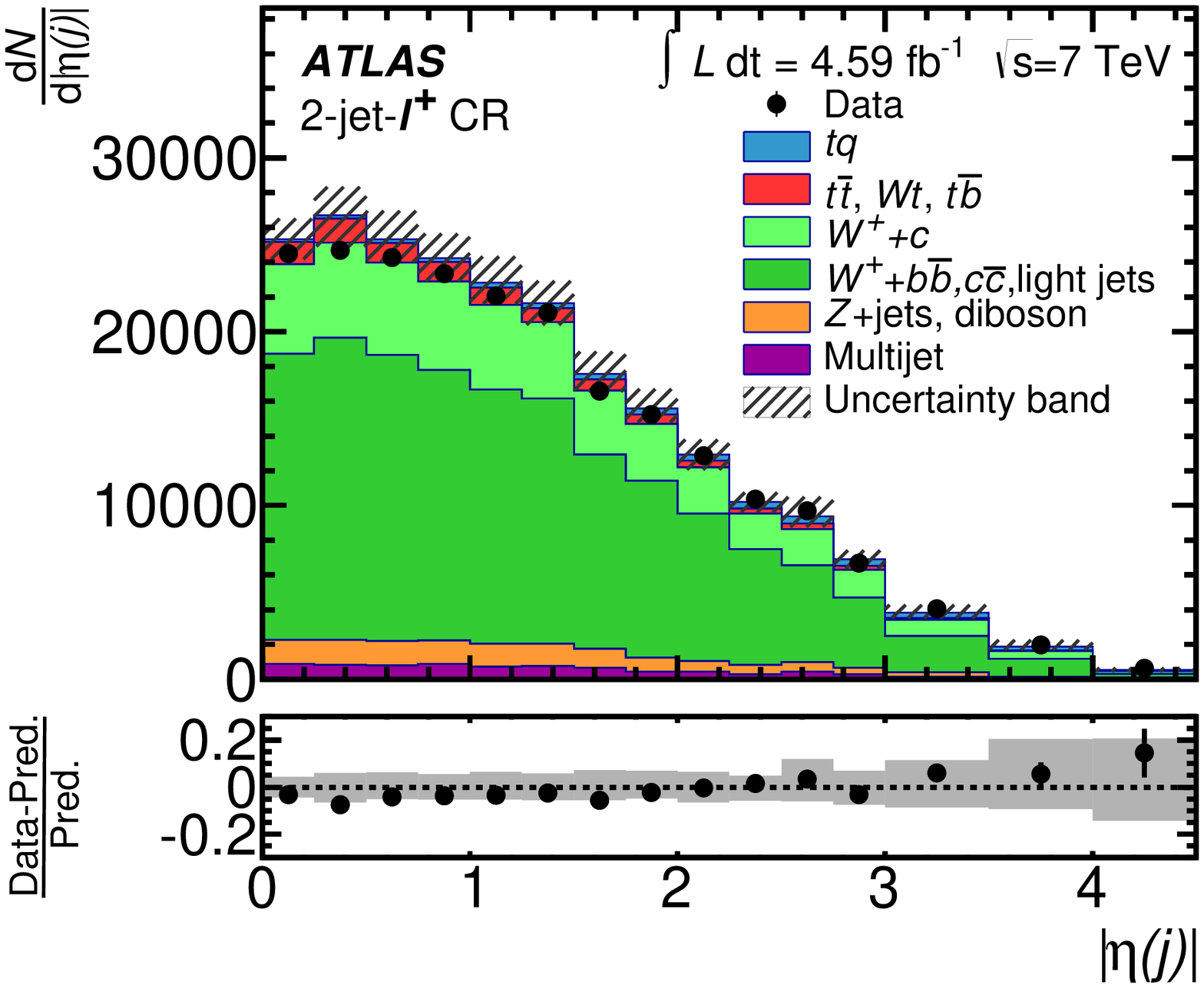}
     \label{subfig:2j_cr_plus_lighteta}
  }
  \subfigure[]{
     \includegraphics[width=0.45\textwidth]{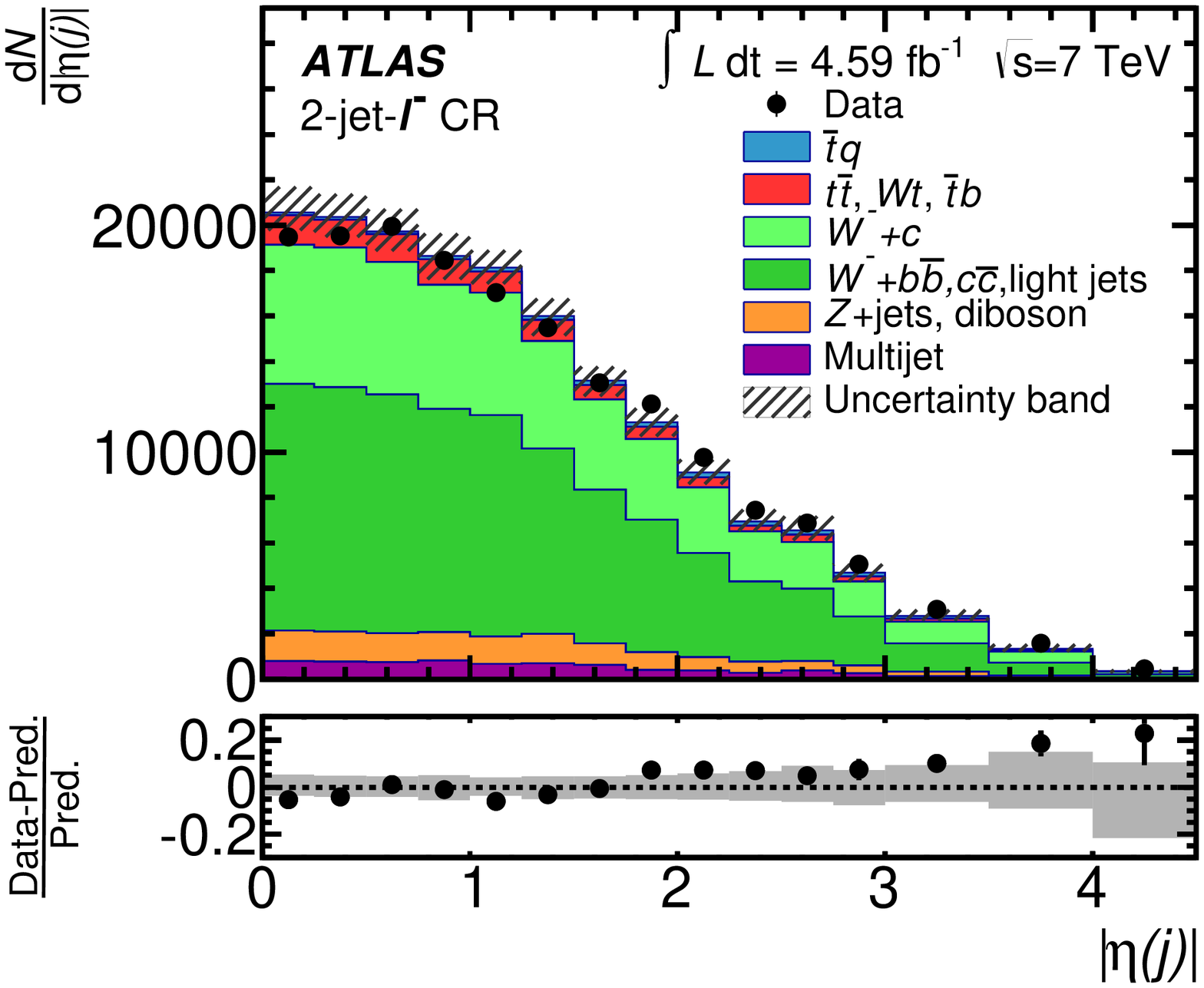}
     \label{subfig:2j_cr_minus_lighteta}
  }
  \subfigure[]{
     \includegraphics[width=0.45\textwidth]{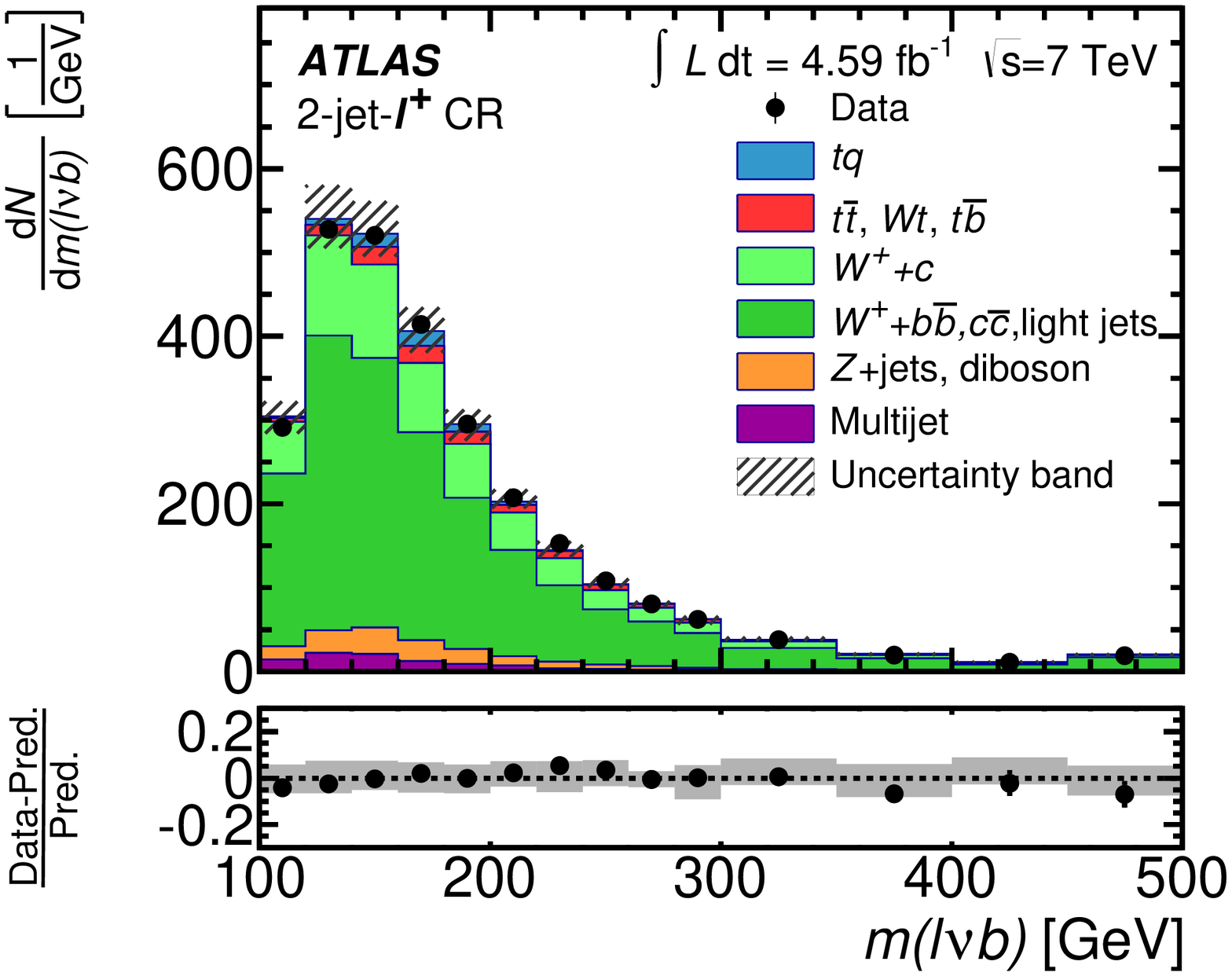}
     \label{subfig:2j_cr_plus_mlnub}
  }
  \subfigure[]{
     \includegraphics[width=0.45\textwidth]{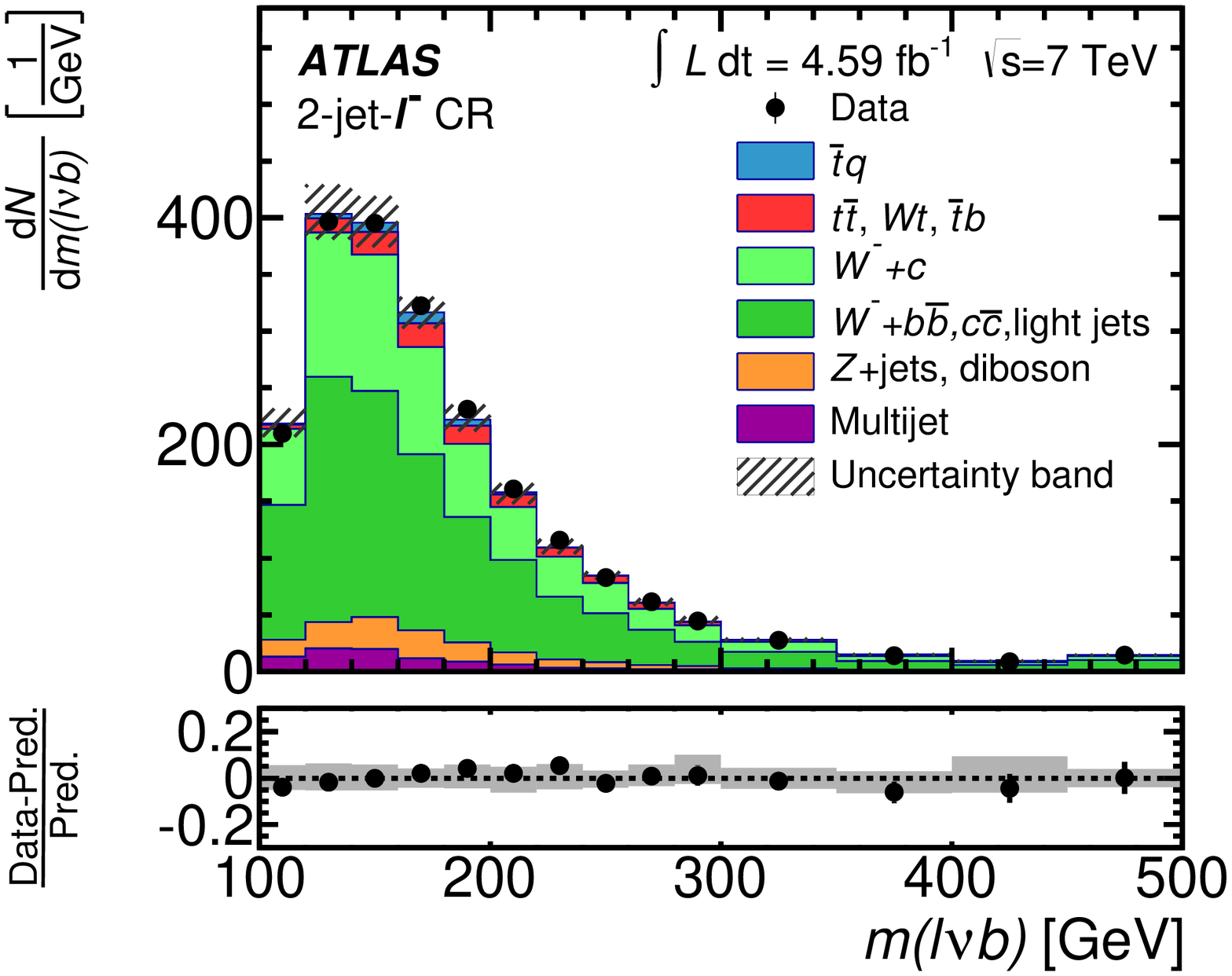}
     \label{subfig:2j_cr_minus_mlnub}
  }
  \subfigure[]{
     \includegraphics[width=0.45\textwidth]{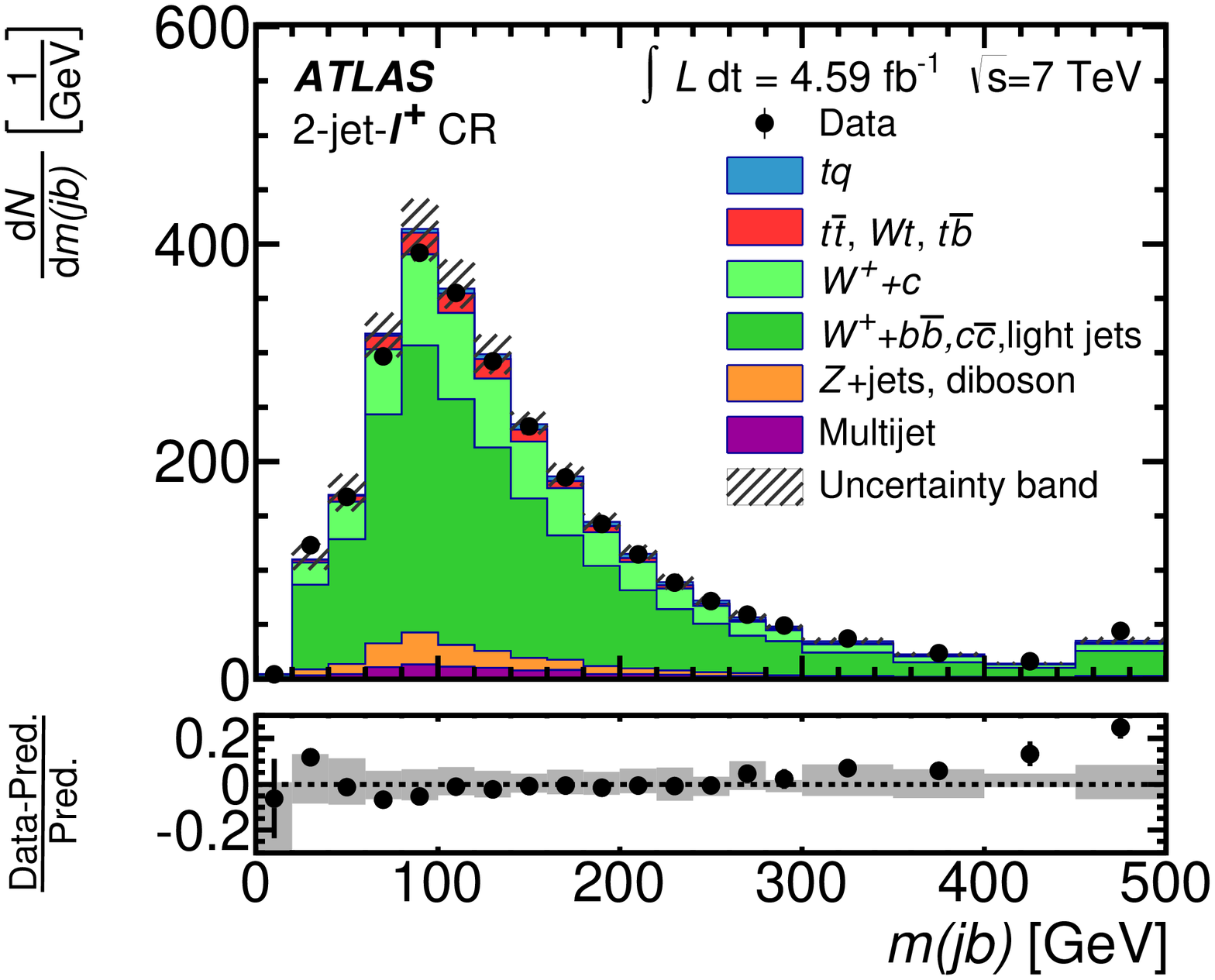}
     \label{subfig:2j_cr_plus_mj1j2}  
  }
  \subfigure[]{
     \includegraphics[width=0.45\textwidth]{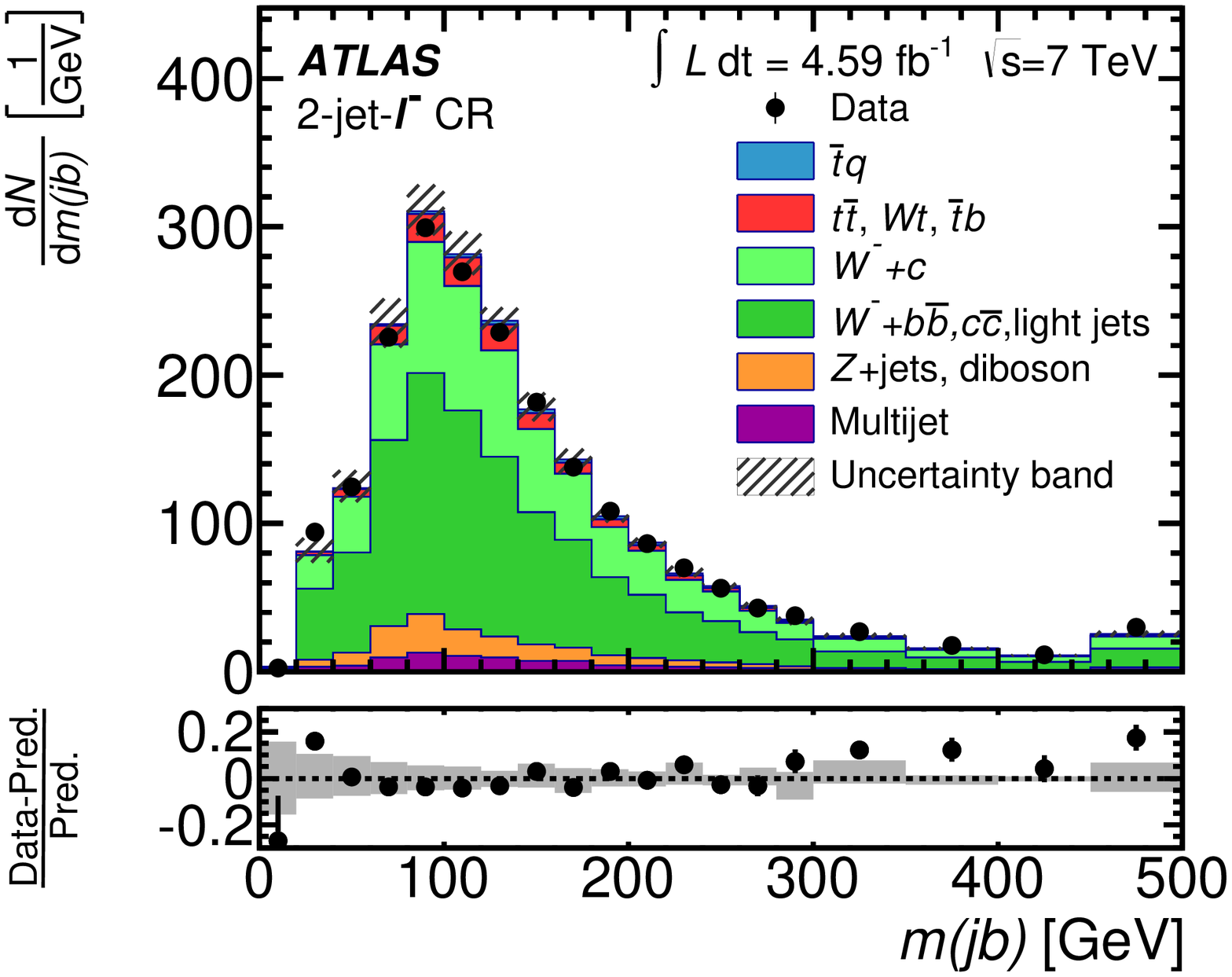}
     \label{subfig:2j_cr_minus_mj1j2}  
  }
  \caption{\label{fig:ctrl_reg_2j_vars} Distributions of the three most important discriminating 
    variables in the 2-jet-$\ell^+$ and 2-jet-$\ell^-$ channels in the control region~(CR). 
    Figures~\subref{subfig:2j_cr_plus_lighteta} and \subref{subfig:2j_cr_minus_lighteta} display the
    absolute value of the pseudorapidity of the untagged jet $|\eta(j)|$. 
    Figures~\subref{subfig:2j_cr_plus_mlnub} and \subref{subfig:2j_cr_minus_mlnub} show the invariant mass of the reconstructed top quark 
    $m(\ell\nu b)$,
    \subref{subfig:2j_cr_plus_mj1j2} and
    \subref{subfig:2j_cr_minus_mj1j2} the invariant mass of the untagged and the $b$-tagged jet
    $m(jb)$.
    The last histogram bin includes overflows. 
    The multijet and the $W+$jets event yields are determined by a fit to the $\MET$ distribution
    as described in Sec.~\ref{sec:QCDestimate}.
    The uncertainty band represents the normalization uncertainty due to the uncertainty on the
    jet energy scale and the Monte Carlo statistical uncertainty. The relative difference between the observed and expected number of events 
    in each bin is shown in the lower panels.
    }
\end{figure*}

\begin{figure*}[p]
  \centering
  \subfigure[]{
     \includegraphics[width=0.45\textwidth]{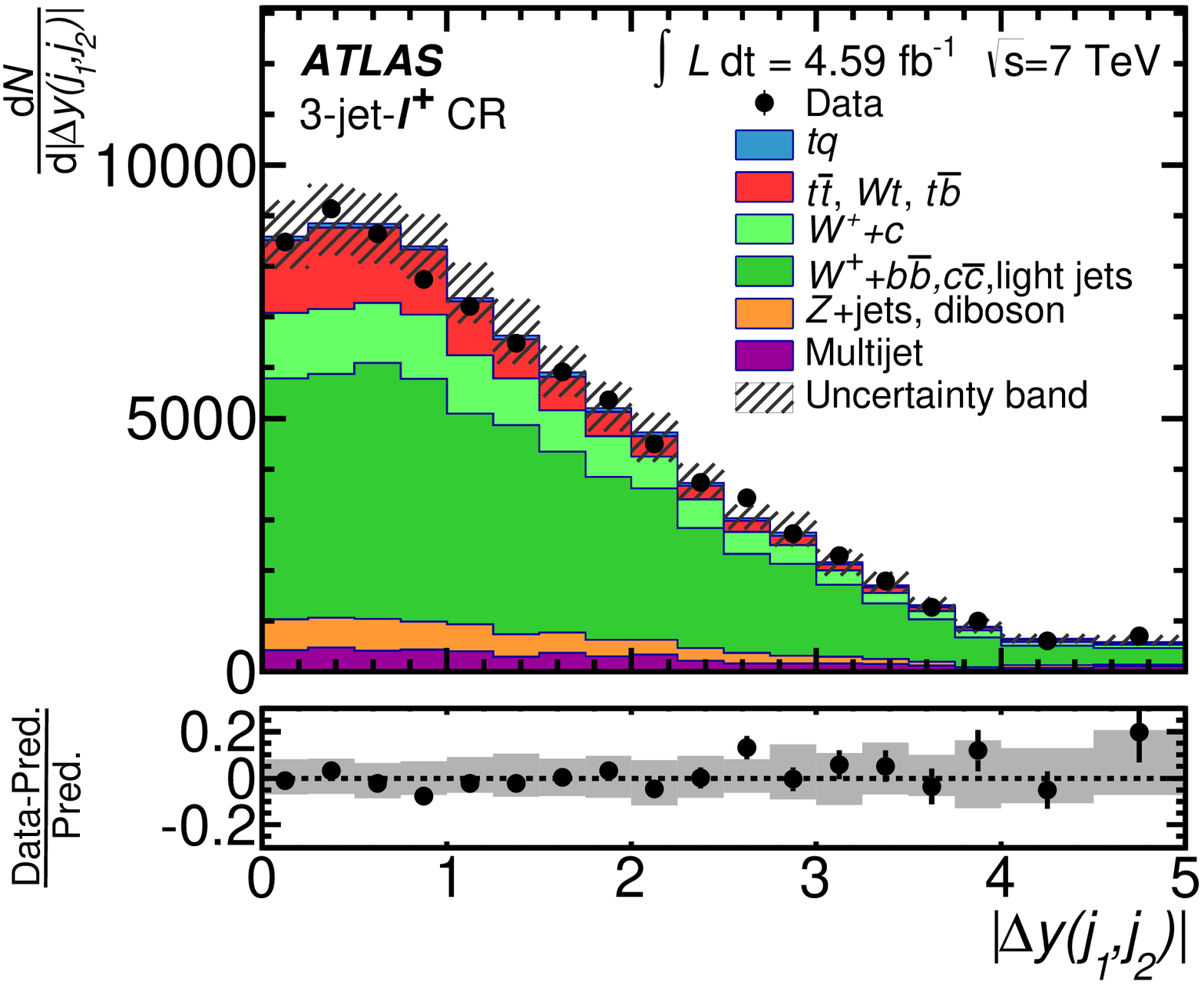}
     \label{subfig:3j_cr_plus_dyj1j2}
  }
  \subfigure[]{
     \includegraphics[width=0.45\textwidth]{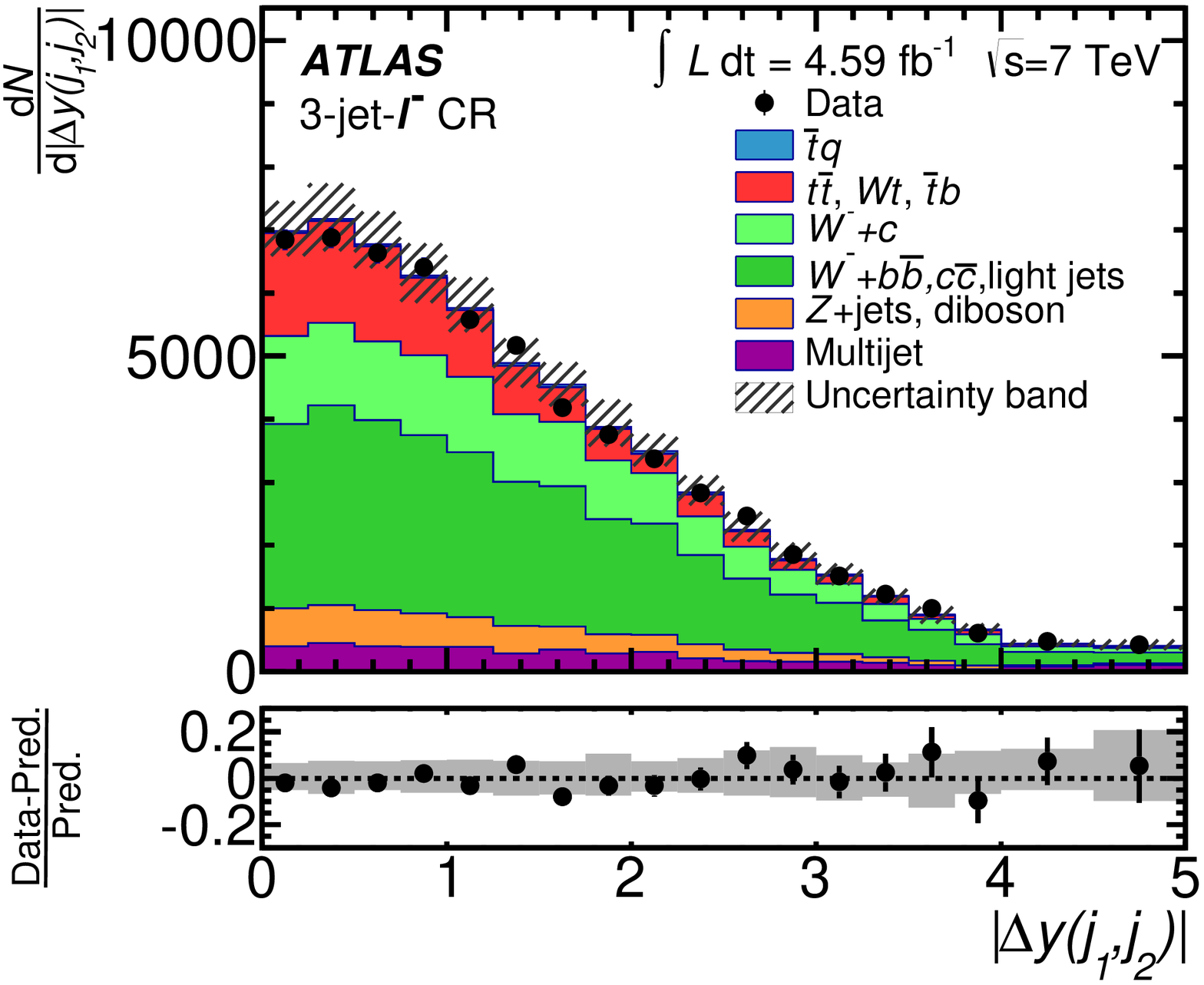}
     \label{subfig:3j_cr_minus_dyj1j2}
  }
  \subfigure[]{
     \includegraphics[width=0.45\textwidth]{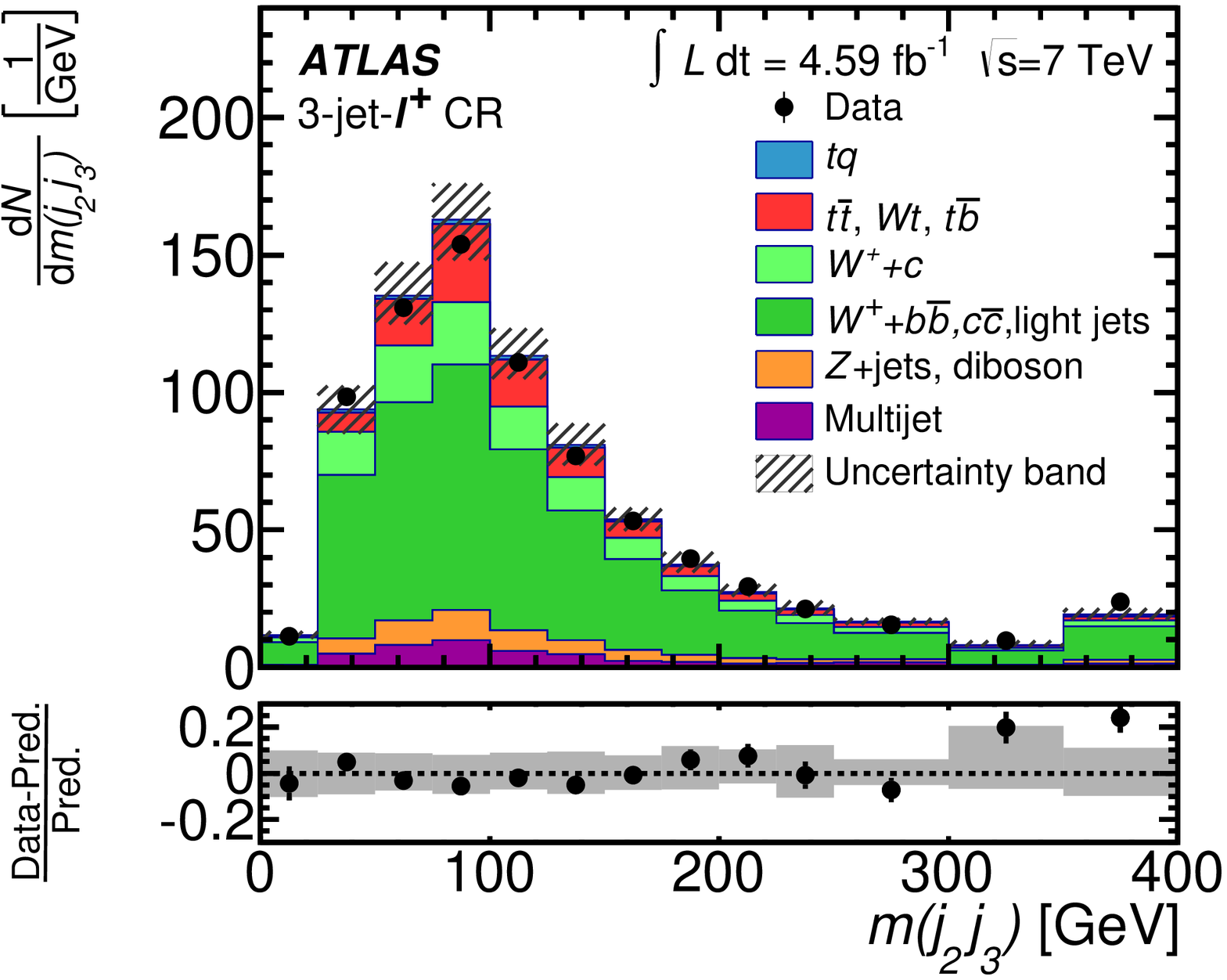}
     \label{subfig:3j_cr_plus_mj2j3}  
  }
  \subfigure[]{
     \includegraphics[width=0.45\textwidth]{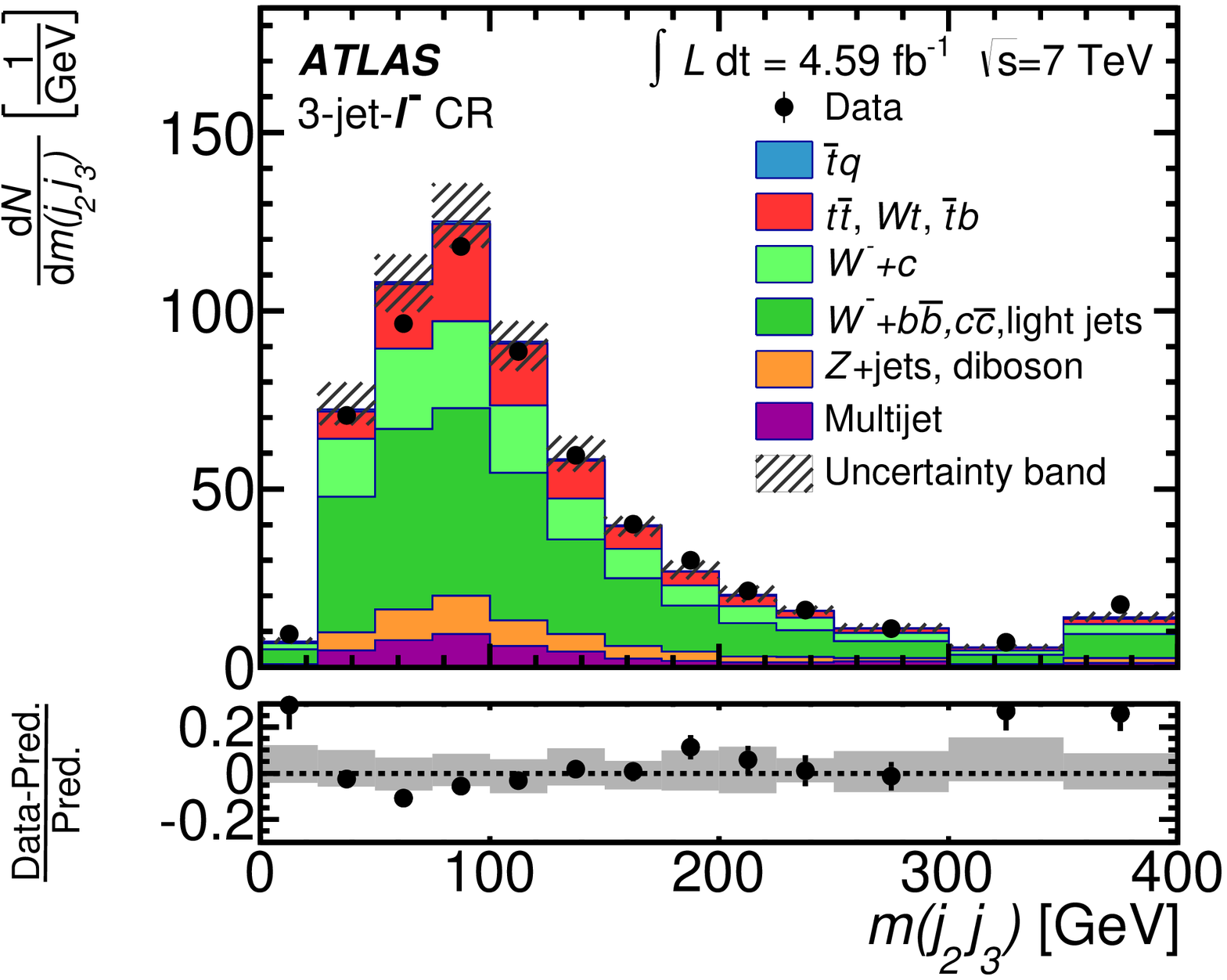}
     \label{subfig:3j_cr_minus_mj2j3}  
  }
  \subfigure[]{
     \includegraphics[width=0.45\textwidth]{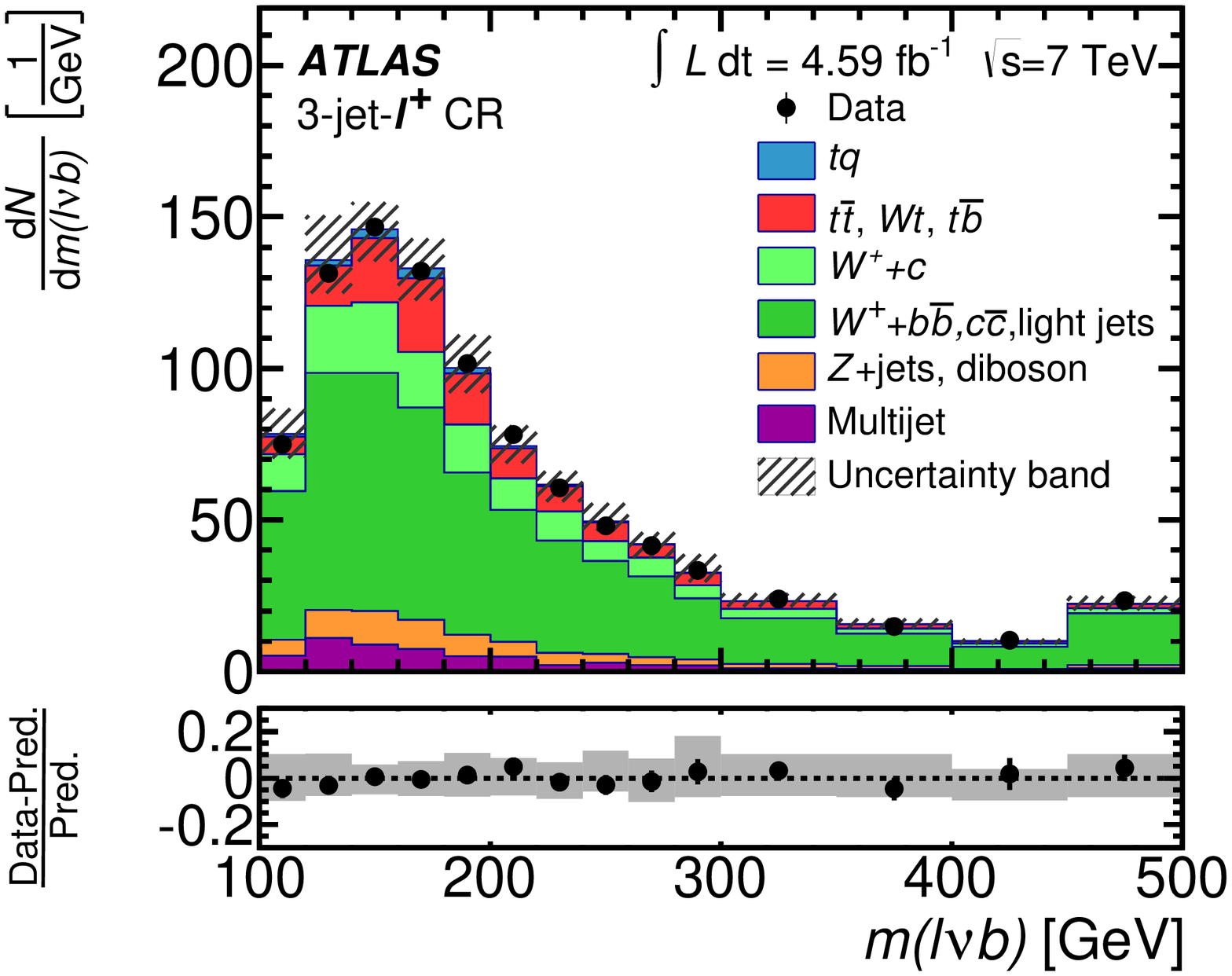}
     \label{subfig:3j_cr_plus_mlnub}
  }
  \subfigure[]{
     \includegraphics[width=0.45\textwidth]{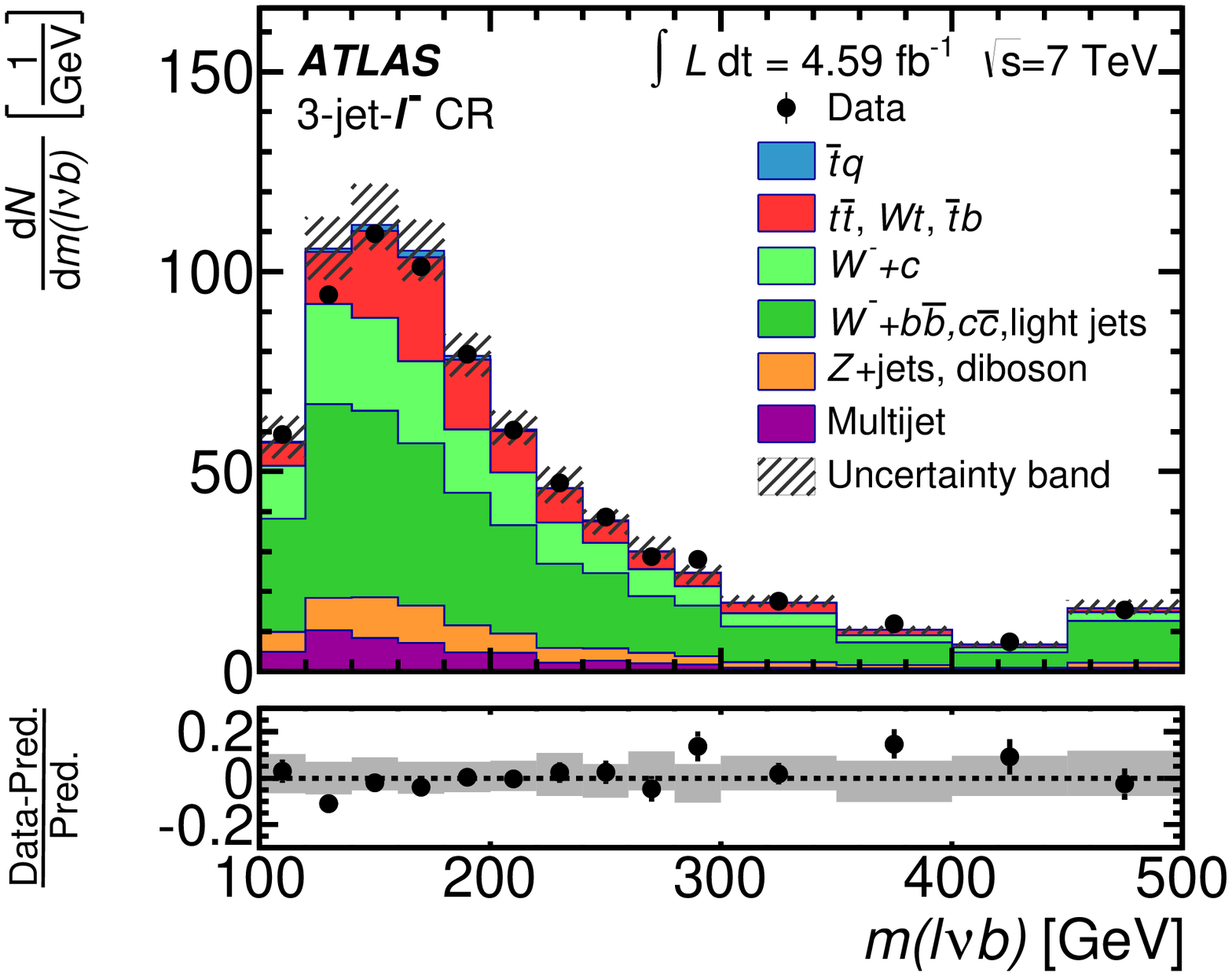}
     \label{subfig:3j_cr_minus_mlnub}
  }
  \caption{\label{fig:ctrl_reg_3j_vars} Distributions of the three most important discriminating 
    variables in the 3-jet-$\ell^+$ and 3-jet-$\ell^-$ channels in the control region~(CR).
    Figures \subref{subfig:3j_cr_plus_dyj1j2} and \subref{subfig:3j_cr_minus_dyj1j2} display the absolute
    value of the rapidity difference of the leading and 2$^\mathrm{nd}$ leading jet 
    $|\Delta y\left(j_{1},j_{2}\right)|$,
    \subref{subfig:3j_cr_plus_mj2j3} and \subref{subfig:3j_cr_minus_mj2j3} the invariant mass of 
    the 2$^\mathrm{nd}$ leading jet and the 3$^\mathrm{rd}$ jet $m\left(j_{2}j_{3}\right)$, and 
    \subref{subfig:3j_cr_plus_mlnub} and \subref{subfig:3j_cr_minus_mlnub} show the invariant mass 
    of the reconstructed top quark $m(\ell\nu b)$.
    The last histogram bin includes overflows. 
    The multijet and the $W+$jets event yields are determined by a fit to the $\MET$ distribution
    as described in Sec.~\ref{sec:QCDestimate}.
    The uncertainty band represents the normalization uncertainty due to the uncertainty on the
    jet energy scale and the Monte Carlo statistical uncertainty. The relative difference between the observed and expected number of events 
    in each bin is shown in the lower panels. 
    } 
\end{figure*}

The modeling of the input variables is checked in a control region (see Sec.~\ref{sec:selection}
for the definition) that is enriched in $W+$jets events. 
Figures~\ref{fig:ctrl_reg_2j_vars} and~\ref{fig:ctrl_reg_3j_vars} show the three most discriminating variables
in the 2-jet-$\ell^{\pm}$ and 3-jet-$\ell^{\pm}$-1-tag channels, respectively. 
Good modeling of the variables is observed.

\renewcommand\arraystretch{1.0} 

\subsection{Neural network training}
After choosing a set of variables based on the criteria outlined above, 
the analysis proceeds with the training of the NNs using 
a three-layer feed-forward architecture. 
The number of hidden nodes was chosen to be 15 for both networks.
Samples of simulated events are used for the training process, the
size of the signal samples in the $2$-jet channel being about 37,000 events for top-quark and 
about 40,000 events for top-antiquark $t$-channel production.
In the 3-jet channel the sizes of the training samples are 14,000 and 13,000 events,
respectively.
All background processes are used in the training, except for the multijet background
whose modeling is associated with large uncertainties.
The total number of simulated background events used in the training is about 89,000 in
the 2-jet channel and about 57,000 in the 3-jet channel. 
The ratio of signal events to background events in the training is 
chosen to be 1:1, while the different background processes are weighted relative to each other 
according to the number of expected events. 

Regularization techniques are applied in the training process to dampen statistical 
fluctuations in the training sample and to avoid overtraining. 
At the preprocessing stage mentioned above (Sec. \ref{subsec:prepro}), the input variables 
are transformed in several steps to define new input variables that are optimally 
prepared to be fed into an NN. First, the variables are transformed,
such that they populate a finite interval and are distributed according 
to a uniform distribution. The influence of outliers is thereby strongly reduced. 
The distributions of the transformed variables are discretized using 100 bins, and
the distributions for signal events are divided by the sum of signal and background 
events bin-by-bin, yielding the purity distributions in each variable. Next, these purity curves
are fitted with a regularized spline function, thereby yielding a continuous 
transformation from the original input variables to the purities. 
By means of the spline fit statistical fluctuations in the input variables are 
significantly reduced. Applying the continuous purity functions to the input variables yields
purity distributions that are further transformed, such that the distributions of the 
resulting variables are centered at zero and have an RMS of one. These variables are input
to the NNs.
In the training process, the network structure
is pruned to arrive at a minimal topology, i.e. statistically insignificant network 
connections and nodes are removed. 

\begin{figure*}[!t]
  \centering
  \subfigure[]{
     \includegraphics[width=0.45\textwidth]{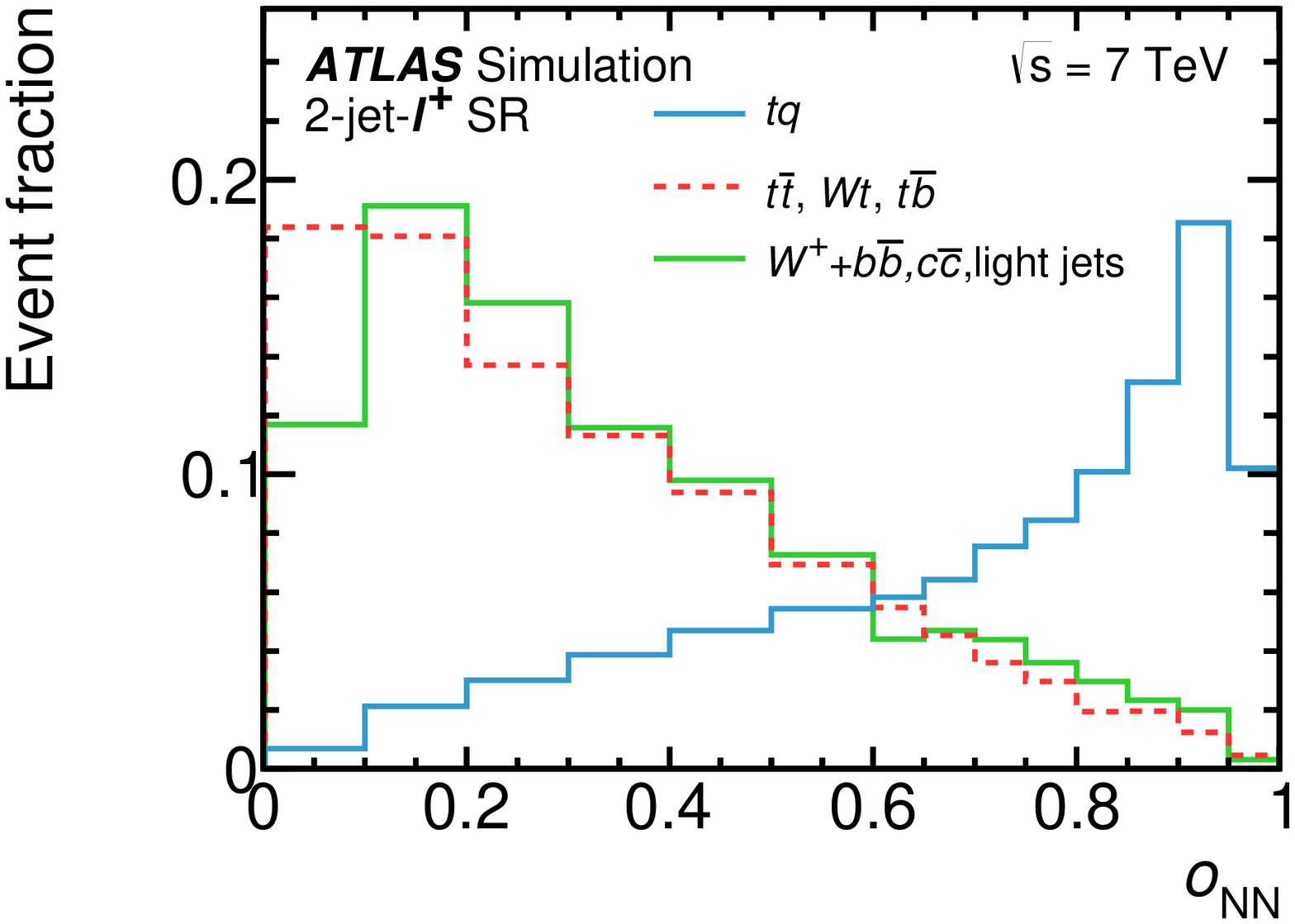}
     \label{subfig:2j_plus_NNshape}
  }
  \subfigure[]{
     \includegraphics[width=0.45\textwidth]{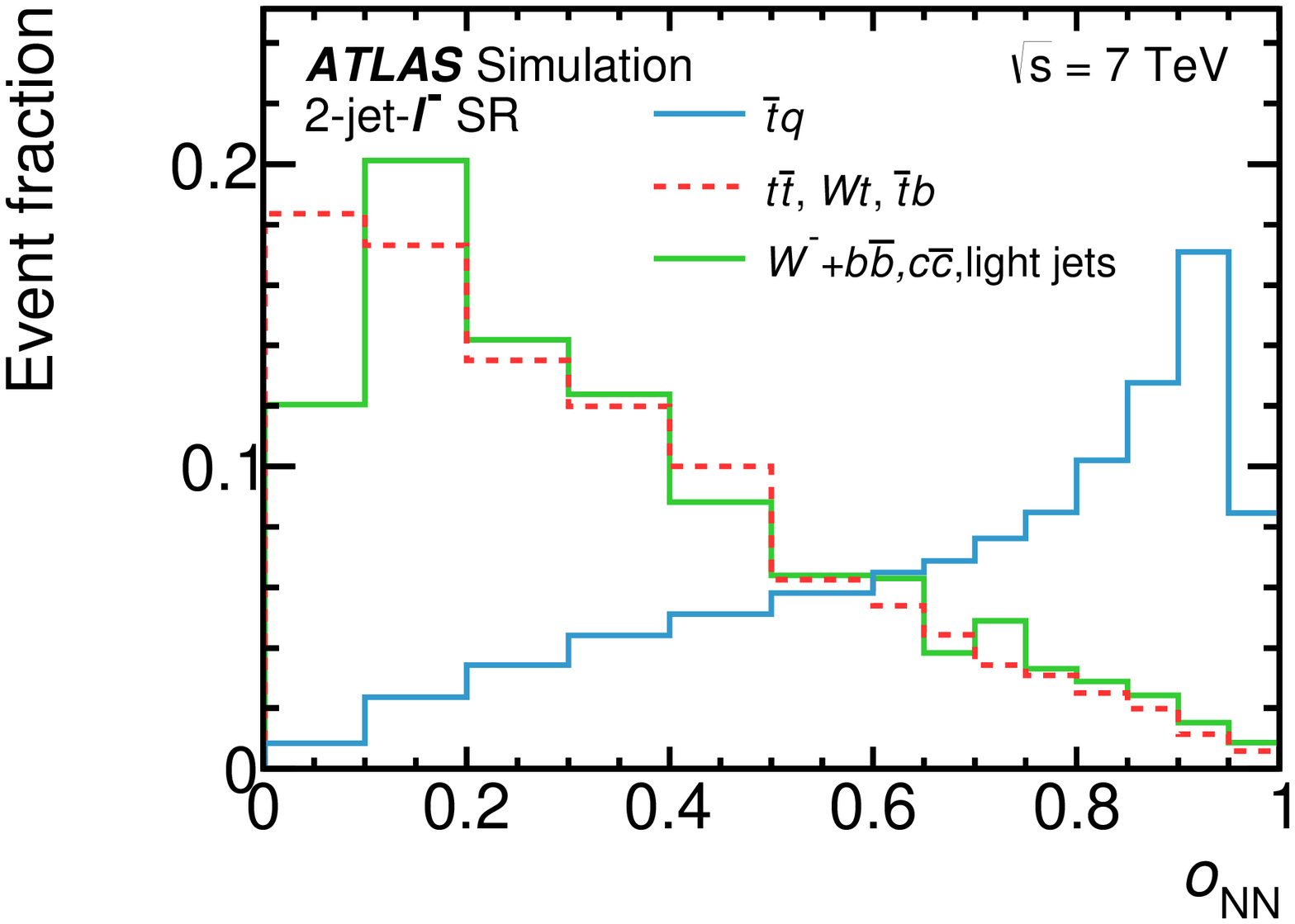}
     \label{subfig:2j_minus_NNshape}
  }
  \subfigure[]{
     \includegraphics[width=0.45\textwidth]{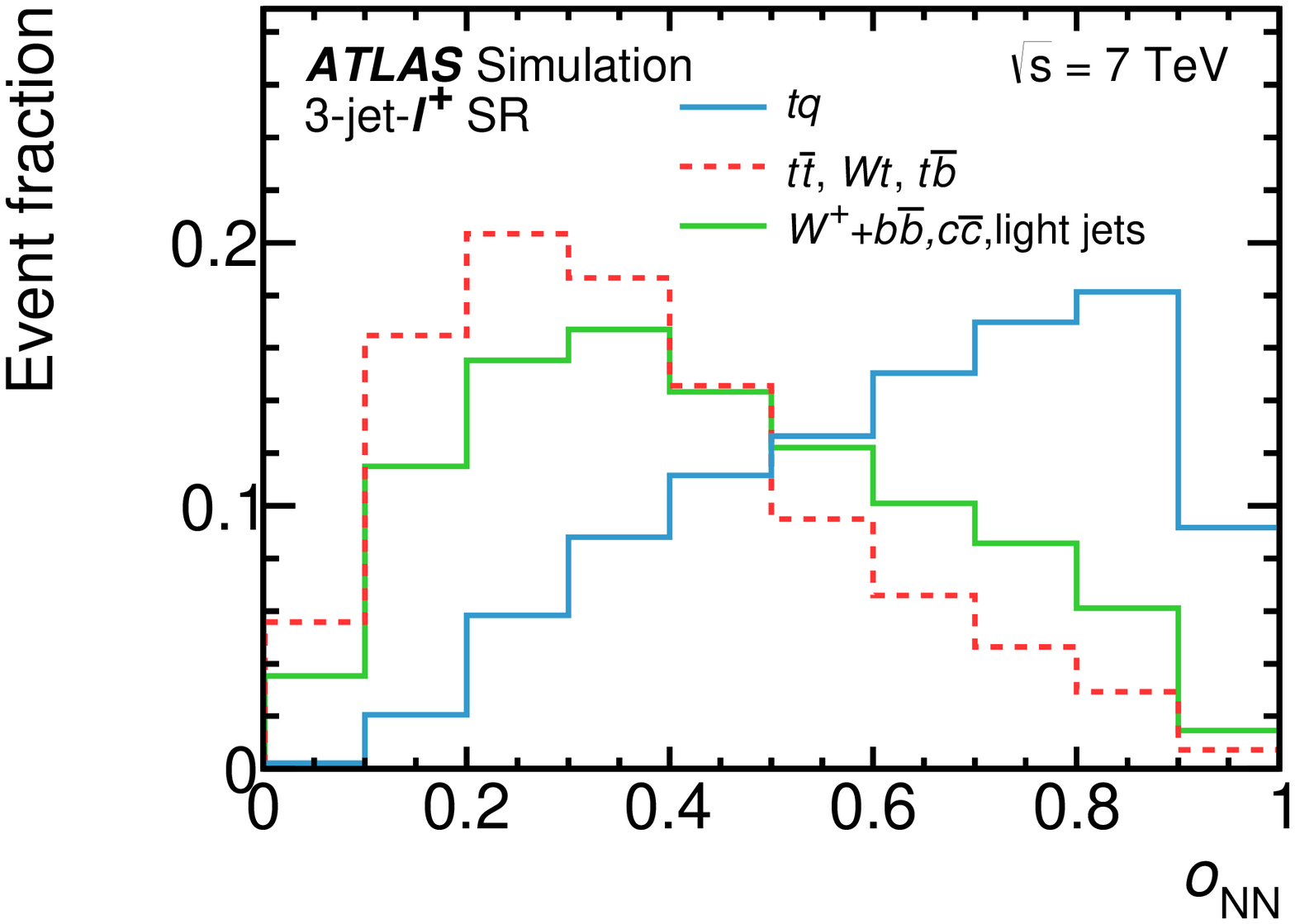}
     \label{subfig:3j_plus_NNshape}
  }
  \subfigure[]{
     \includegraphics[width=0.45\textwidth]{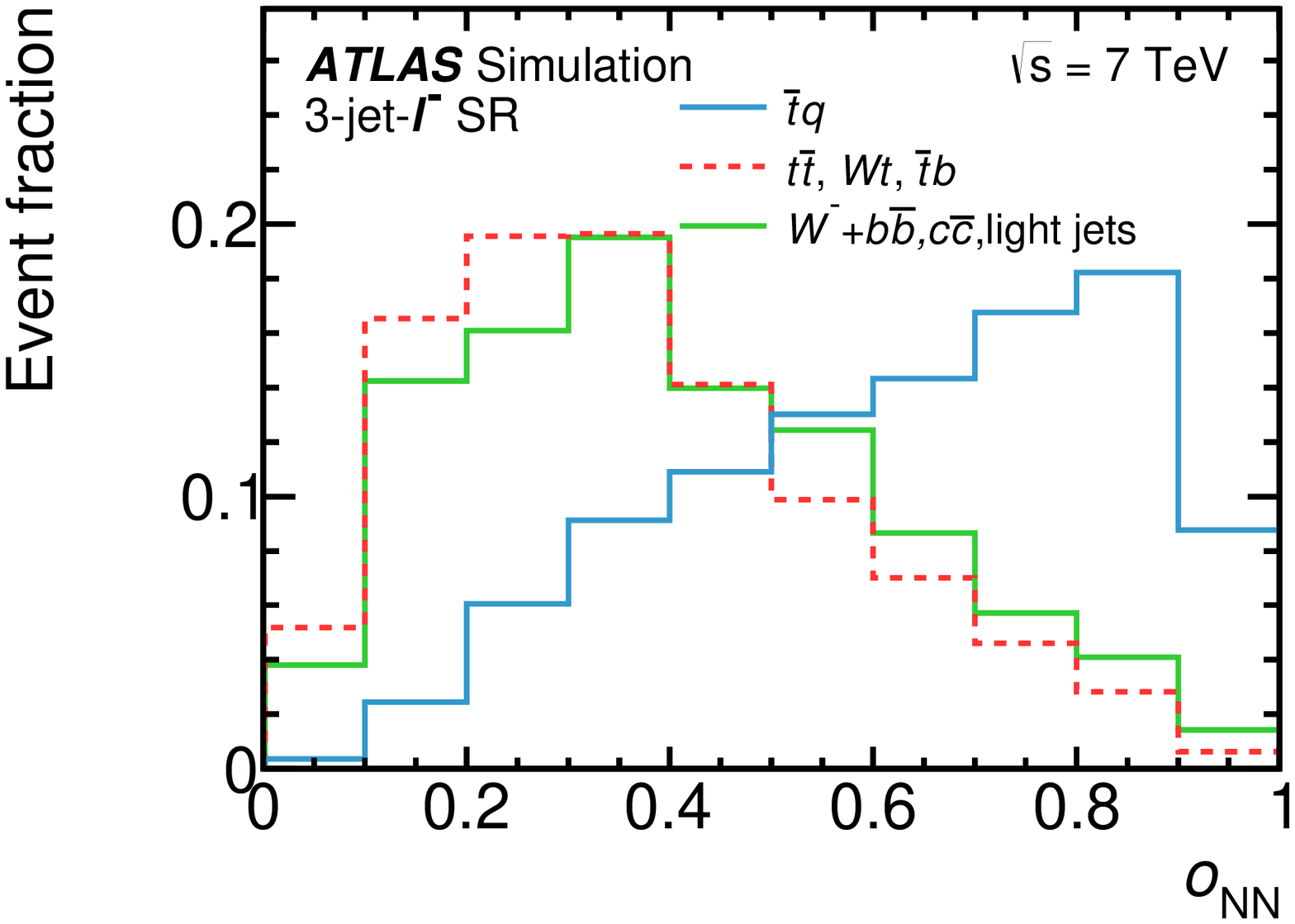}
     \label{subfig:3j_minus_NNshape}
  }
  \caption{\label{fig:nn_templates} Probability densities of the NN discriminants
     in the 2-jet channels and 3-jet channels in the signal region (SR): 
     \subref{subfig:2j_plus_NNshape} 2-jet-$\ell^+$ channel, 
     \subref{subfig:2j_minus_NNshape} 2-jet-$\ell^-$ channel,
     \subref{subfig:3j_plus_NNshape} 3-jet-$\ell^+$ channel, and
     \subref{subfig:3j_minus_NNshape} 3-jet-$\ell^-$ channel.
     The distributions are normalized to unit area.
  }
\end{figure*}

In Fig.~\ref{fig:nn_templates}, the probability densities of the resulting NN 
discriminants are shown for the signal, the top-quark backgrounds, and the combined 
$W$+light-jets, $W$+$c\bar{c}$, and $W$+$b\bar{b}$ background.
The separation between signal and backgrounds is equally good for the positive and the negative
charge channels, which demonstrates that the choice of training the NNs with a charge-combined 
sample is appropriate.

\subsection{Extraction of the signal yield}
\label{sec:fitresult}
The cross sections $\sigma(tq)$ and $\sigma(\bar{t}q)$ are extracted by performing a binned 
maximum-likelihood fit to the NN discriminant distributions in the 
2-jet-$\ell^+$, 2-jet-$\ell^-$, 3-jet-$\ell^+$-1-tag, and 3-jet-$\ell^-$-1-tag channels and 
to the event yield in the 3-jet-2-tag channel, treating 
$t$-channel top-quark and $t$-channel top-antiquark production as independent processes.
The signal rates, the rate of the combined top-quark background ($t\bar{t}$, $Wt$, $t\bar{b}$, 
and $\bar{t}b$), the rate of the combined 
$W$+light-jets, $W$+$c\bar{c}$, and $W$+$b\bar{b}$ background,
and the $b$-tagging efficiency correction factor (discussed in Sec.~\ref{sec:bjets}) are fitted in all channels simultaneously. 
The event yields of the multijet background and the $W$+$c$ background are not allowed 
to vary in the fit, but instead are fixed to the estimates given in Table~\ref{tab:evtyield}.
The cross-section ratio is subsequently computed as $R_t = \sigma(tq)/\sigma(\bar{t}q)$. 

The maximum-likelihood function is given by the product of Poisson probability terms for 
the individual histogram bins (see Ref.~\cite{Aad:2012ux}). 
Gaussian priors are added multiplicatively to the maximum-likelihood function to constrain 
the background rates subject to the fit and the correction factor of the $b$-tagging efficiency to their
predictions within the associated uncertainties. 

The sensitivity to the background rates is mostly given by the background-dominated region 
close to zero in the NN discriminant distributions, while the sensitivity
to the $b$-tagging efficiency stems from the event yield in the 3-jet-2-tag channel with respect
to the event yields in the 1-tag channels.

\begin{figure*}[p]
  \centering
  \subfigure[]{
     \includegraphics[width=0.45\textwidth]{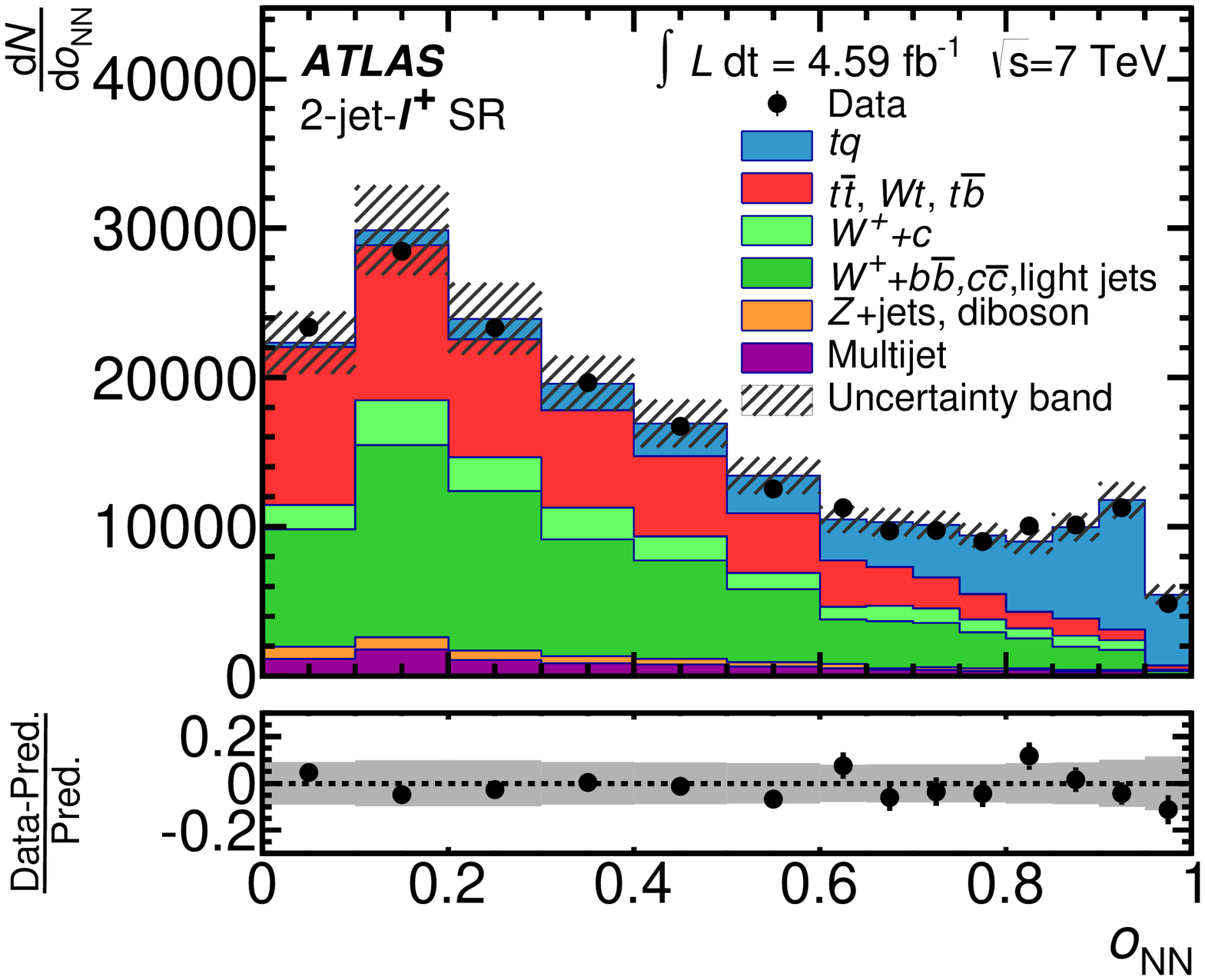}
     \label{fig:NN_fit_2jets_plus}
  }
  \subfigure[]{
    \includegraphics[width=0.45\textwidth]{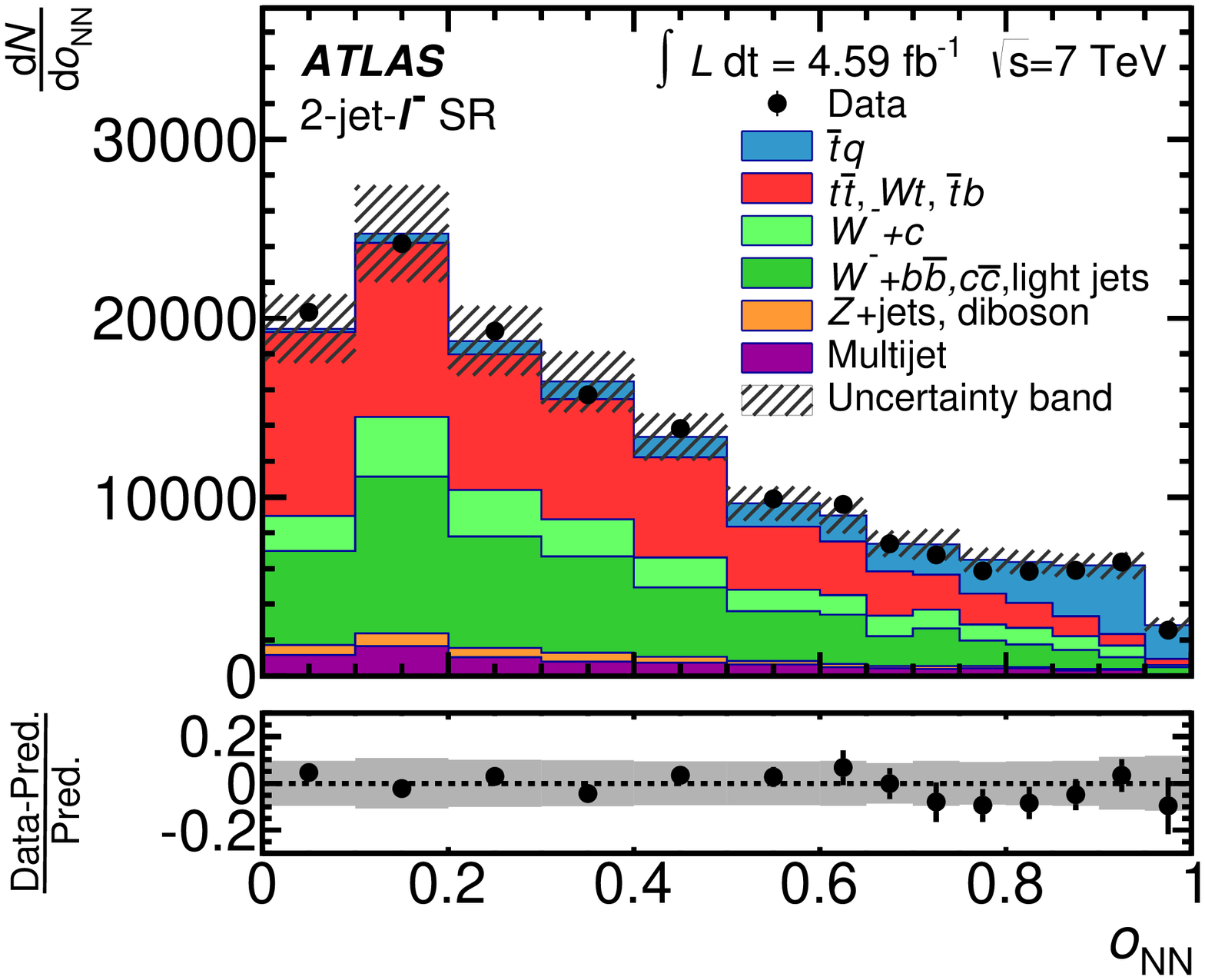}
     \label{fig:NN_fit_2jets_minus}
  }
  \subfigure[]{
    \includegraphics[width=0.45\textwidth]{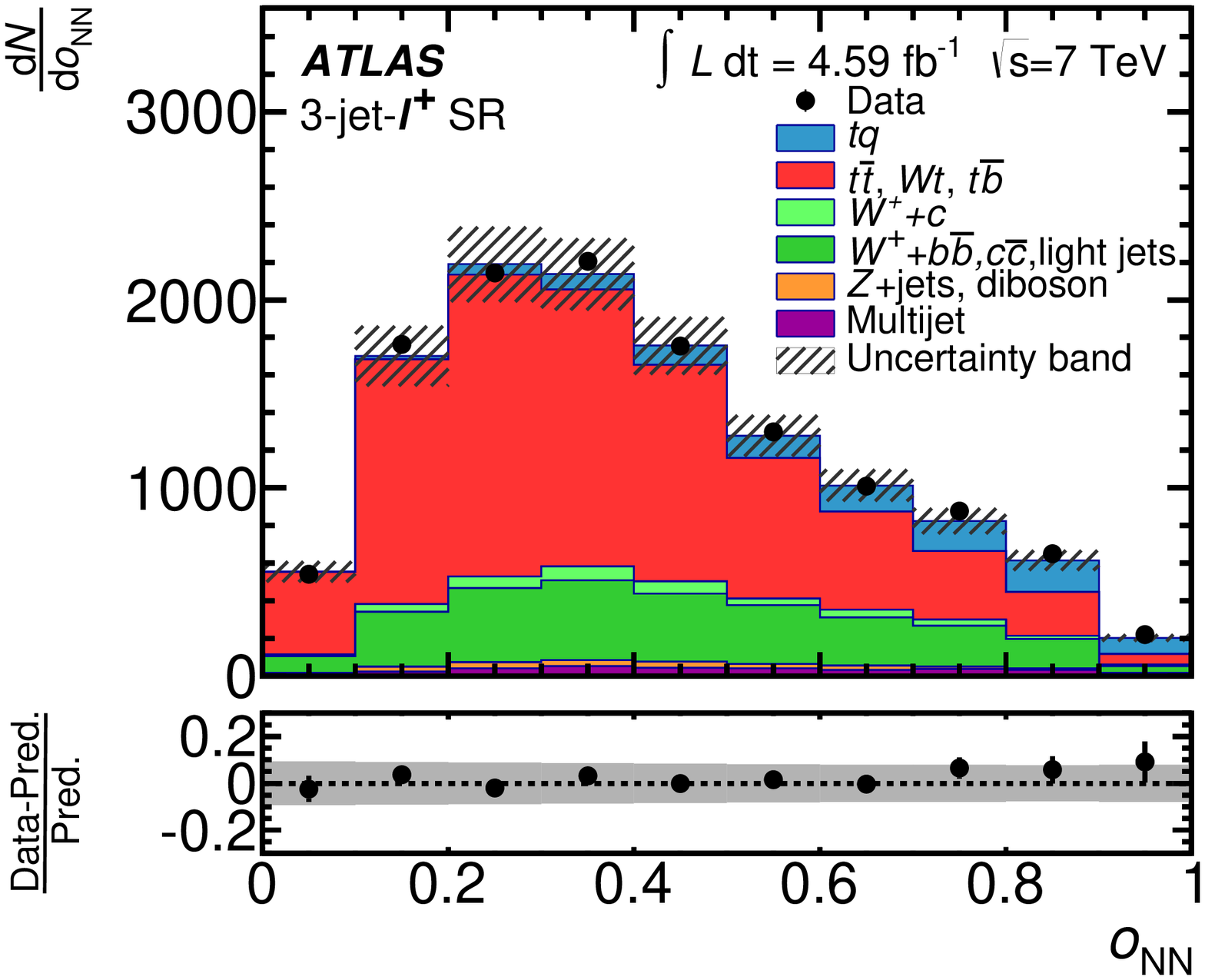}
     \label{fig:NN_fit_3jets_plus}
  }
  \subfigure[]{
    \includegraphics[width=0.45\textwidth]{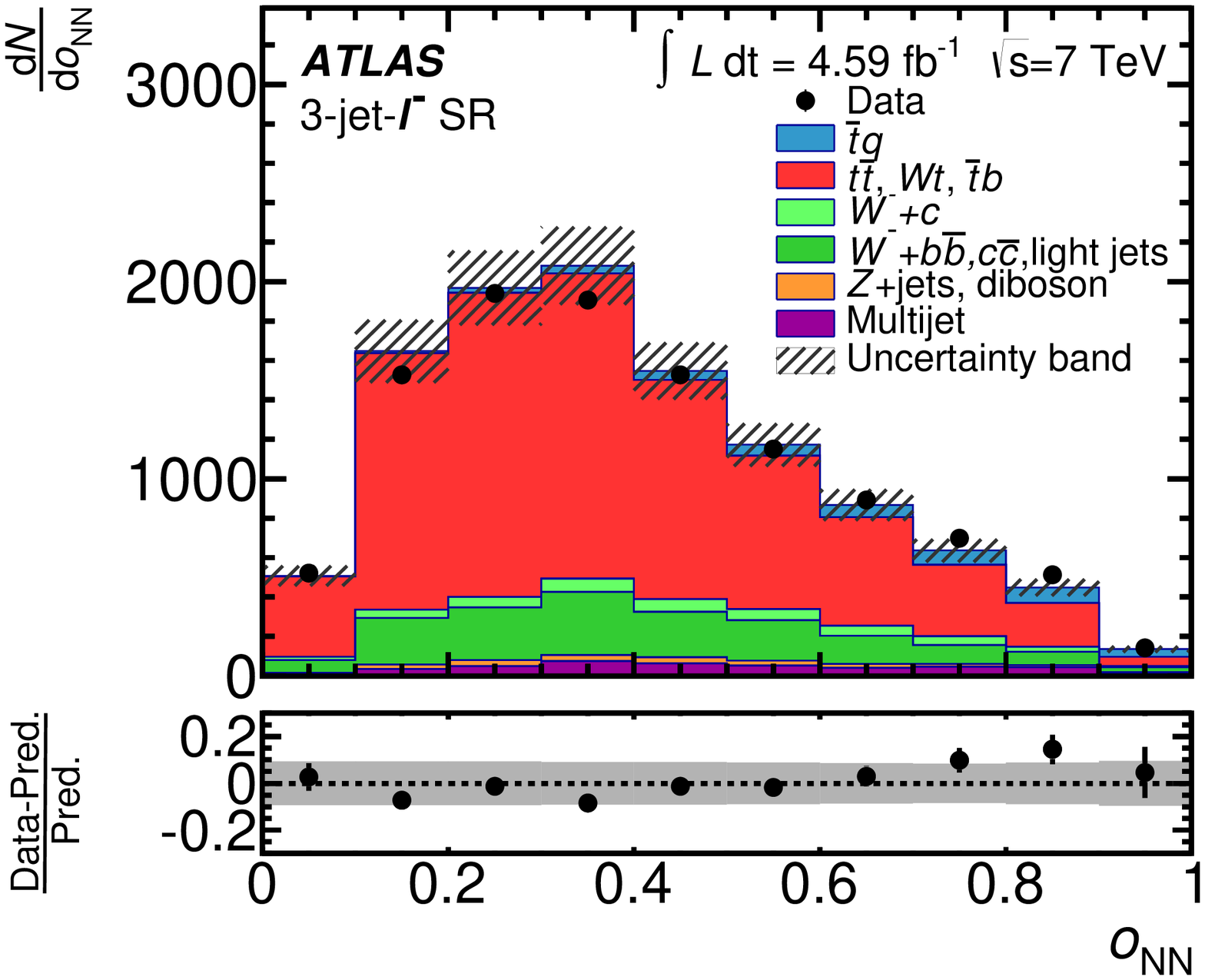}
     \label{fig:NN_fit_3jets_minus}
  }
  \caption{\label{fig:nnout_fit_results} 
  Neural network discriminant distributions normalized to the result of the binned 
  maximum-likelihood fit in \subref{fig:NN_fit_2jets_plus} the 2-jet-$\ell^+$ channel, 
  \subref{fig:NN_fit_2jets_minus} the 2-jet-$\ell^-$ channel, 
  \subref{fig:NN_fit_3jets_plus} the 3-jet-$\ell^+$ channel, and 
  \subref{fig:NN_fit_3jets_minus} the 3-jet-$\ell^-$ channel.
  The uncertainty band represents the normalization uncertainty of all processes 
  after the fit and the Monte Carlo statistical uncertainty, added in quadrature. The relative difference
  between the observed and expected number of events in each bin is shown in the lower panels.
  }
\end{figure*}

\begin{figure*}[p]
  \centering
  \subfigure[]{
  \includegraphics[width=0.45\textwidth]{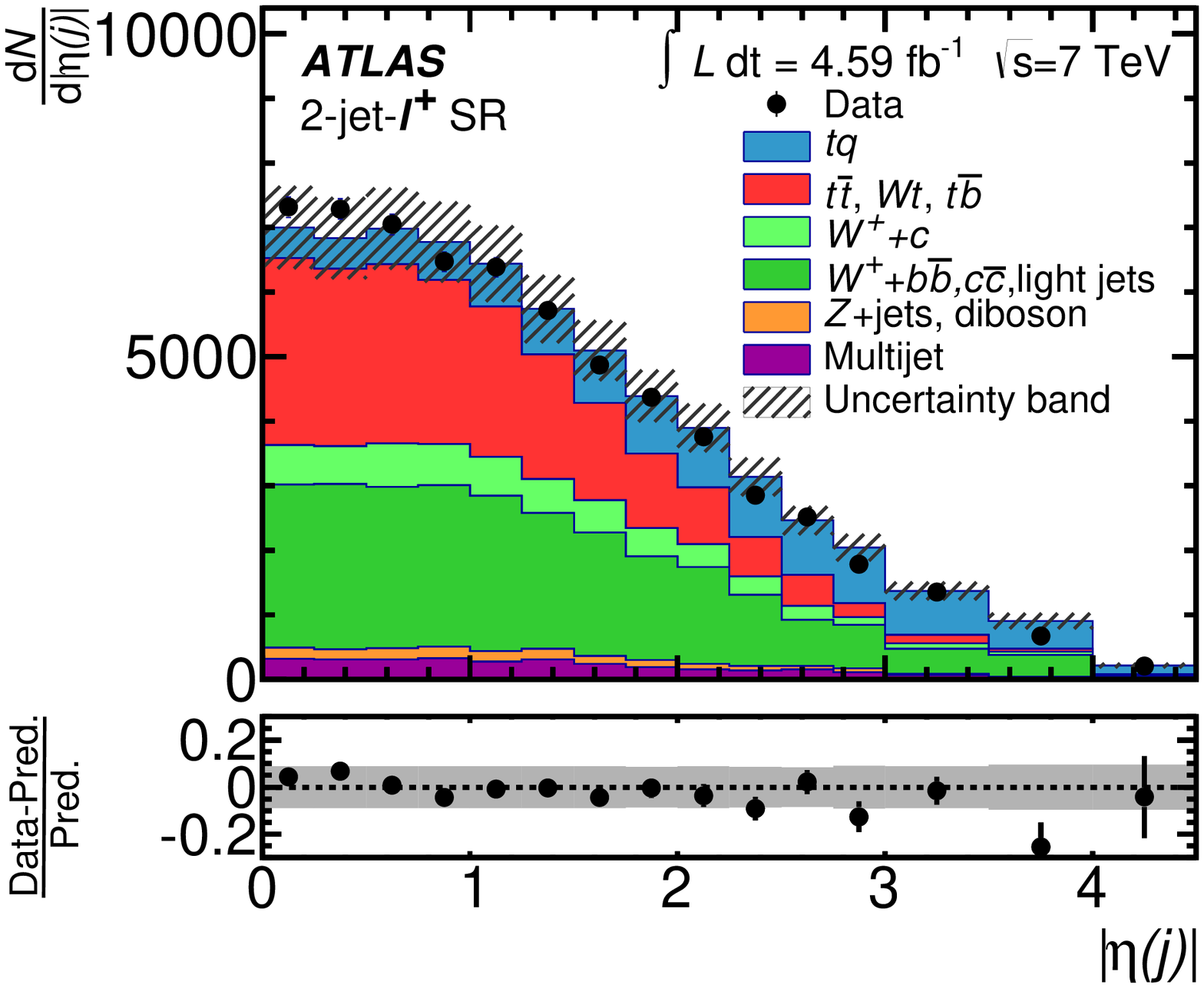}
     \label{subfig:2j_sr_plus_lighteta}
  }
  \subfigure[]{
  \includegraphics[width=0.45\textwidth]{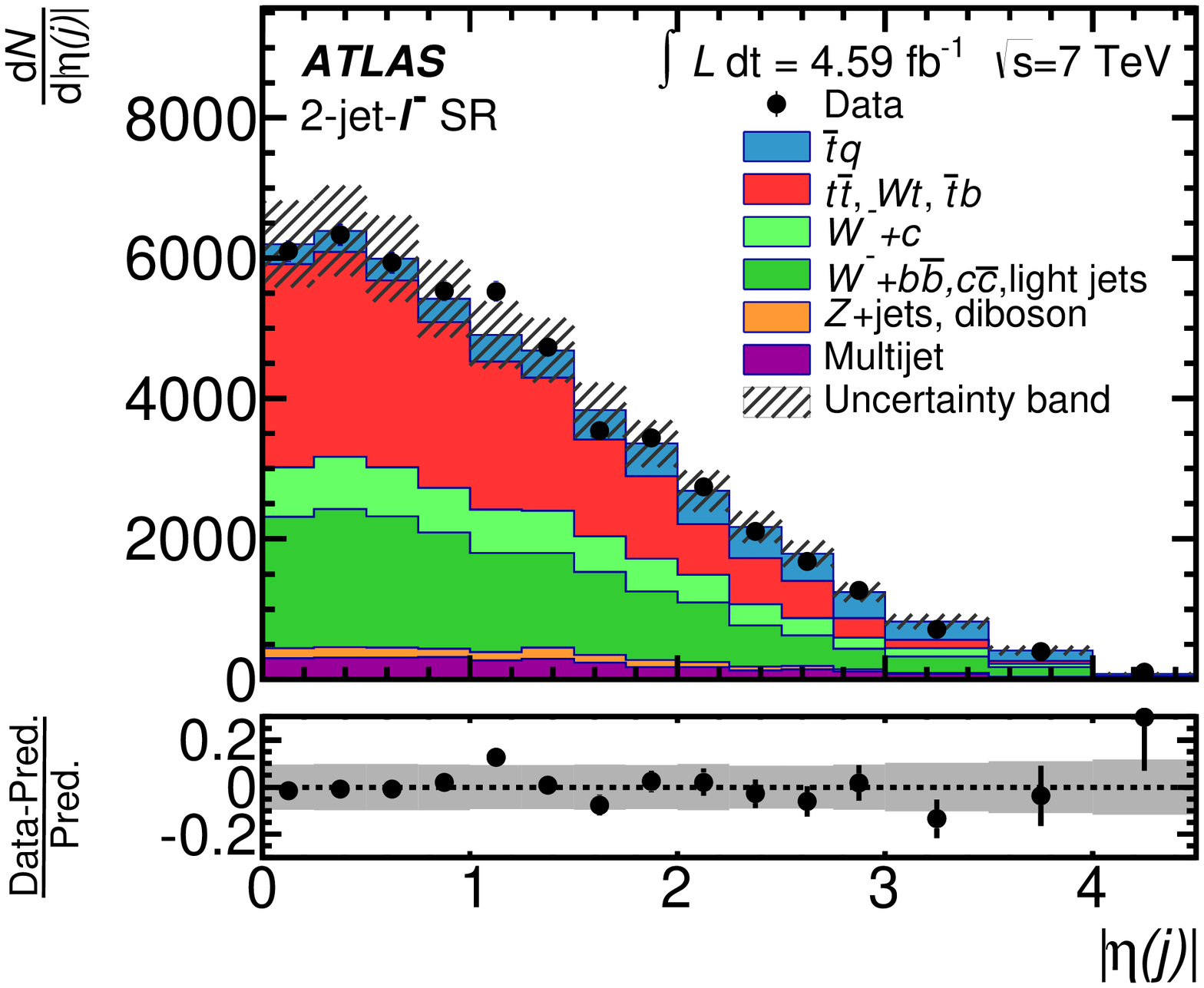}
     \label{subfig:2j_sr_minus_lighteta}
  }
  \subfigure[]{
  \includegraphics[width=0.45\textwidth]{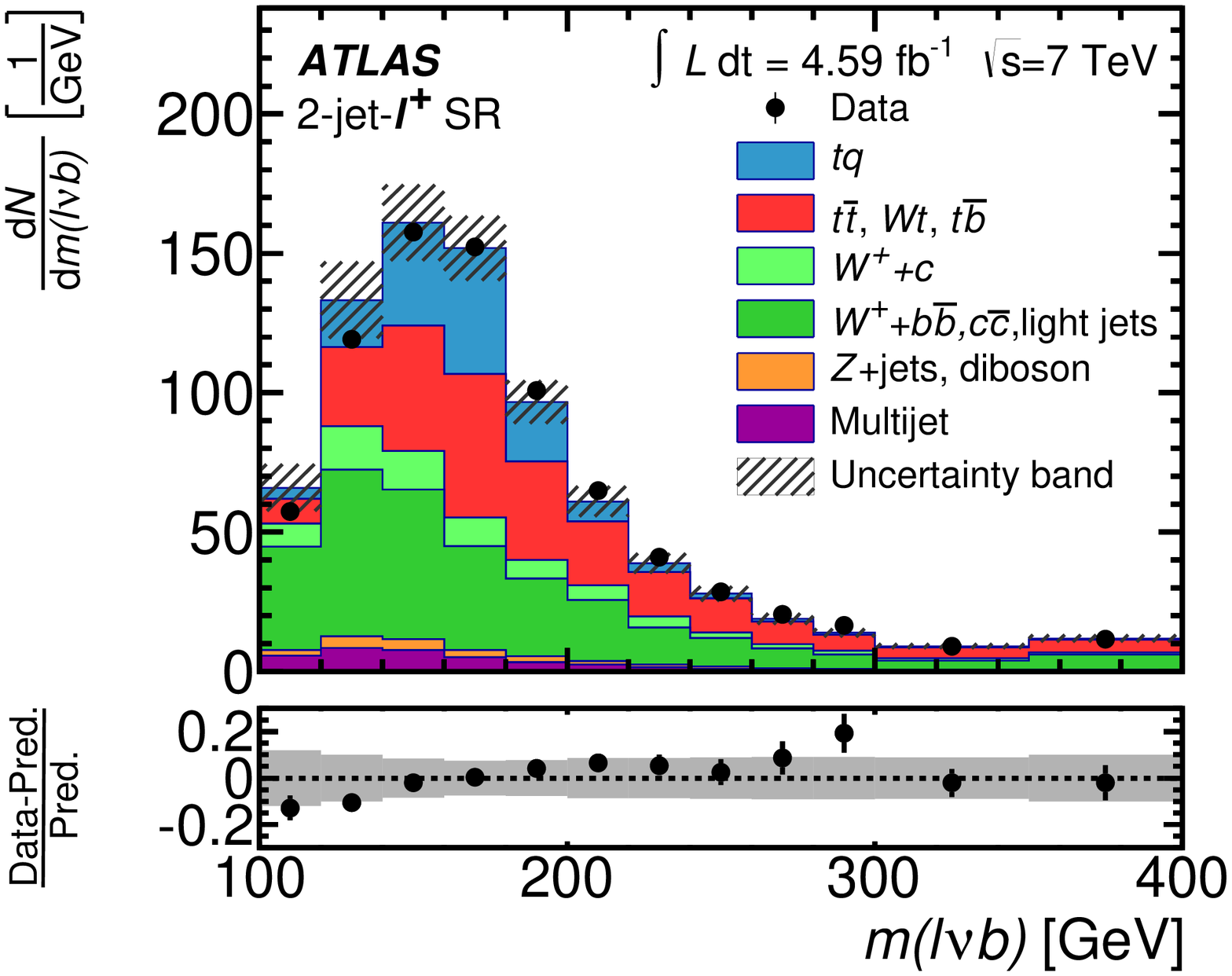}
     \label{subfig:2j_sr_plus_mlnub}
  }
  \subfigure[]{
  \includegraphics[width=0.45\textwidth]{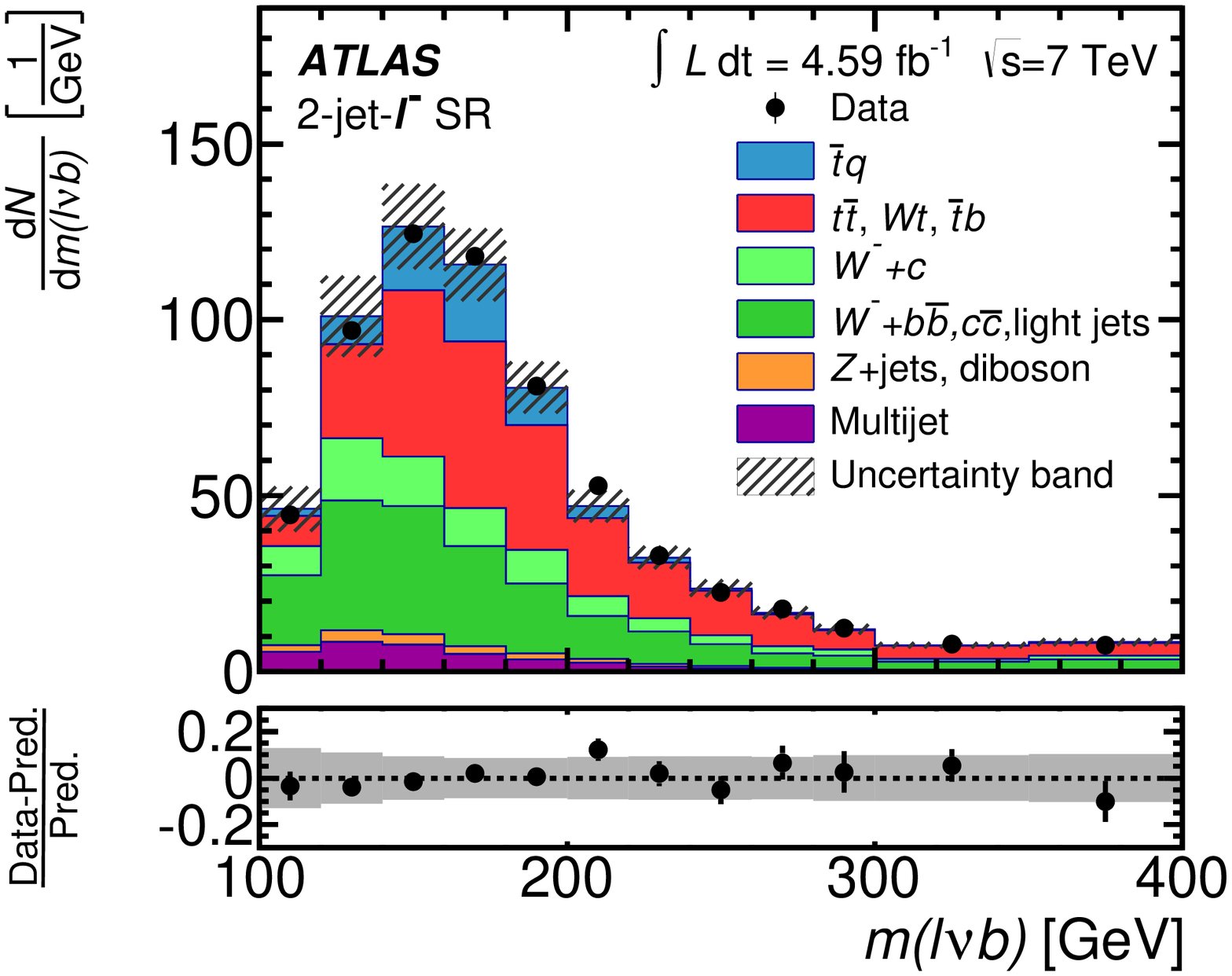}
     \label{subfig:2j_sr_minus_mlnub}
  }
  \subfigure[]{
  \includegraphics[width=0.45\textwidth]{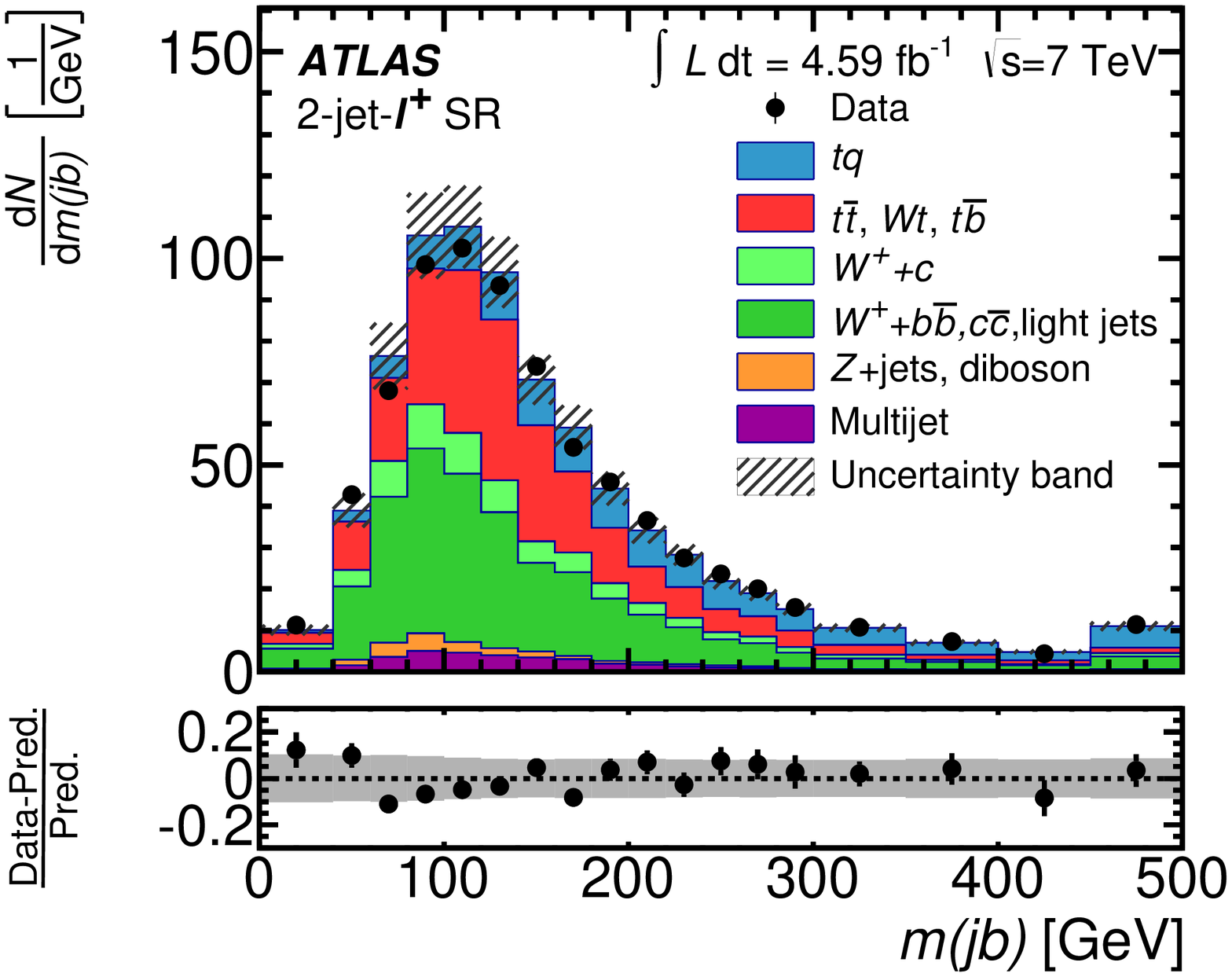}
     \label{subfig:2j_sr_plus_mj1j2}  
  }
  \subfigure[]{
  \includegraphics[width=0.45\textwidth]{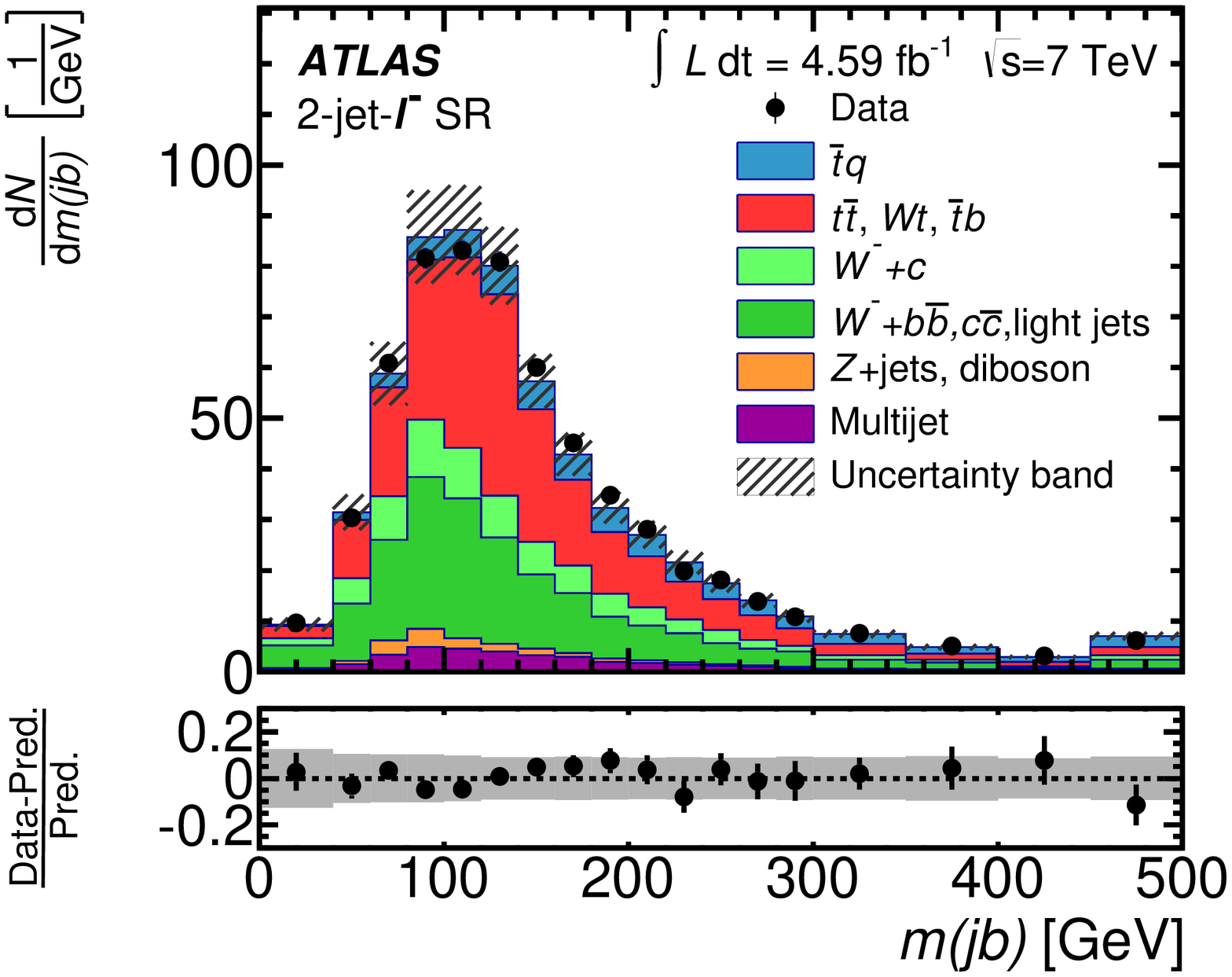}
     \label{subfig:2j_sr_minus_mj1j2}  
  }
  \caption{\label{fig:postfit_sigreg_2j_vars} Distributions of the three most important discriminating 
    variables in the 2-jet-$\ell^+$ and 2-jet-$\ell^-$ channels in the signal region normalized to the result of the binned 
  maximum-likelihood fit to the NN discriminant as described in Sec.~\ref{sec:fitresult}. 
    Figures~\subref{subfig:2j_sr_plus_lighteta} and \subref{subfig:2j_sr_minus_lighteta} display the
    absolute value of the pseudorapidity of the untagged jet $|\eta(j)|$.
    Figures~\subref{subfig:2j_sr_plus_mlnub} and \subref{subfig:2j_sr_minus_mlnub} show the invariant mass of the 
    reconstructed top quark $m(\ell\nu b)$, 
    \subref{subfig:2j_sr_plus_mj1j2} and
    \subref{subfig:2j_sr_minus_mj1j2} the invariant mass of the $b$-tagged and the untagged jet
    $m(jb)$.
    The last histogram bin includes overflows. 
    The uncertainty band represents the normalization uncertainty  
    of all processes after the fit and the Monte Carlo statistical uncertainty, added in quadrature. The relative difference
    between the observed and expected number of events in each bin is shown in the lower panels.
    }
\end{figure*}

\begin{figure*}[p]
  \centering
  \subfigure[]{
   \includegraphics[width=0.45\textwidth]{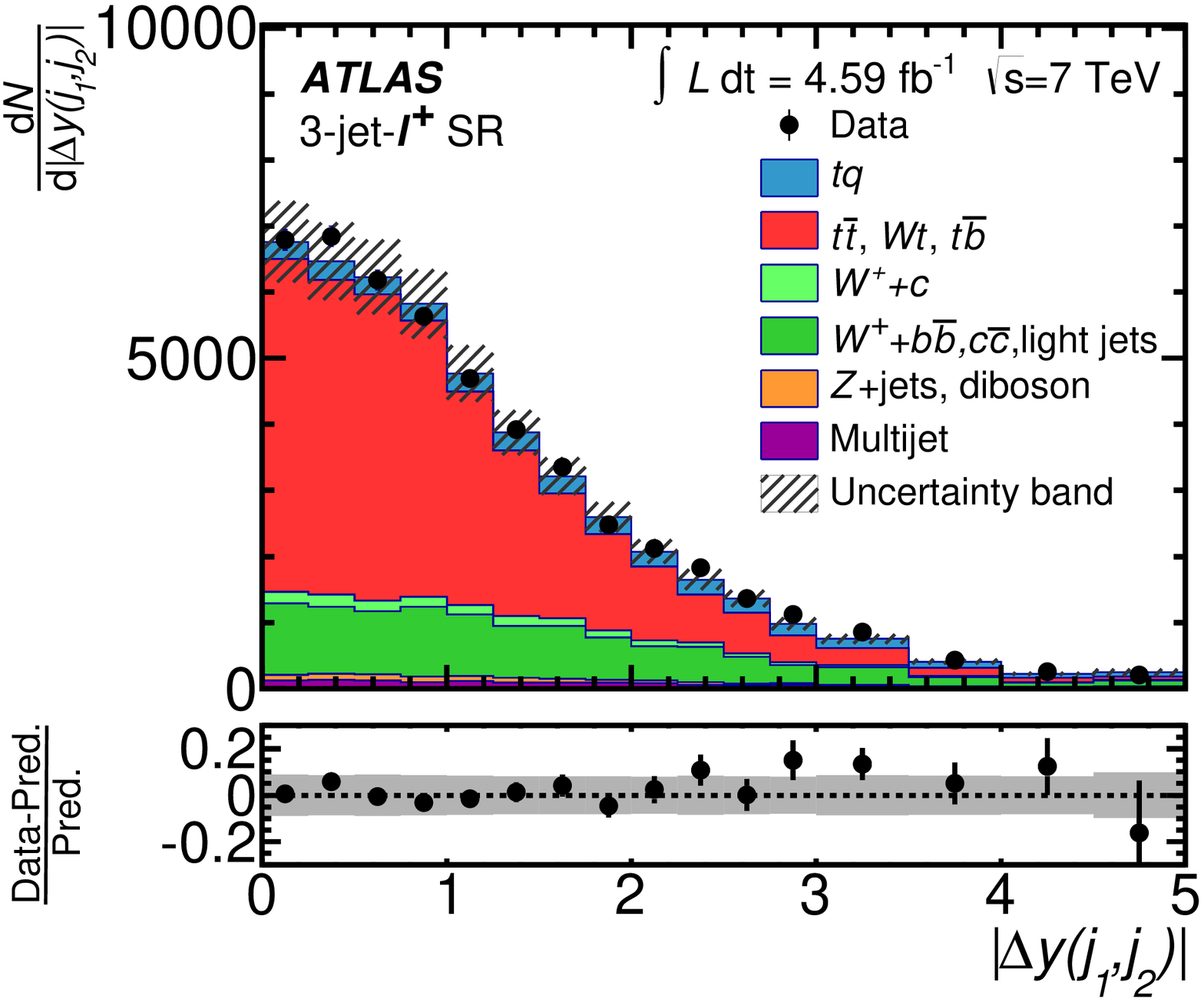}
   \label{subfig:3j_sr_plus_dyj1j2}
  }
  \subfigure[]{
 \includegraphics[width=0.45\textwidth]{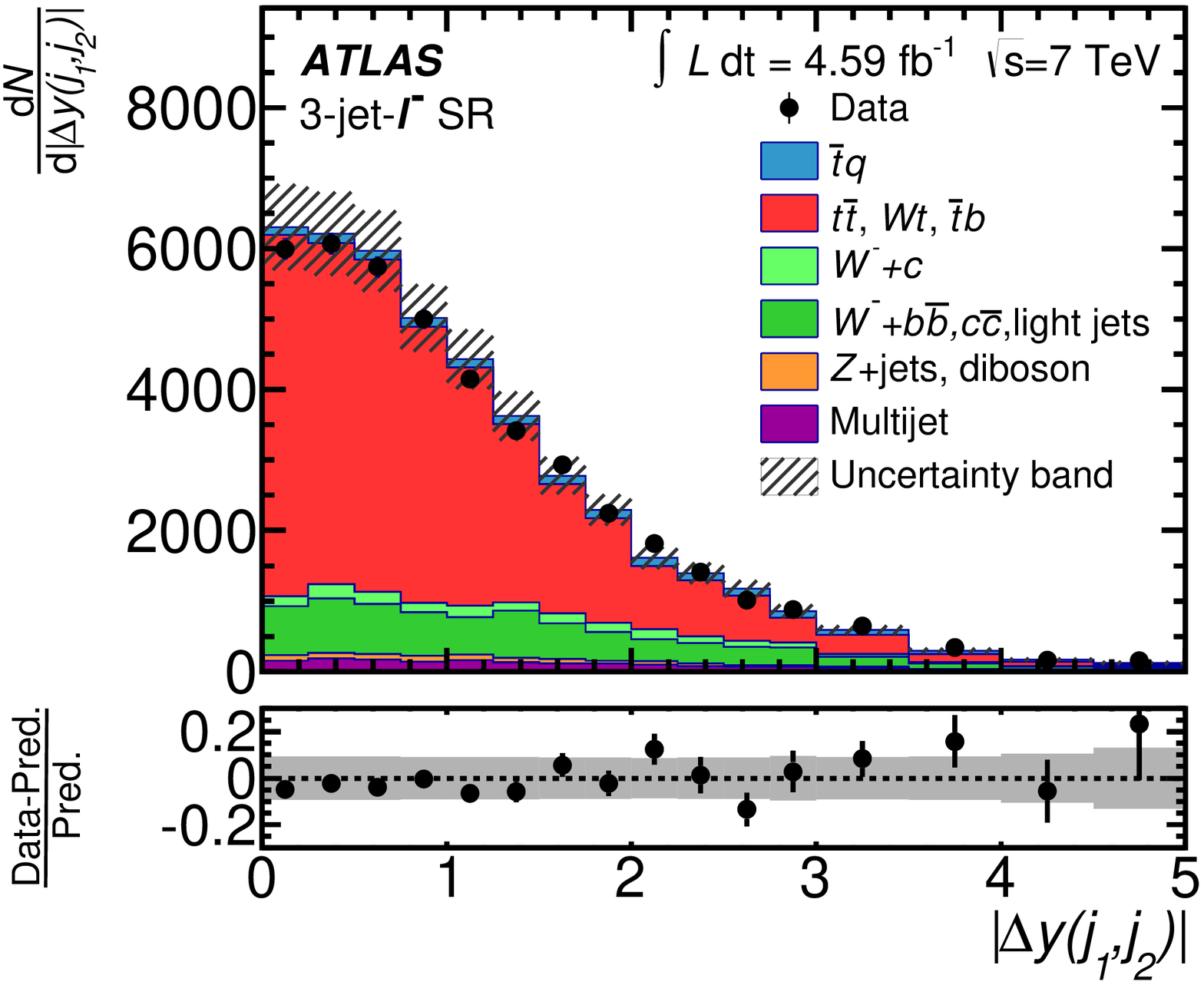}
     \label{subfig:3j_sr_minus_dyj1j2}
  }
  \subfigure[]{
 \includegraphics[width=0.45\textwidth]{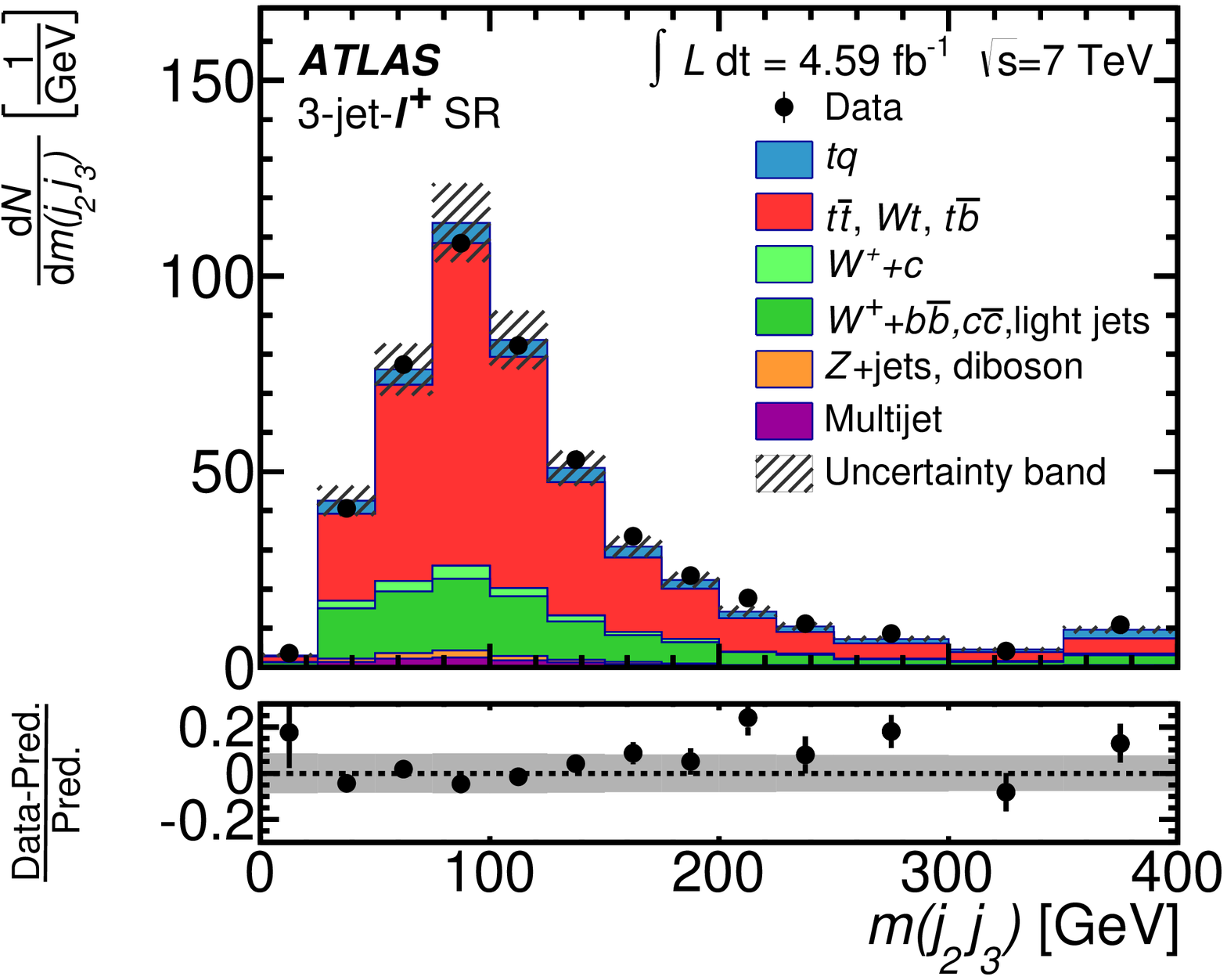}
     \label{subfig:3j_sr_plus_mj2j3}  
  }
  \subfigure[]{
 \includegraphics[width=0.45\textwidth]{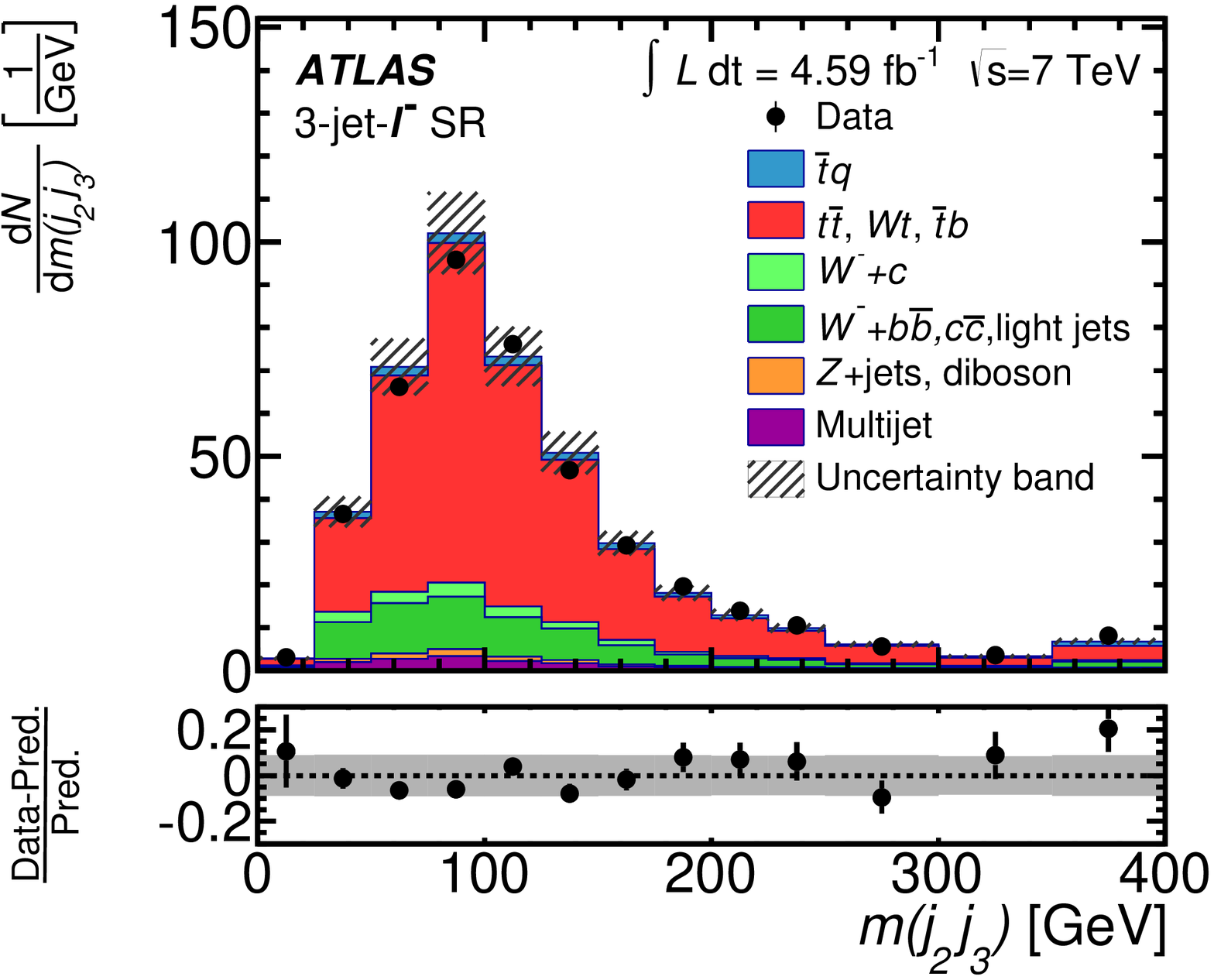}
     \label{subfig:3j_sr_minus_mj2j3}  
  }
  \subfigure[]{
 \includegraphics[width=0.45\textwidth]{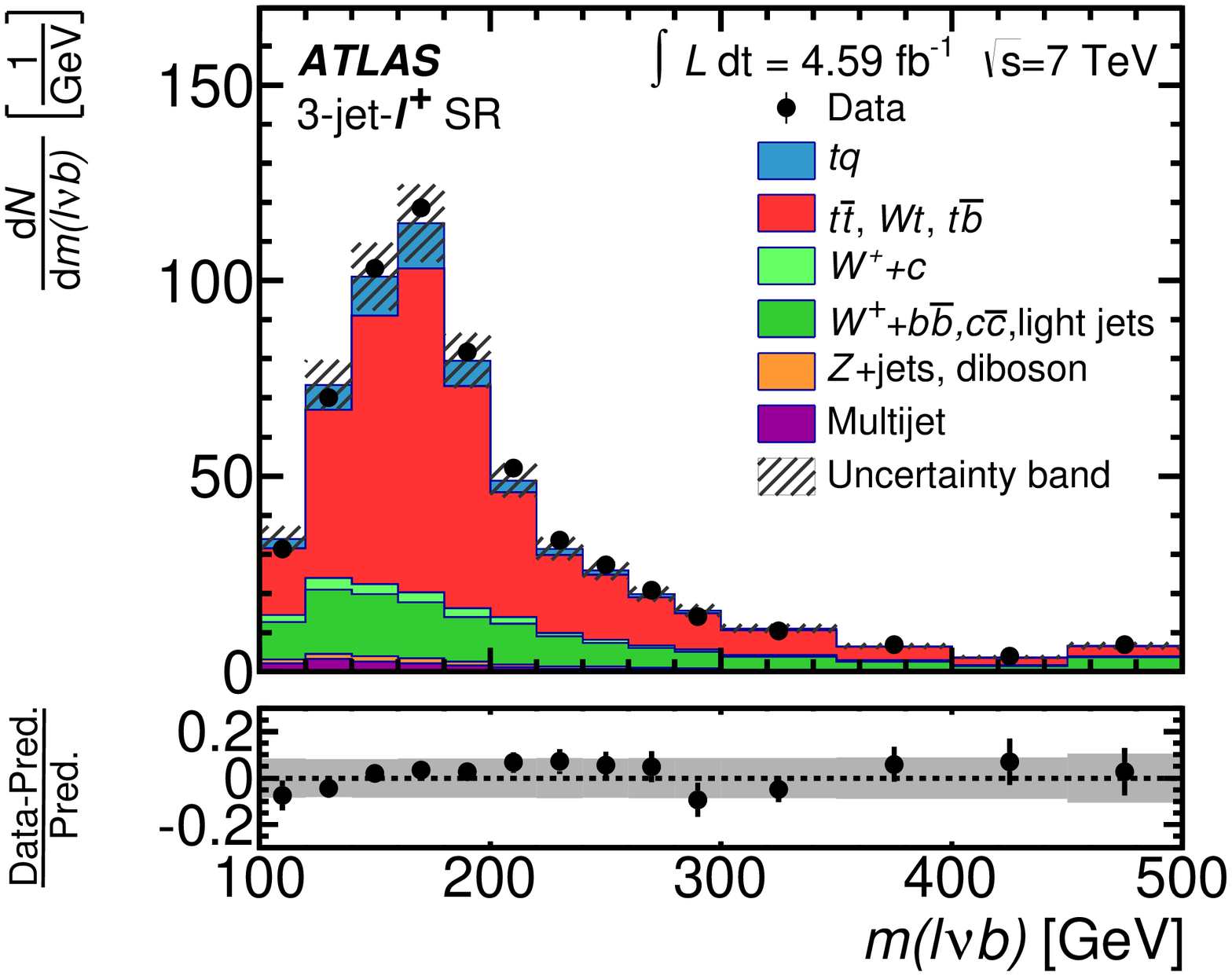}
     \label{subfig:3j_sr_plus_mlnub}
  }
  \subfigure[]{
 \includegraphics[width=0.45\textwidth]{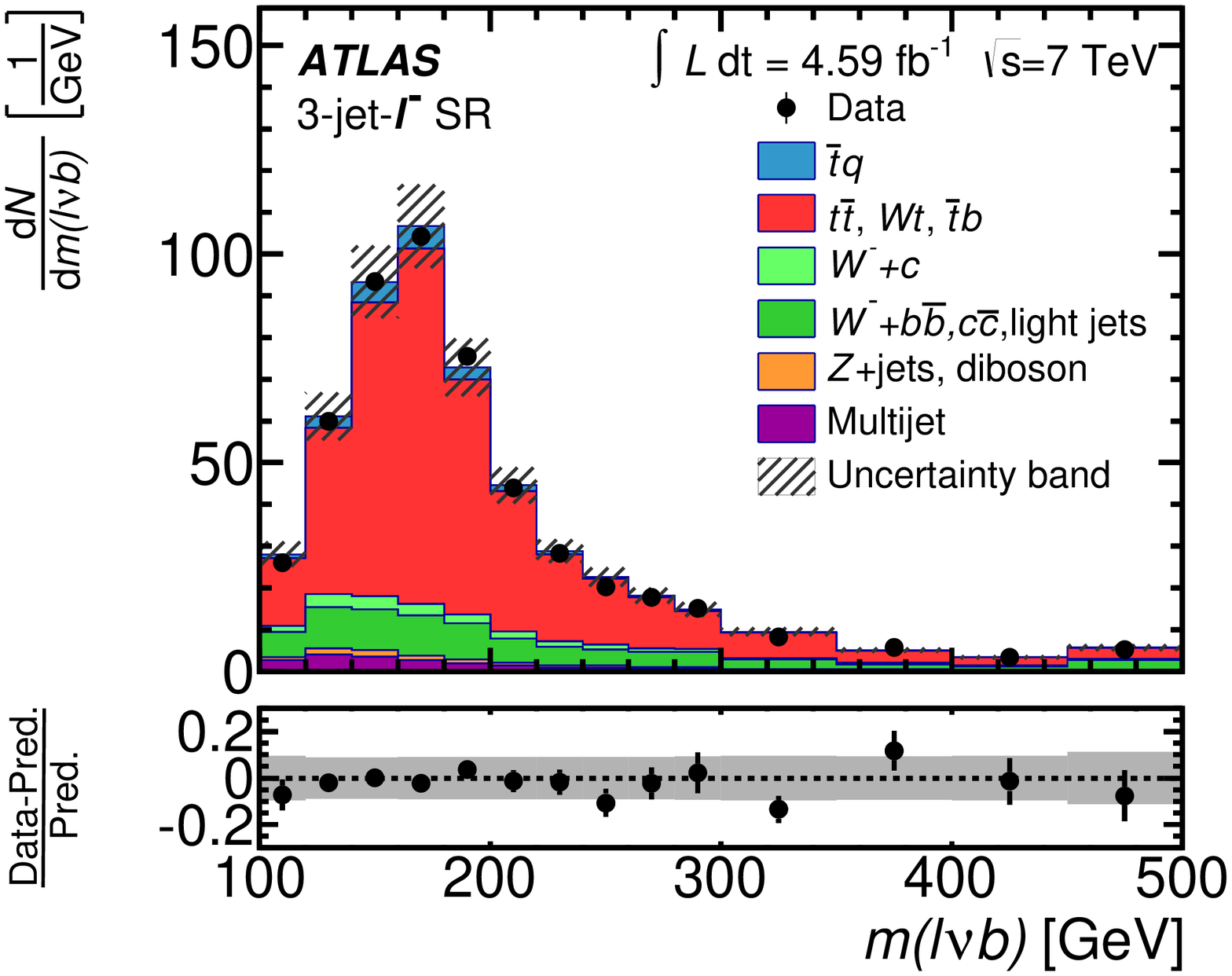}
     \label{subfig:3j_sr_minus_mlnub}
  }
  \caption{\label{fig:postfit_sigreg_3j_vars} Distributions of the three most important discriminating 
    variables in the 3-jet-$\ell^+$ and 3-jet-$\ell^-$ channels in the signal region normalized to the result of the binned 
  maximum-likelihood fit to the NN discriminant as described in Sec.~\ref{sec:fitresult}.
    Figures~\subref{subfig:3j_sr_plus_dyj1j2} and \subref{subfig:3j_sr_minus_dyj1j2} display the absolute
    value of the rapidity difference of the leading and 2$^\mathrm{nd}$ leading jet 
    $|\Delta y\left(j_{1},j_{2}\right)|$,
    \subref{subfig:3j_sr_plus_mj2j3} and \subref{subfig:3j_sr_minus_mj2j3} the invariant mass of 
    the 2$^\mathrm{nd}$ leading jet and the 3$^\mathrm{rd}$ jet $m\left(j_{2}j_{3}\right)$, and 
    \subref{subfig:3j_sr_plus_mlnub} and \subref{subfig:3j_sr_minus_mlnub} show the invariant mass 
    of the reconstructed top quark $m(\ell\nu b)$.
    The last histogram bin includes overflows. 
    The uncertainty band represents the normalization uncertainty 
    of all processes after the fit and the Monte Carlo statistical uncertainty, added in quadrature. The relative difference between the
    observed and expected number of events in each bin is shown in the lower panels.
  }
\end{figure*}

In Fig.~\ref{fig:nnout_fit_results} the observed NN discriminant distributions are shown compared to the
compound model of signal and background normalized to the fit results. Figures~\ref{fig:postfit_sigreg_2j_vars} 
and~\ref{fig:postfit_sigreg_3j_vars} show the three most discriminating variables normalized to the fit results 
in the 2-jet-$\ell^{\pm}$ and 3-jet-$\ell^{\pm}$-1-tag channels, respectively.
Differences between data and prediction are covered by the normalization uncertainty  
of the different processes after the fit. 

\subsection{High-purity region}
\label{sec:hpr}
\begin{figure*}[p]
  \centering
  \subfigure[]{
\includegraphics[width=0.45\textwidth]{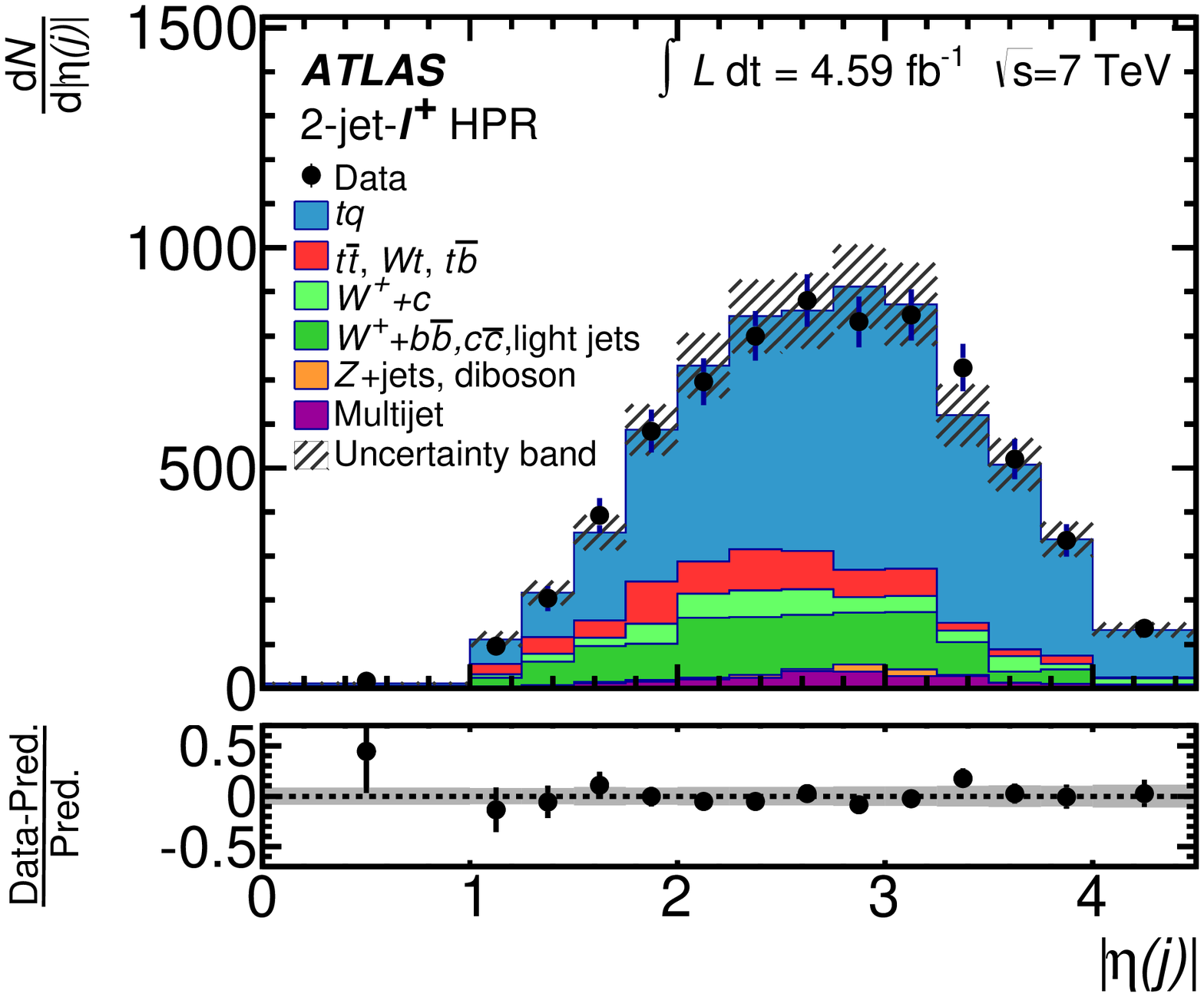}
     \label{subfig:2j_hpr_plus_lighteta}
  }
  \subfigure[]{
\includegraphics[width=0.45\textwidth]{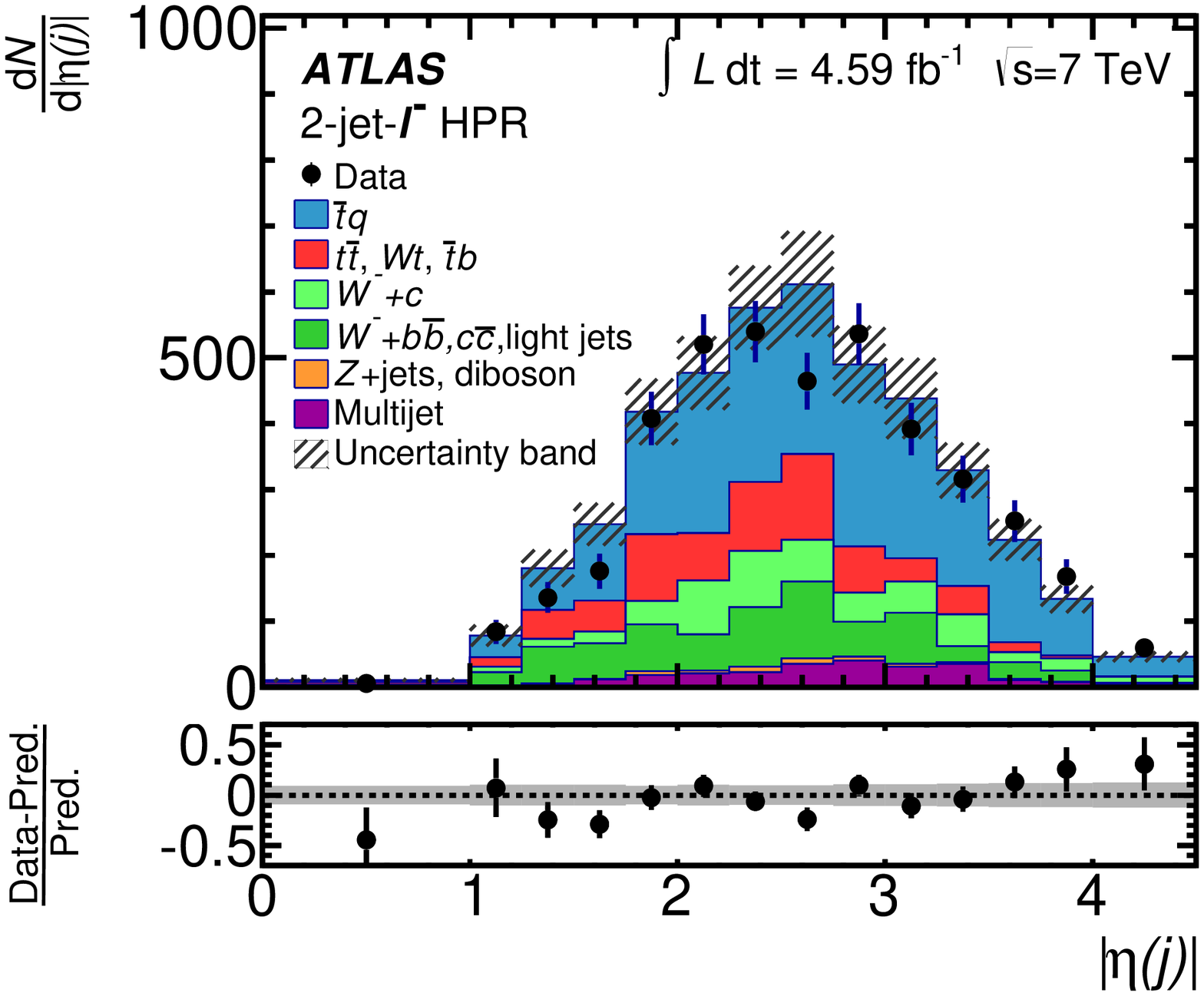}
     \label{subfig:2j_hpr_minus_lighteta}
  }
  \subfigure[]{
\includegraphics[width=0.45\textwidth]{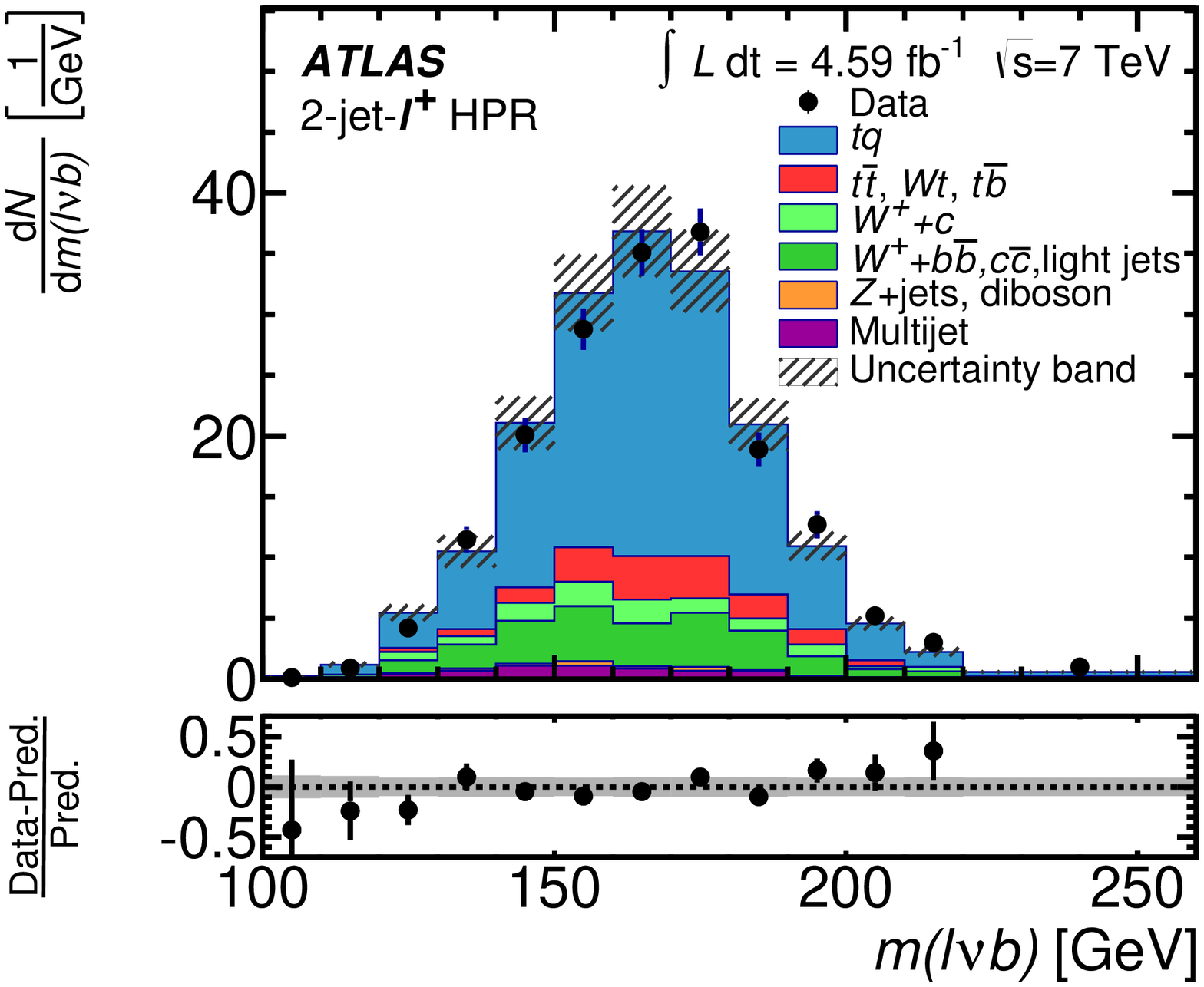}
     \label{subfig:2j_hpr_plus_mlnub}
  }
  \subfigure[]{
\includegraphics[width=0.45\textwidth]{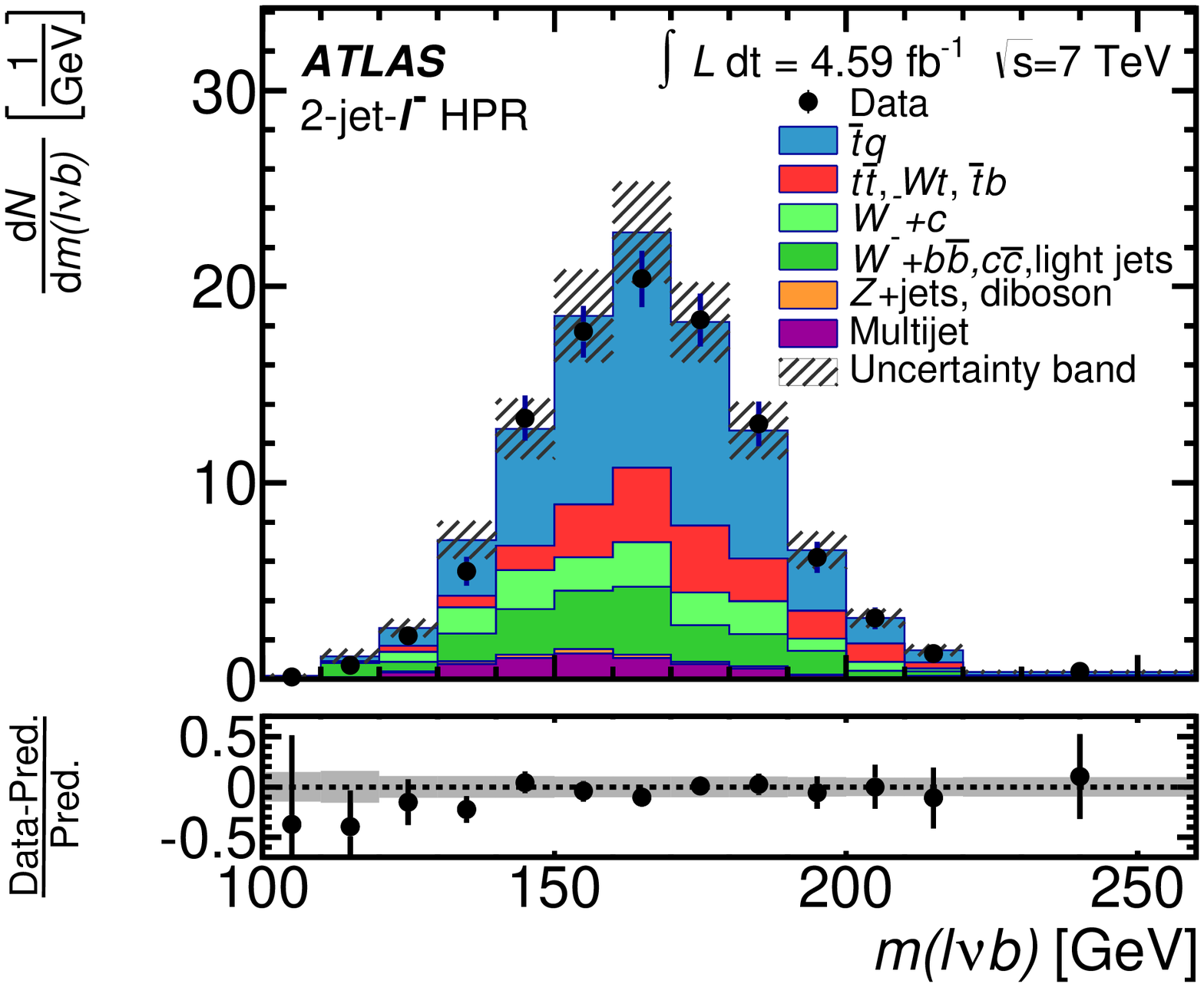}
     \label{subfig:2j_hpr_minus_mlnub}
  }
  \subfigure[]{
\includegraphics[width=0.45\textwidth]{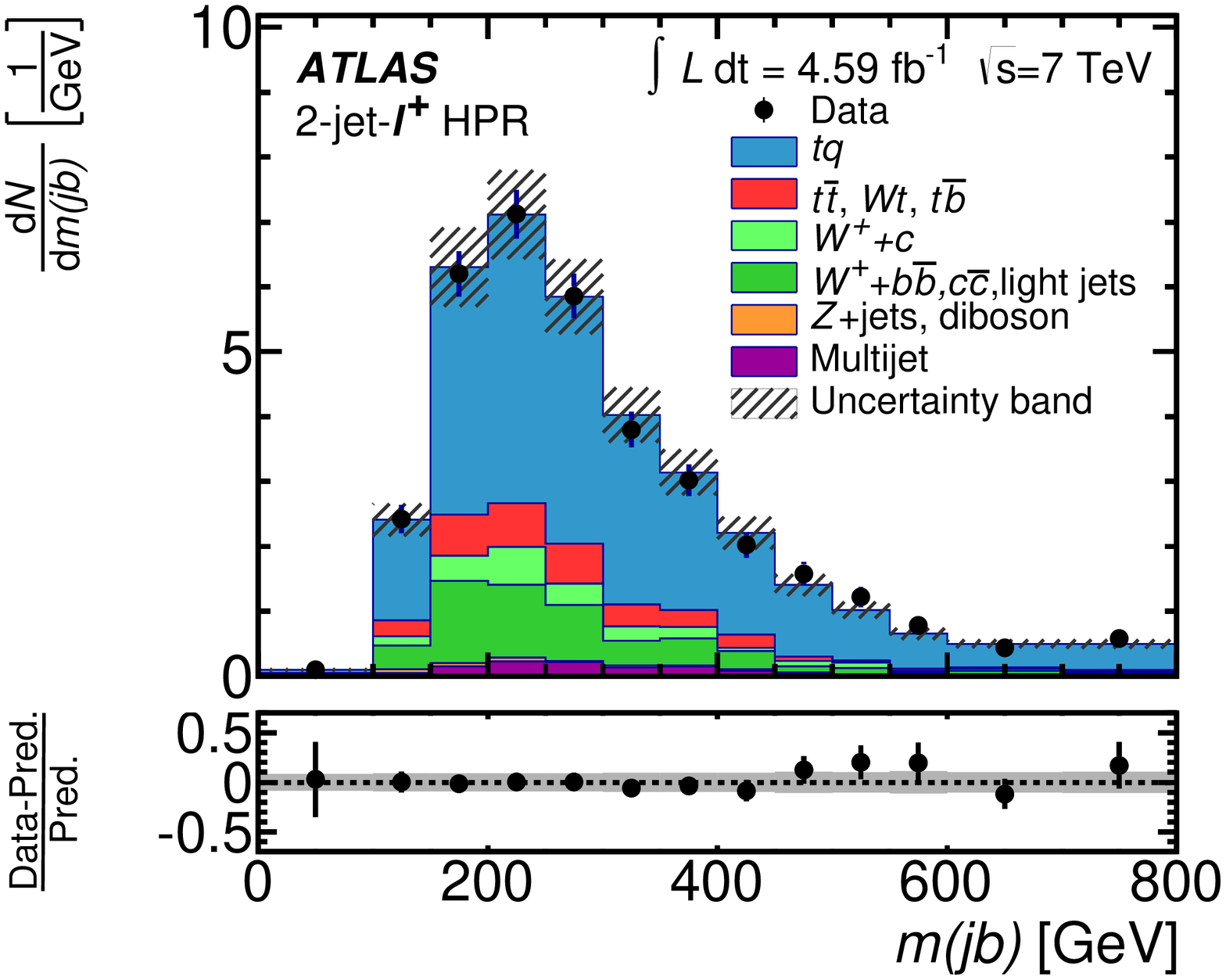}
     \label{subfig:2j_hpr_plus_mj1j2}  
  }
  \subfigure[]{
\includegraphics[width=0.45\textwidth]{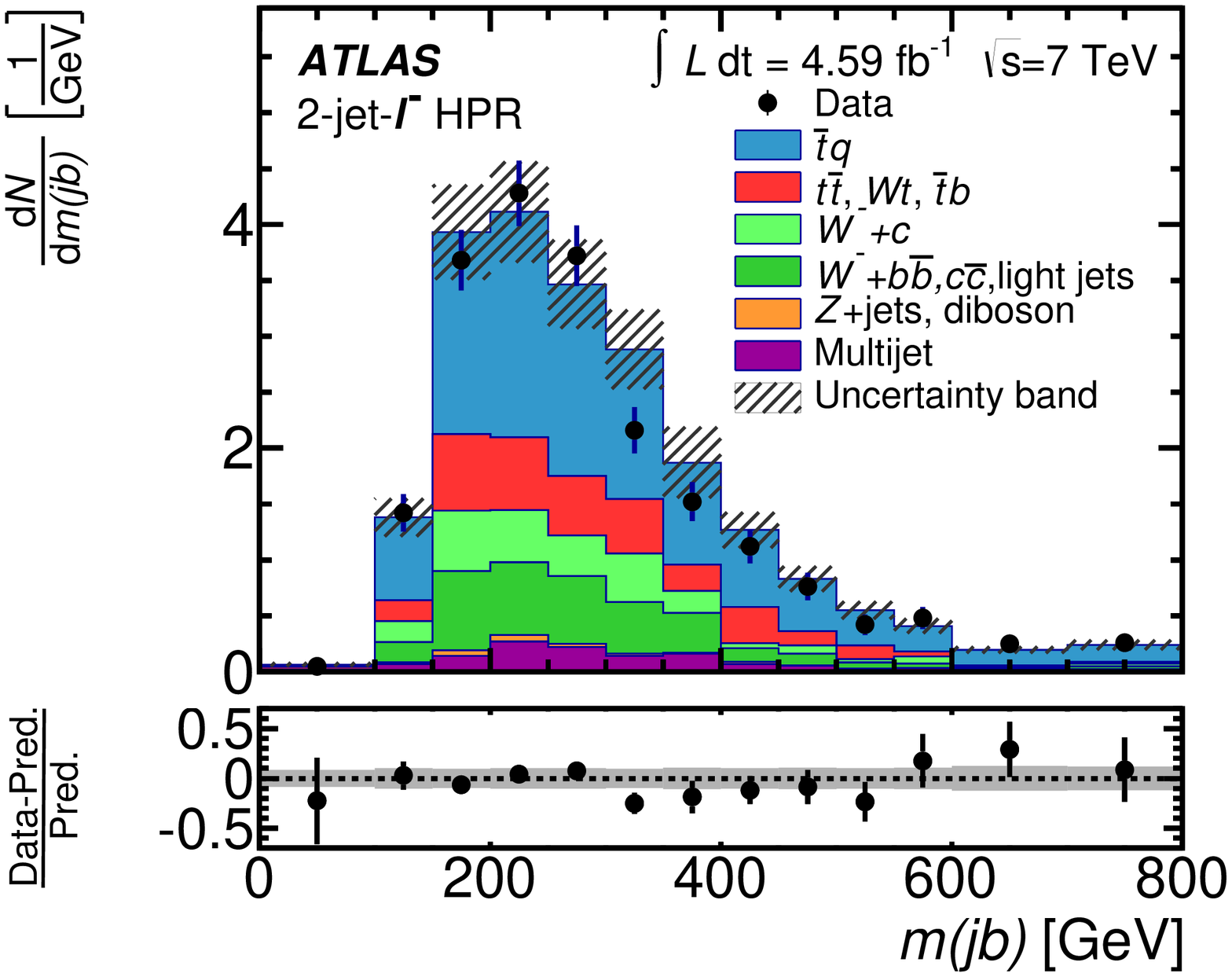}
     \label{subfig:2j_hpr_minus_mj1j2}  
  }
  \caption{\label{fig:highpurity_2j_vars} Distributions of the three most important discriminating 
    variables in the 2-jet-$\ell^+$ and 2-jet-$\ell^-$ channels in the high-purity region (HPR) 
    normalized to the result of the binned maximum-likelihood fit to the NN discriminant as described in Sec.~\ref{sec:fitresult}. 
    Figures~\subref{subfig:2j_hpr_plus_lighteta} and \subref{subfig:2j_hpr_minus_lighteta} display the
    absolute value of the pseudorapidity of the untagged jet $|\eta(j)|$.
    Figures~\subref{subfig:2j_hpr_plus_mlnub} and \subref{subfig:2j_hpr_minus_mlnub} show the invariant mass of the 
    reconstructed top quark $m(\ell\nu b)$,
    \subref{subfig:2j_hpr_plus_mj1j2} and
    \subref{subfig:2j_hpr_minus_mj1j2} the invariant mass of the $b$-tagged and the untagged jet
    $m(jb)$.
    The last histogram bin includes overflows. 
    The uncertainty band represents the normalization uncertainty  
    of all processes after the fit and the Monte Carlo statistical uncertainty, added in quadrature. The relative difference between the observed
    and expected number of events in each bin is shown in the lower panels.}
\end{figure*}

A high-purity region (HPR) is defined to measure the differential cross sections in the 2-jet-$\ell^+$ 
and 2-jet-$\ell^-$ channels, by requiring the NN discriminant to be larger than~$0.8$. 
In the 2-jet-$\ell^+$ HPR the signal contribution is twice as large as the background contribution.
The signal and background contributions in the 2-jet-$\ell^-$ HPR are of approximately the same size.
The result of the fit described above is used to normalize the background in the HPR. 
Figure~\ref{fig:highpurity_2j_vars} shows the three most discriminating variables in the 2-jet-$\ell^+$ and 2-jet-$\ell^-$ 
high-purity channels, normalized to the fit results. The data are well described by the predicted compound model.

\clearpage

\section{Systematic uncertainties}
\label{sec:systematics}

For both the physical object definitions and the background estimations, systematic uncertainties
are assigned to account for detector calibration and resolution uncertainties, as well as the  
uncertainties of theoretical predictions. 
These variations affect both normalization and shape of distributions for signal and backgrounds.
The uncertainties can be split into the following categories: physics object modeling, Monte Carlo generators,
PDFs, theoretical cross-section normalization, and luminosity.

\subsection{Physics object modeling}
Systematic uncertainties
on the reconstruction and energy calibration of 
jets, electrons and muons are propagated through the entire analysis.
The main source of object modeling uncertainty comes from the jet energy scale (JES). 
The JES uncertainty has been evaluated for the in situ jet calibration~\cite{Aad:2011he,Aad:2014bia}, which uses $Z$+jet, $\gamma$+jet, and dijet  
$\pt$-balance measurements in data. 
The JES uncertainty is evaluated in several different categories: 
\begin{itemize}
  \item Detector: The different \pT-balance measurements have uncertainties due to the jet energy resolution, the electron and photon energy scale and the photon purity.
  \item Physics modeling: The uncertainties in the in situ calibration techniques due to the choice of Monte Carlo generator, radiation modeling, and the extrapolation of $\Delta\phi$ between the jet and the $Z$ boson.
  \item Statistics: The uncertainty due to the limited size of the data sets of the in situ jet calibration measurements. 
  \item Mixed detector and modeling: In this category the uncertainty due to the modeling of the underlying event and soft radiation as well as modeling of the jet fragmentation are considered.
  \item $\eta$ intercalibration modeling: The uncertainty in the dijet-\pT-balance technique due to the modeling of additional parton radiation is estimated by comparing dijet events simulated with {\sc PYTHIA} and {\textsc HERWIG}. 
  This JES category is the largest contribution from the jet energy scale to the cross-section measurements.  
  \item Close-by jets: The jet calibration can be affected by the presence of close-by jets, located at radii $\Delta R < 1.0$. 
  \item Pile-up: Uncertainties due to the modeling of the large pile-up effects in data are included as a function of jet \pT~ and $\eta$.
  \item Flavor composition: This uncertainty covers effects due to the difference in quark--gluon composition between the jets used in the calibration and the jets used in this analysis. Since the response to quark and gluon jets is different, the uncertainty on the quark--gluon composition in a given data sample leads to an uncertainty in the jet calibration.
  \item Flavor response: In this category an uncertainty is considered due to imperfect knowledge of the calorimeter response to light-quark jets and gluon jets.
  \item $b$-JES: An additional JES uncertainty is evaluated for $b$-quark jets by varying the modeling of $b$-quark fragmentation.
\end{itemize}

The uncertainty due to the jet energy resolution is modeled
by varying the \pT~ of the jets according to the systematic uncertainties of the resolution measurement performed on data using the dijet-balance method~\cite{JetResolution2013}.
The effect of uncertainties associated with the jet vertex fraction is also considered for each jet.

The tagging efficiencies of $b$-jets, $c$-jets, and light jets are derived from 
data~\cite{btaggingSF2012,ATLAS-CONF-2012-039,ATLAS-CONF-2012-040} and parameterized as a function 
of \pT~ and $\eta$ of the jet. The corresponding efficiencies in simulated events are corrected to
be the same as those observed in data, and the uncertainties in the calibration method are propagated 
to the analysis.
The difference in the $b$-tagging efficiency between jets initiated by $b$-quark and $b$-antiquark 
is $\sim$1\%, estimated from simulated $tq$ and $\bar{t}q$ events. To account for a possible uncertainty 
in the modeling of the detector response the full difference is taken as a systematic uncertainty.  
In Table~\ref{tab:xs_uncertainties} this uncertainty is called $b/\bbar$ acceptance.

The uncertainties due to lepton reconstruction, 
identification and trigger efficiencies are evaluated using tag-and-probe methods in $Z \rightarrow \ell\ell$ events.
Uncertainties due to the energy scale and resolution are considered for electrons and muons. 
Additionally, the lepton charge misidentification is taken into account and was evaluated to be about 0.1\%.
All lepton uncertainties are summarized in Table~\ref{tab:xs_uncertainties} in one item.

Other minor uncertainties are assigned to the reconstruction of \met\ and to account
for the impact of pile-up collisions on the calculation of \met. The uncertainties on $\met$
are summarized under $\met$ modeling in Table~\ref{tab:xs_uncertainties}.

\subsection{Monte Carlo generators} 
Systematic uncertainties arising from the modeling of the single top-quark signal,
the $\ttbar$ background, and the $W$+jets background are taken into account.

The uncertainty due to the choice of single top-quark $t$-channel generator and parton shower 
model is estimated by comparing 
events generated with {\textsc POWHEG-BOX} interfaced to \textsc{PYTHIA} and events generated with the NLO matrix-element generator
{\textsc MG5\_}a{\sc MC@NLO}~\cite{Alwall:2014hca} and showered with \textsc{HERWIG} and \textsc{JIMMY}.
Again the fixed four-flavor PDF set {\textsc CT10f4}~\cite{Lai:2010vv} is used, and the renormalization and factorization scales 
are set to $\mu_\mathrm{R} = \mu_\mathrm{F} = 4\cdot \sqrt{m_b^2 +p_{\mathrm{T},b}^2}$, 
where $m_b = 4.75 \gev$ is the $b$-quark mass,
and $p_{\mathrm{T},b}$ is the transverse momentum of the $b$-quark.
The uncertainty on the choice of $\mu_\mathrm{R}$ and $\mu_\mathrm{F}$ is estimated using events 
generated with {\textsc POWHEG-BOX} interfaced to \textsc{PYTHIA}. Factorization and renormalization scales
are varied independently by factors of $0.5$ and $2.0$, while the scale of the parton shower is varied consistently with the 
renormalization scale. The uncertainty related to scale variations is then given by the envelope of all variations. 
 
The modeling uncertainty for the $\ttbar$ background is evaluated by comparing events simulated with
the NLO generator {\textsc POWHEG-BOX} interfaced to {\sc PYTHIA} and the multileg LO generator {\textsc ALPGEN} interfaced to {\textsc HERWIG}.
An additional uncertainty for the top-quark background processes comes from the amount of initial-state and final-state radiation, estimated
using dedicated \textsc{AcerMC} samples interfaced to \textsc{PYTHIA}
where parameters controlling initial-state and final-state radiation (ISR/FSR) emission are varied.
The variations of the parameters are constrained by a measurement of $\ttbar$ production with a veto on additional central jet activity~\cite{jetVeto}. 

A shape uncertainty is assigned to the $W$+jets background, based on variation of
the choices of the matching scale and of the functional form of the factorization scale in \textsc{ALPGEN}.

The impact of using simulation samples of limited size is also taken into account.

\subsection{Parton distribution function}
The systematic uncertainties related to the PDFs
are taken into account for the acceptance of all single top-quark processes 
and $t\bar{t}$ production.
The simulated events are reweighted according to each of the PDF
uncertainty eigenvectors. The uncertainty is calculated following the recommendation of the
respective PDF group. The final PDF uncertainty is the envelope of the estimated uncertainties for the {\textsc CT10} PDF set, 
the {\textsc MSTW2008nlo}~\cite{Martin:2009bu} PDF set and the {\textsc NNPDF2.3}~\cite{Ball:2012cx} PDF set. 
For all PDFs the variable flavor number scheme~\cite{Thorne:2010pa} is used.

\subsection{Theoretical cross-section normalization}
In Sec.~\ref{sec:bgestimation} the theoretical cross sections and their uncertainties are quoted for each background 
process.
Since the $\ttbar$, single top-quark $Wt$ and $s$-channel processes are grouped together in the statistical analysis, 
their uncertainties are added in proportion to their relative fractions, leading to a combined uncertainty
of 6.7\%. 
The uncertainty on the combined $Z$+jets and diboson background is 60\%
including a conservative estimate of the uncertainty of the heavy-flavor fraction of 50\%, while
the uncertainties of the $W$+jets backgrounds are 24\% for $W$+$c$ and 36\% for the combined $W$+$b\bbar$, $c\cbar$ and light jets including the same heavy-flavor-fraction uncertainty on the $b\bbar$ and $c\cbar$ contributions.
Additionally, an uncertainty on the relative fraction of 2-jet to 3-jet events of 5\% for events with light-flavor jets 
and 7\% for events with heavy-flavor jets is applied for the $W$+jets estimation. This uncertainty was estimated by 
varying the following input parameters of the generation with {\textsc ALPGEN} by a factor of two: the hard scattering scale, the 
coupling of the hard interaction, and the minimum $\pT$ and $\Delta R$ separation of the partons.

\subsection{Luminosity} 
The luminosity measurement is calibrated using dedicated beam-separation scans, referred to as van der Meer scans,
where the absolute luminosity can be inferred from the measurement of the beam parameters~\cite{LumiPaper2012}.
The resulting uncertainty is 1.8\%.

\subsection{Uncertainties on the cross-section measurements}
\label{sec:unc_est}
\begin{figure}[!t]
\begin{center}
  \includegraphics[width=0.45\textwidth]{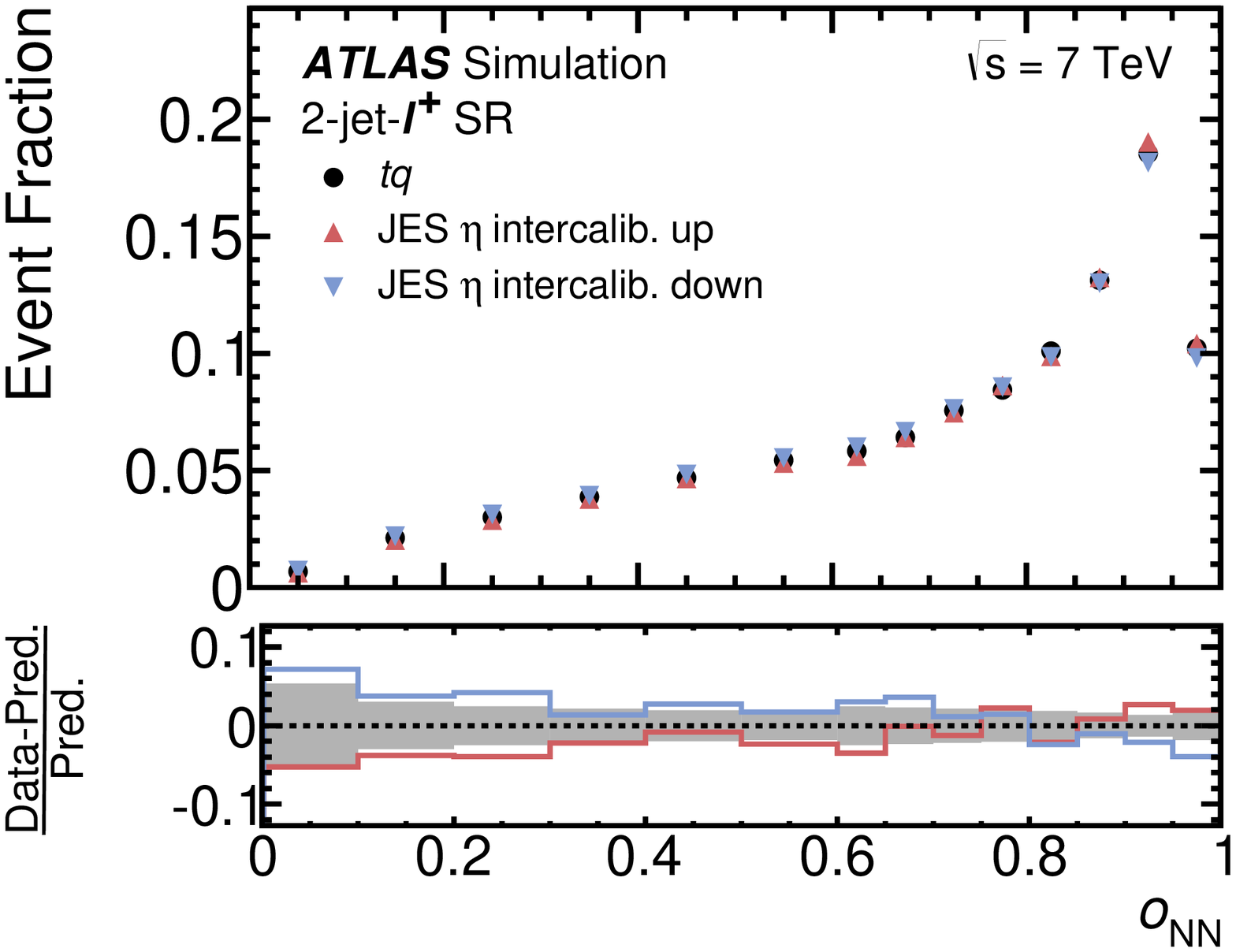}
\caption{\label{fig:NNShapeUnc}The normalized shape variation of the NN discriminant for the JES variation due to the uncertainty on the $\eta$ intercalibration
in the 2-jet-$\ell^+$ channel, shown for the $tq$ sample. 
The nominal shape is shown by the black points. Red denotes the JES shift-up and blue the NN response for JES shift-down. 
In the lower panel the relative difference between the number of expected events in the systematic variation
and the nominal distribution is shown for each bin. 
The grey uncertainty band in the lower histogram represents the normalization uncertainty 
due to the Monte Carlo statistical uncertainty.
}
\end{center}
\end{figure}

The systematic uncertainties on the individual top-quark and top-antiquark 
cross-section measurements and their ratio are determined using pseudo-experiments
that account for variations of the
signal acceptance, the background rates, and the shape of the NN discriminant due
to all sources of uncertainty described above. 
As an example, Fig.~\ref{fig:NNShapeUnc} shows the shape variation of the NN discriminant for $t$-channel 
single top-quark signal events 
due to the variation of the JES because of the uncertainty on the $\eta$ intercalibration.
The correlations between the different channels and the physics processes are fully accounted for.
The probability densities of all possible outcomes of the measurements of $\sigma(tq)$, 
$\sigma(\bar{t}q)$ and $R_t$ are obtained by performing the measurements 
on the pseudo-data. The values measured in data are used as central values when generating 
the pseudo-experiments. The root mean squares of the estimator distributions of the
measured quantities are estimators of the measurement uncertainties. 

Table~\ref{tab:xs_uncertainties} summarizes the contributions of the various sources 
of systematic uncertainty to the uncertainties on the
measured values of $\sigma(tq)$, $\sigma(\bar{t}q)$, $R_t$ and $\sigma(tq+\bar{t}q)$. 
The dominant systematic uncertainty on the cross sections is the JES $\eta$-intercalibration
uncertainty since one of the prominent features of $tq$ production is a jet in the
forward region.  
\begin{table*}[!thbp]
\begin{center}
\begin{ruledtabular} 
\caption{Detailed list of the contribution of each source of uncertainty to the  
   total uncertainty on the measured values of $\sigma(tq)$, $\sigma(\bar{t}q)$, $R_t$, and $\sigma(tq+\bar{t}q)$. The
   evaluation of the systematic uncertainties has a statistical uncertainty of 0.3\%. Uncertainties contributing less than
   1.0\% are marked with ``$<1$''.}
\label{tab:xs_uncertainties}
\begin{tabular}{lcccc}
    Source & $\Delta \sigma(tq)/ \sigma(tq)$ [\%] & 
             $\Delta \sigma(\tbar q)/ \sigma(\bar{t}q)$ [\%] & 
             $\Delta R_{t}/ R_{t}$ [\%] &
             $\Delta \sigma(tq+\tbar q)/ \sigma(tq+\bar{t}q)$ [\%] \\
    \hline
    Data statistical           & $\pm$3.1    & $\pm$5.4    & $\pm$6.2   & $\pm$2.7  \\
    Monte Carlo statistical    & $\pm$1.9    & $\pm$3.2    & $\pm$3.6   & $\pm$1.9  \\
    \multicolumn{5}{c}{ } \\
    Multijet normalization     & $\pm$1.1    & $\pm$2.0    & $\pm$1.6   & $\pm$1.4  \\
    Other background normalization   & $\pm$1.1    & $\pm$2.8    & $\pm$1.9  & $\pm$1.6  \\ 
    \multicolumn{5}{c}{ } \\
    JES detector               & $\pm$1.6    & $\pm$1.4    & $<1$       & $\pm$1.4  \\
    JES statistical            & $<1$        & $<1$        & $<1$       & $<1$  \\
    JES physics modeling       & $<1$        & $<1$        & $<1$       & $<1$  \\
    JES $\eta$ intercalibration     & $\pm$6.9    & $\pm$8.4    & $\pm$1.8   & $\pm$7.3  \\
    JES mixed detector and modeling & $<1$        & $<1$        & $<1$       & $<1$  \\
    JES close-by jets          & $<1$        & $<1$        & $<1$       & $<1$  \\
    JES pile-up                & $<1$        & $<1$        & $<1$       & $<1$  \\
    JES flavor composition           & $\pm$1.4    & $\pm$1.4    & $\pm$1.2   & $\pm$1.6  \\
    JES flavor response        & $<1$        & $<1$        & $\pm$1.0   & $<1$  \\
    $b$-JES    		             & $<1$        & $<1$        & $<1$       & $<1$  \\
    \multicolumn{5}{c}{ } \\
    Jet energy resolution      & $\pm$2.1    & $\pm$1.6    & $\pm$1.0   & $\pm$1.9  \\
    Jet vertex fraction        & $<1$        & $<1$        & $<1$       & $<1$  \\
    $b$-tagging efficiency     & $\pm$3.8    & $\pm$4.1    & $<1$       & $\pm$3.9  \\
    $c$-tagging efficiency     & $<1$        & $\pm$1.4    & $<1$       & $<1$  \\
    Mistag efficiency          & $<1$        & $<1$        & $<1$       & $<1$  \\
    $b/\bbar$ acceptance       & $\pm$1.0    & $<1$        & $<1$       & $--$  \\
    $\MET$ modeling            & $\pm$2.3    & $\pm$3.4    & $\pm$1.6       & $\pm$2.6  \\
    Lepton uncertainties       & $\pm$2.8    & $\pm$3.0    & $\pm$1.0   & $\pm$2.8  \\
    \multicolumn{5}{c}{ } \\
    PDF                        & $\pm$3.2    & $\pm$5.8    & $\pm$2.5   & $\pm$3.2  \\
    $W$+jets shape variation   & $<1$        & $<1$        & $<1$       & $<1$  \\
    $tq$ generator + parton shower          & $\pm$1.9        & $\pm$1.6    & $<1$      & $\pm$1.9 \\
    $tq$ scale variations     & $\pm$2.6        & $\pm$3.0  & $<1$  & $\pm$2.6   \\
    $\ttbar$ generator + parton shower      & $<1$        & $\pm$2.1    & $\pm$1.6       & $<1$  \\
    $\ttbar$ ISR / FSR       & $<1$        & $<1$    & $\pm$1.0   & $<1$  \\
    \multicolumn{5}{c}{ } \\
    Luminosity                 & $\pm$1.8    & $\pm$1.8    & $\pm$0.5   & $\pm$1.8  \\
    \multicolumn{5}{c}{ } \\
    Total systematic           & $\pm$12.0   & $\pm$14.9   & $\pm$6.1   & $\pm$12.1  \\ 
    Total                      & $\pm$12.4   & $\pm$15.9   & $\pm$8.7   & $\pm$12.4  \\ 
   \end{tabular}
\end{ruledtabular}
\end{center}
\end{table*}

\section{Total cross-section measurements}
\label{sec:XS_Rt_measure}

\begin{figure*}[ht]
  \centering
  \subfigure[]{
     \includegraphics[width=0.45\textwidth]{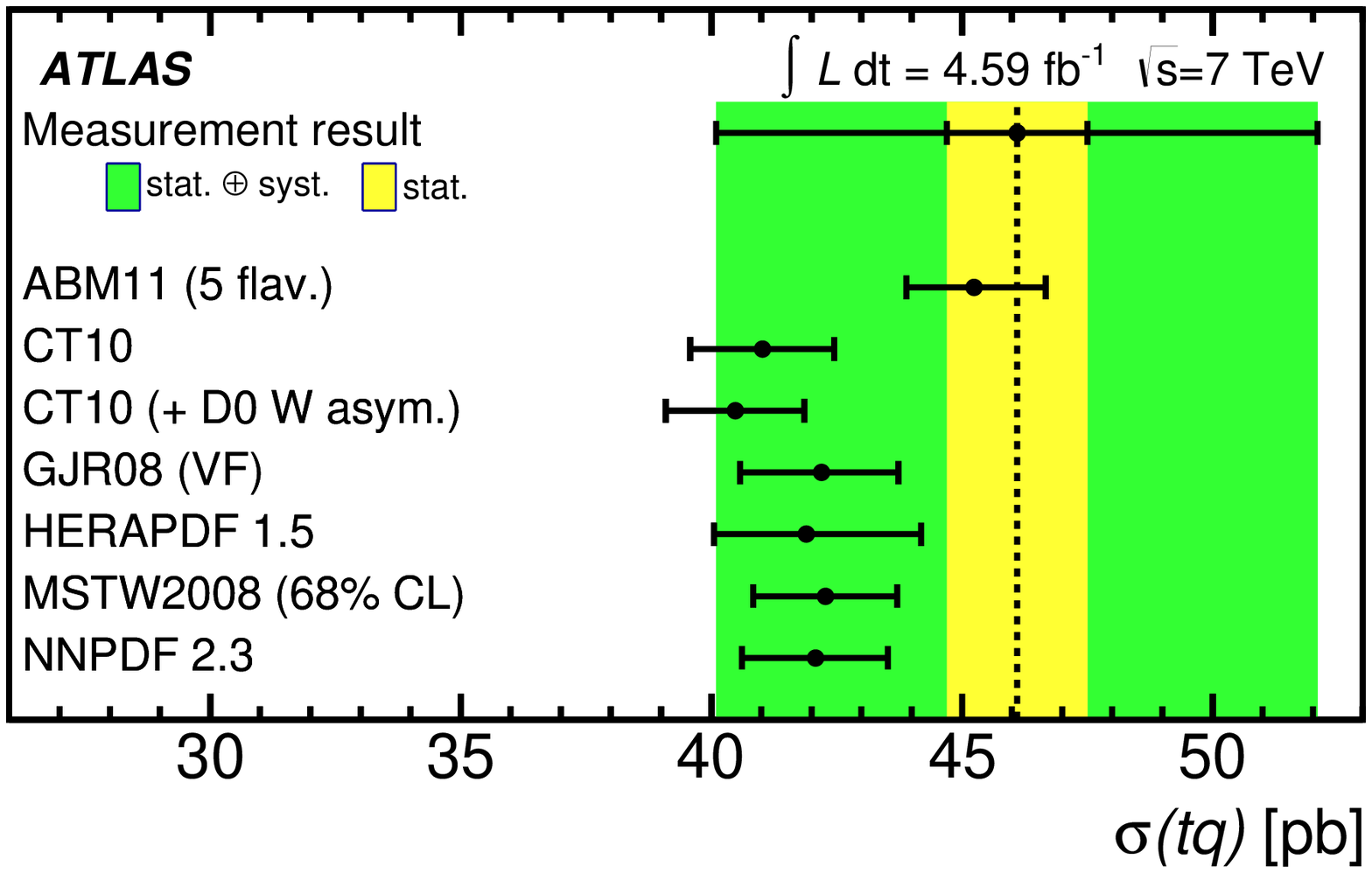}
     \label{subfig:topPDFcomp}
  }
  \subfigure[]{
     \includegraphics[width=0.45\textwidth]{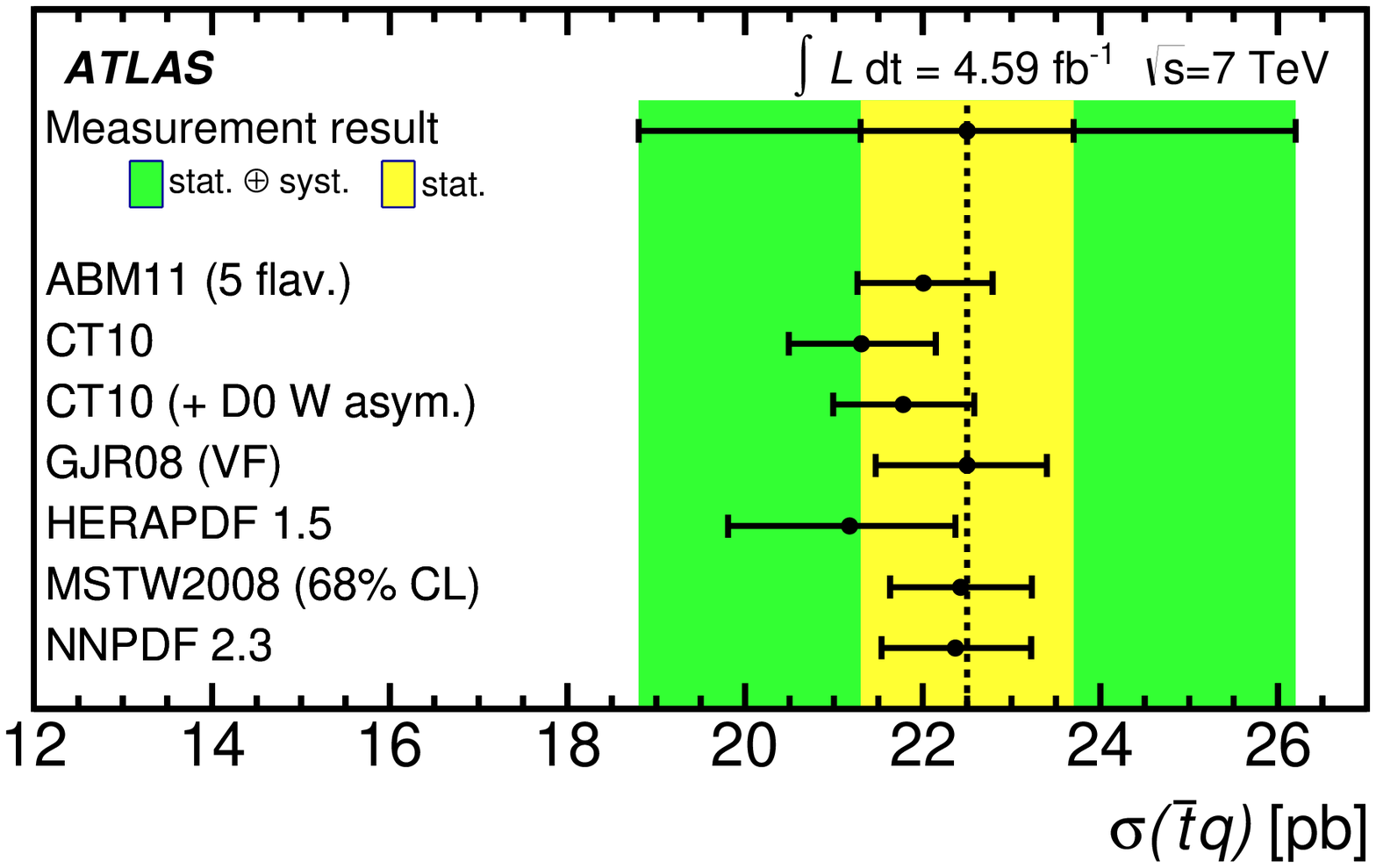}
     \label{subfig:antitopPDFcomp}
  }
  \subfigure[]{
     \includegraphics[width=0.45\textwidth]{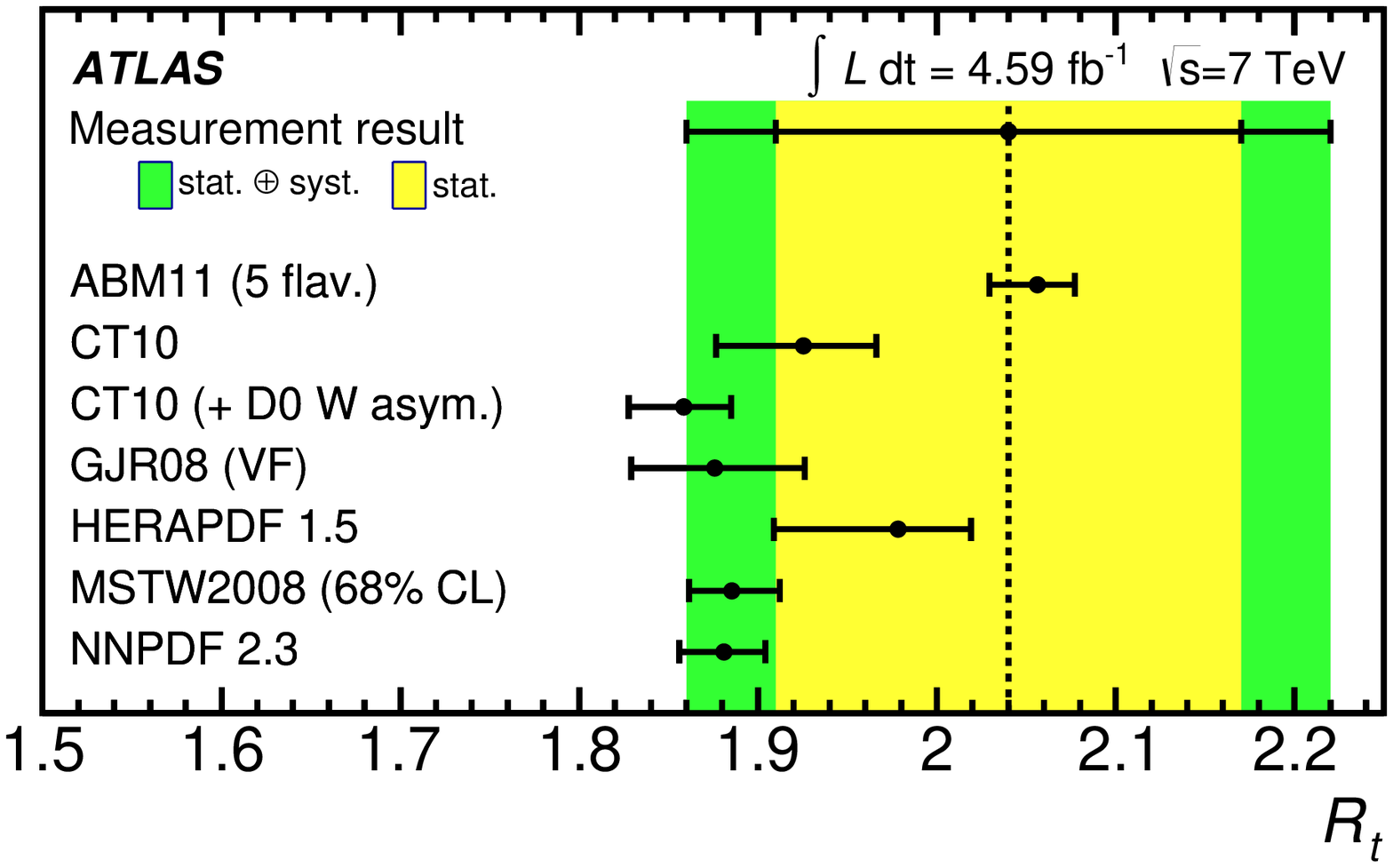}
     \label{subfig:RtPDFcomp}
  }
  \subfigure[]{
     \includegraphics[width=0.45\textwidth]{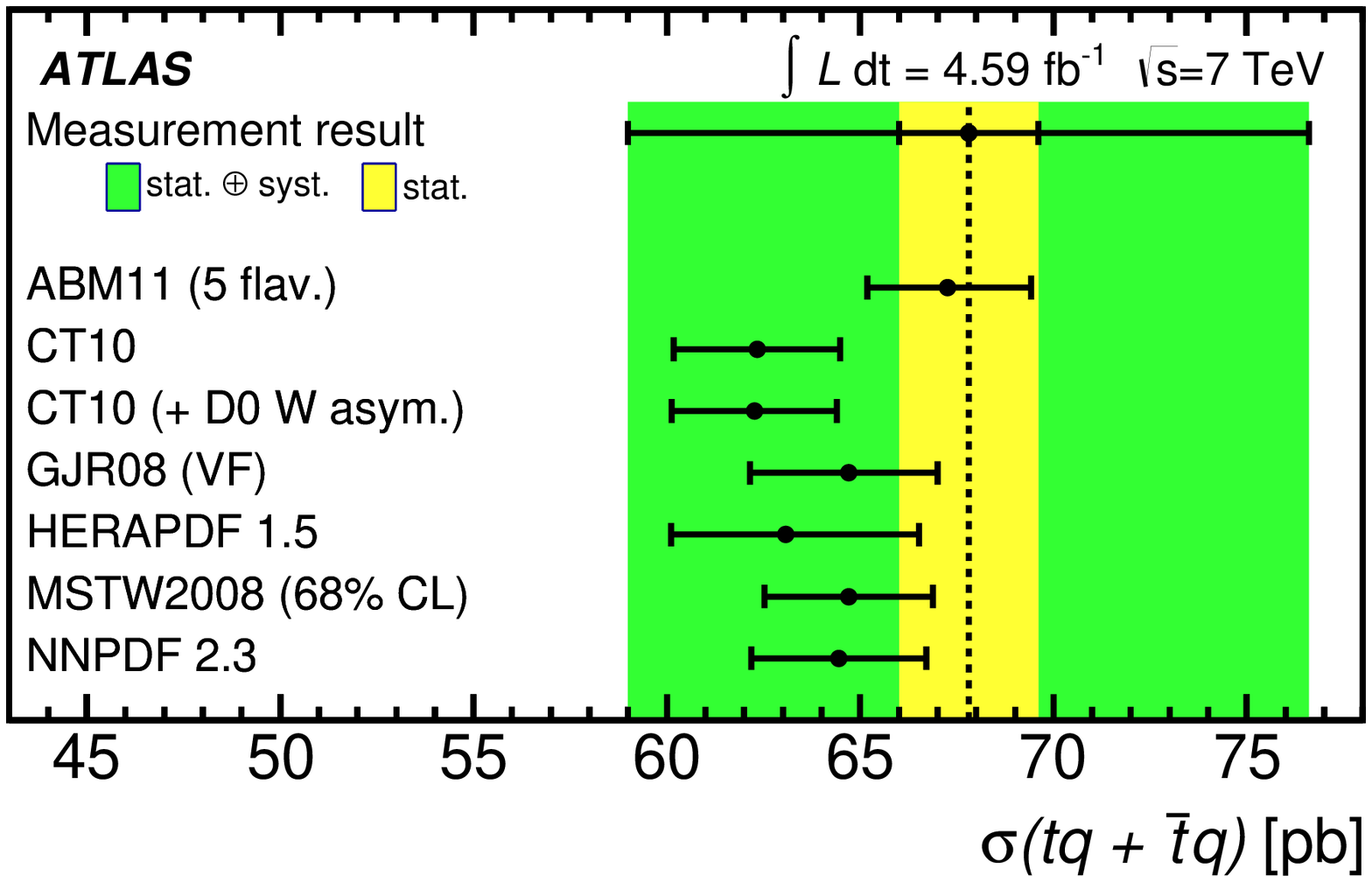}
     \label{subfig:combPDFcomp}
  }
  \caption{\label{fig:rtop} Comparison between observed and predicted values of \subref{subfig:topPDFcomp}~$\sigma(tq)$, 
  \subref{subfig:antitopPDFcomp}~$\sigma(\tbar q)$, \subref{subfig:RtPDFcomp}~$R_t$, and \subref{subfig:combPDFcomp}~$\sigma(tq+\tbar q)$. 
  The predictions are calculated at NLO precision~\cite{Campbell:2009ss,Kant:2014oha} in the five-flavor scheme and 
  given for different NLO PDF sets~\cite{Gluck:2008gs,Aaron:2009aa,Aaron:2009kv} and the uncertainty includes the uncertainty on the renormalization and factorization scales, 
  the combined internal PDF and $\alpha_{\mathrm{s}}$ uncertainty. 
  The dotted black line indicates the central value of the measured value. 
  The combined statistical and systematic uncertainty of the measurement is shown in green, 
  while the statistical uncertainty is represented by the yellow error band. }
\end{figure*}

After performing the binned maximum-likelihood fit and estimating the total uncertainty, the cross sections 
of top-quark and top-antiquark production in the $t$-channel and their cross-section ratio $R_t$ are measured to be:
$$\setlength\arraycolsep{0.1em}
 \begin{array}{rclcl}
  \sigma(tq)       & = & 46 \pm 1\, (\mathrm{stat.}) \pm 6\, (\mathrm{syst.})\, \mathrm{pb} & = & 46\pm 6\, \mathrm{pb}, \\
  \sigma(\bar{t}q) & = & 23 \pm 1\, (\mathrm{stat.}) \pm 3\, (\mathrm{syst.})\, \mathrm{pb} & = & 23 \pm 4\, \mathrm{pb} \ \ \ \mathrm{and}  \\
  R_t             & = & 2.04 \pm 0.13\, (\mathrm{stat.})\, \pm 0.12\, (\mathrm{syst.}) & = & 2.04 \pm 0.18,
 \end{array}
$$
assuming a top-quark mass of $m_t = 172.5 \gev$.
Figure~\ref{fig:rtop} compares the measured values of $\sigma(tq)$, $\sigma(\bar{t}q)$, and $R_t$ to NLO predictions from
MCFM~\cite{Campbell:2009ss} and Hathor~\cite{Kant:2014oha} using different PDF
sets. Uncertainties on the predicted values include the uncertainty on the renormalization and factorization scales 
and the combined PDF and $\alpha_{\mathrm{s}}$ uncertainty of the respective PDF set. 

All PDF predictions are in agreement with all measurements. For $\sigma(\bar{t}q)$, the predictions of all PDF sets agree well with each other and with the measured value.
The predictions for $\sigma(tq)$ and $R_t$ with the {\sc ABM11} PDF set~\cite{Alekhin:2012ig} show an offset 
compared to the other predictions. With increasing precision, the measurement of these
observables could provide a way to further constrain the involved PDFs. 

\subsection{Inclusive cross-section measurement}

The inclusive $t$-channel cross section $\sigma(tq+\bar{t}q)$ is extracted by 
using only one scale factor $\beta(tq+\bar{t}q)$ in the likelihood function, scaling
the top-quark and the top-antiquark contributions simultaneously. 
The top-quark-to-antiquark ratio is taken from the approximate NNLO 
prediction~\cite{Kidonakis:2011wy} (see Sec.~\ref{sec:intro}).
The systematic uncertainties on the measured value of inclusive cross-section are 
determined as described in Sec.~\ref{sec:systematics}. A detailed list of the 
uncertainties is given in Table~\ref{tab:xs_uncertainties}.

The binned maximum-likelihood fit yields  
a cross section of 
$$\setlength\arraycolsep{0.1em}
\begin{array}{rcl}
  \sigma_{t}(tq+\bar{t}q) &=& 68 \pm 2\,(\mathrm{stat.})\;\pm 8\,(\mathrm{syst.})\;\mathrm{pb} \\
                          &=& 68\pm 8\;\mathrm{pb,}
\end{array}
$$
assuming $m_t = 172.5 \gev$.
Figure~\ref{subfig:combPDFcomp} compares the measured value for $\sigma(tq+\bar{t}q)$ to NLO
predictions~\cite{Campbell:2009ss,Kant:2014oha} obtained with different PDF sets. All predictions are in agreement with the measurement.

\subsection{Cross-section dependence on the top-quark mass}

The $t$-channel single top-quark cross sections are measured using a signal model with $m_t = 172.5 \gev$. 
The dependence of the cross-section measurements on $m_t$ is mainly due to acceptance effects
and is expressed by the function:
\begin{equation}
\sigma_{t} = \sigma_{t}(172.5 \gev) + p_1 \cdot \Delta m_t  + p_2 \cdot \Delta m_t^{2}
\label{eq:topmass}
\end{equation}
with $\Delta m_t = m_t - 172.5 \gev$. The parameters $p_1$ and $p_2$ are determined using dedicated signal samples with different $m_t$ and are given 
in Table~\ref{tab:paratopmass} 
for $\sigma(tq)$, $\sigma(\tbar q)$ and $\sigma(tq+\tbar q)$.
The cross-section ratio $R_t$ is largely independent of the top-quark mass.

\begin{table}[htbp]
  \centering
  \caption{Parameterization factors for the $m_t$ dependence (see Eq.~(\ref{eq:topmass})) of $\sigma(tq)$, $\sigma(\tbar q)$ and $\sigma(tq+\tbar q)$.
  \label{tab:paratopmass}}
\begin{ruledtabular} 
\begin{tabular}{lcc}
     & p$_1$ [pb/\gev] & p$_2$ [pb/$\gev^{2}$]\\
    \hline 
    $\sigma(tq+\tbar q)$  & $-0.46$ & $-0.06$ \\
    $\sigma(tq)$ & $-0.27$ & $-0.04$ \\
    $\sigma(\tbar q)$ & $-0.19$ & $-0.02$ \\
    \end{tabular}%
\end{ruledtabular}
\end{table}%

\subsection{$V_{tb}$ extraction}

Since $\sigma(tq+\bar{t}q)$ is proportional to $|V_{tb}|^2$, $|V_{tb}|$ can be extracted from the 
measurement. 
The $|V_{tb}|$ measurement is independent of assumptions about the number of 
quark generations and about the unitarity of the CKM matrix.
The only assumptions required are that $|V_{tb}|\gg|V_{td}|,|V_{ts}|$
and that the $Wtb$ interaction is an SM-like left-handed weak coupling.
The $t\bar{t}$-background rate is unaffected by a variation 
of $|V_{tb}|$, since the decay to a quark of a potentially existing higher generation are 
prohibited by kinematics, such that the branching ratio $B(t \rightarrow Wb)\sim 1$.
On the other hand, the rates of single-top quark $Wt$ and $s$-channel backgrounds also scale with $|V_{tb}|^2$,
but their contributions are small in the signal region. The resulting variation of the total top-quark background 
yield is less than its systematic uncertainty and thus considered negligible.

The value of $|V_{tb}|^2$ is extracted by dividing the
measured value of $\sigma(tq+\bar{t}q)$ by the prediction of the approximate NNLO 
calculation~\cite{Kidonakis:2011wy}. 
The experimental and theoretical uncertainties are added in quadrature.
The result obtained is 
$$\setlength\arraycolsep{0.1em}
\begin{array}{rcl}
|V_{tb}| &=& 1.02 \pm 0.01 (\mathrm{stat.}) \pm 0.06\,(\mathrm{syst.})\pm 0.02\,(\mathrm{theo.})\,^{+0.01}_{-0.00}\,(m_t) \\
          &=& 1.02 \pm 0.07. 
\end{array}
$$
A lower limit on $|V_{tb}|$ is extracted in a Bayesian limit computation,
assuming that the likelihood curve of $|V_{tb}|^2$ has a Gaussian shape, 
centered at the measured value.
A flat prior in $|V_{tb}|^2$ is applied, being one in the interval $[0, 1]$ and
zero otherwise. The resulting lower limit is
$|V_{tb}|>0.88$ at the 95\% CL.

\section{Differential cross-section measurements}

Differential cross sections are measured as a function of the $\pT$ and $|y|$ of $t$ and $\bar{t}$ 
in the 2-jet HPR channels, defined in Sec.~\ref{sec:hpr}. 

\subsection{Signal yield and reconstructed variables}

\begin{figure*}[htb]
  \centering
  \subfigure[]{
     \includegraphics[width=0.45\textwidth]{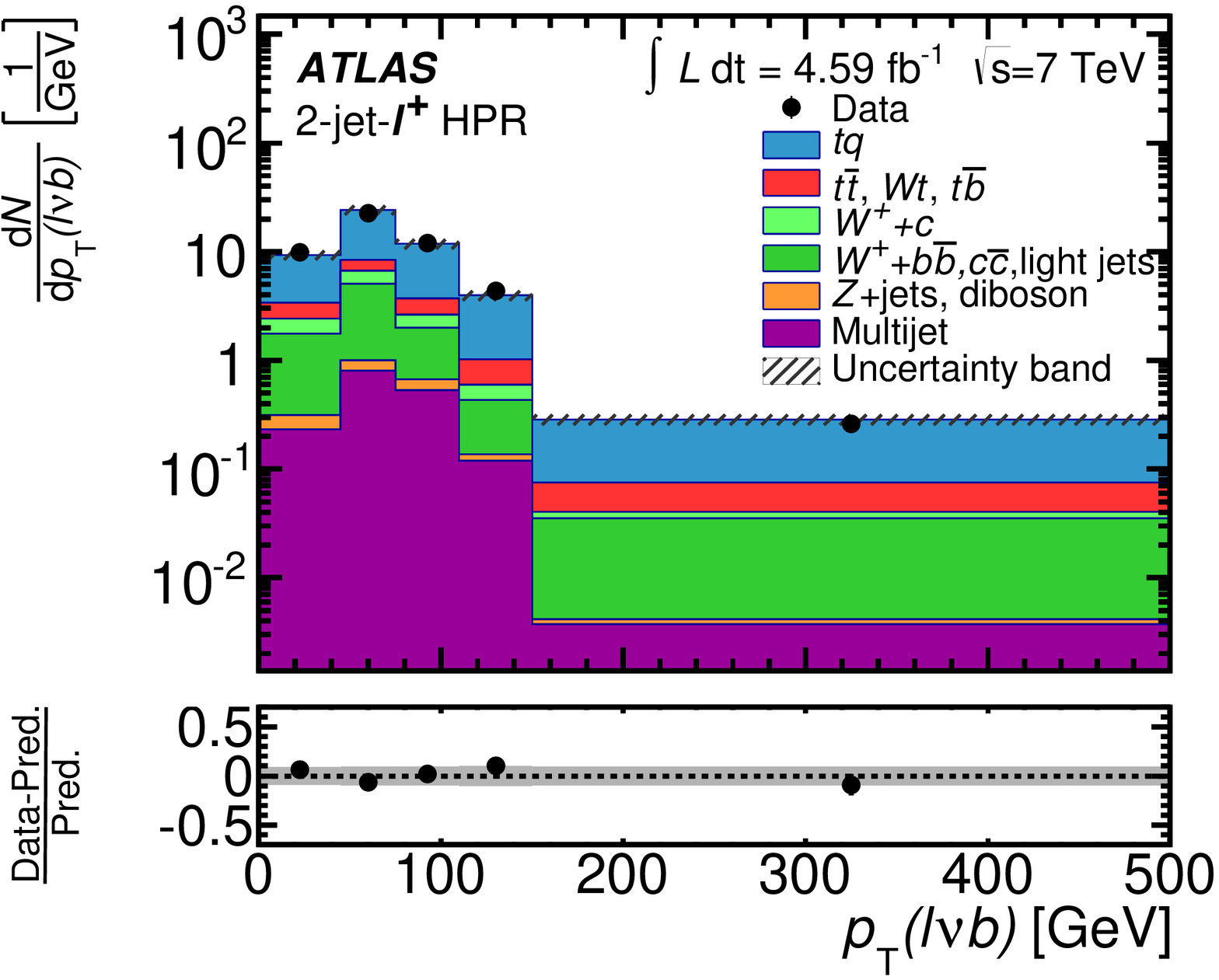}
     \label{subfig:2j_hpr_plus_toppt}  
  }
  \subfigure[]{
     \includegraphics[width=0.45\textwidth]{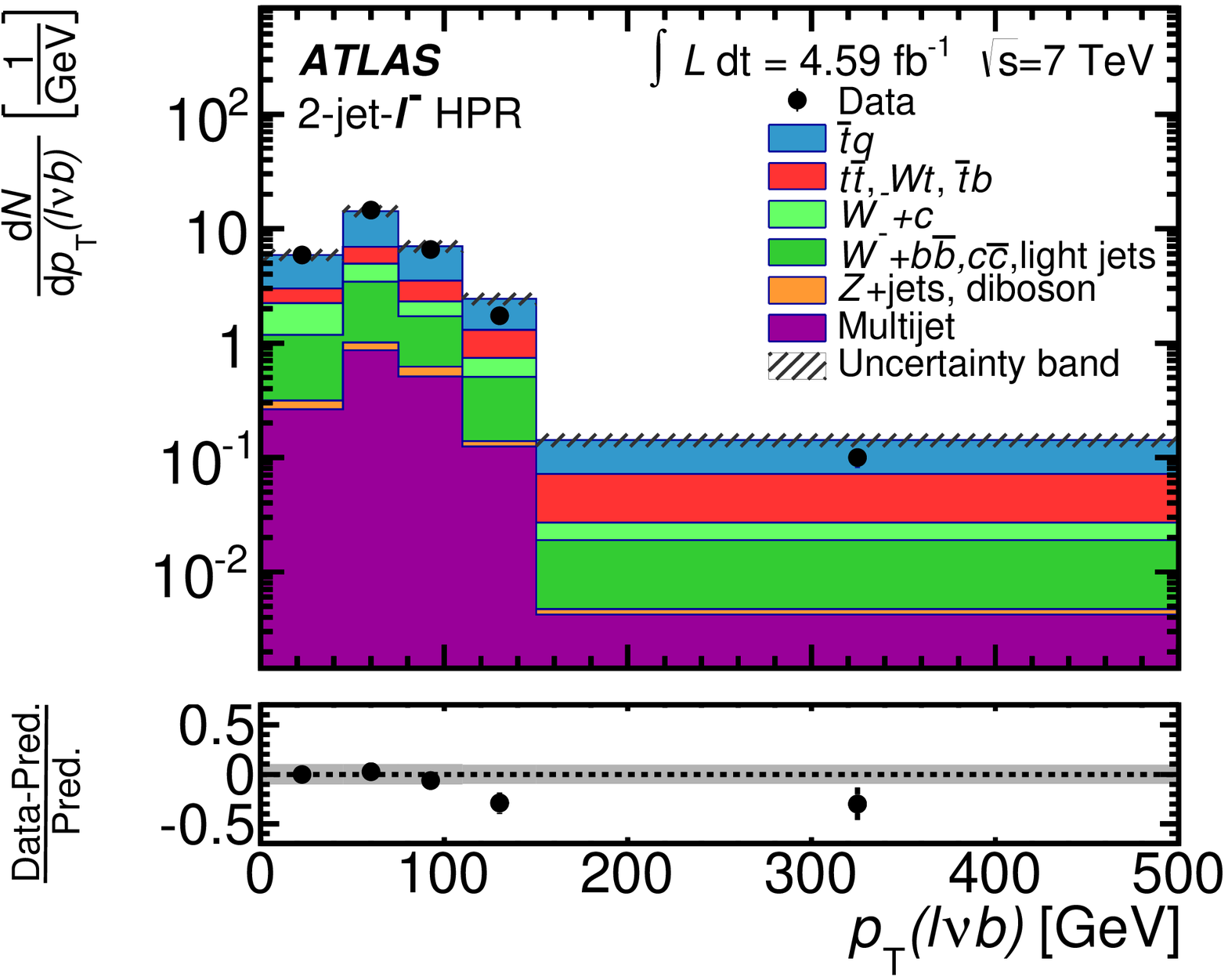}
     \label{subfig:2j_hpr_minus_toppt}  
  }
  \subfigure[]{
     \includegraphics[width=0.45\textwidth]{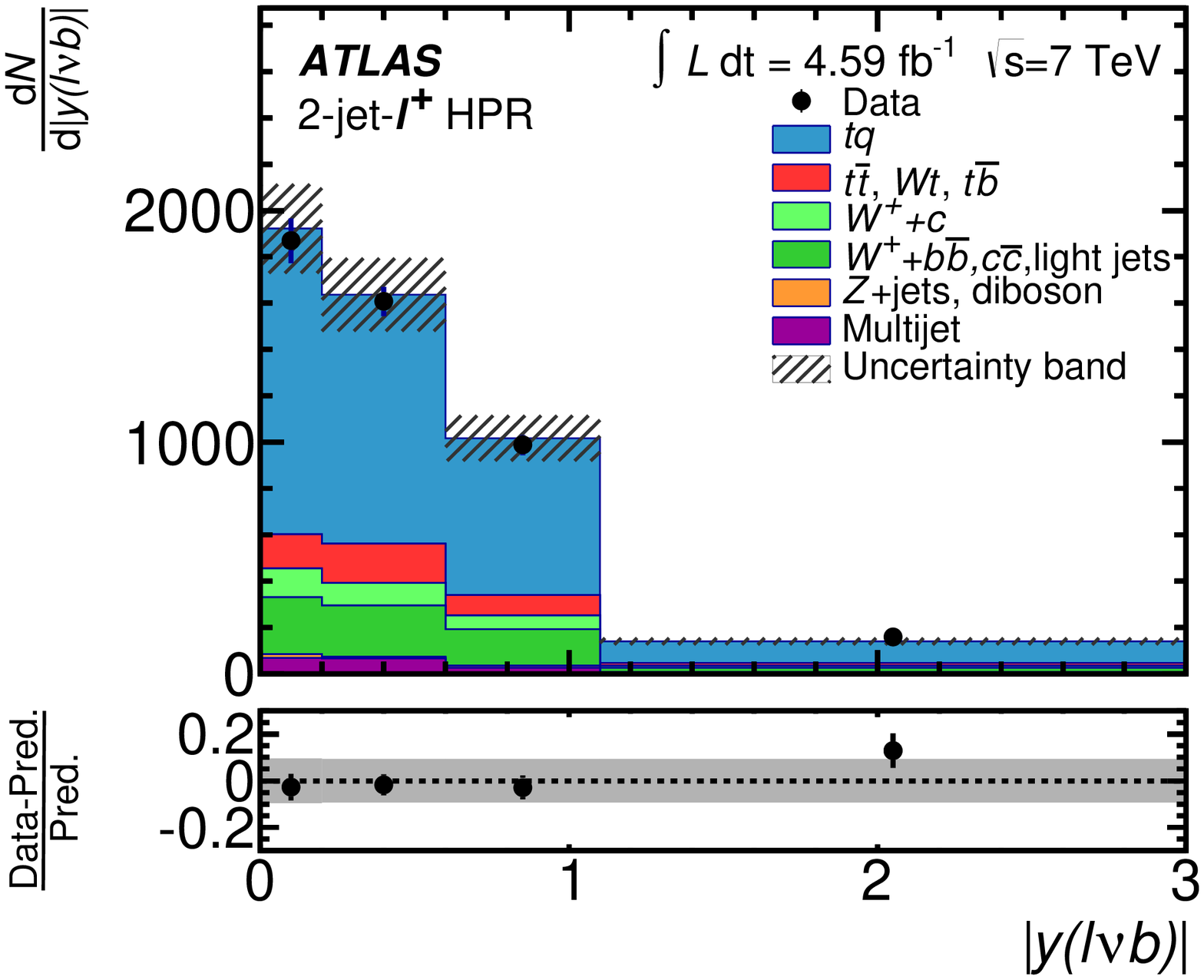}
     \label{subfig:2j_hpr_plus_toprap}  
  }
  \subfigure[]{
     \includegraphics[width=0.45\textwidth]{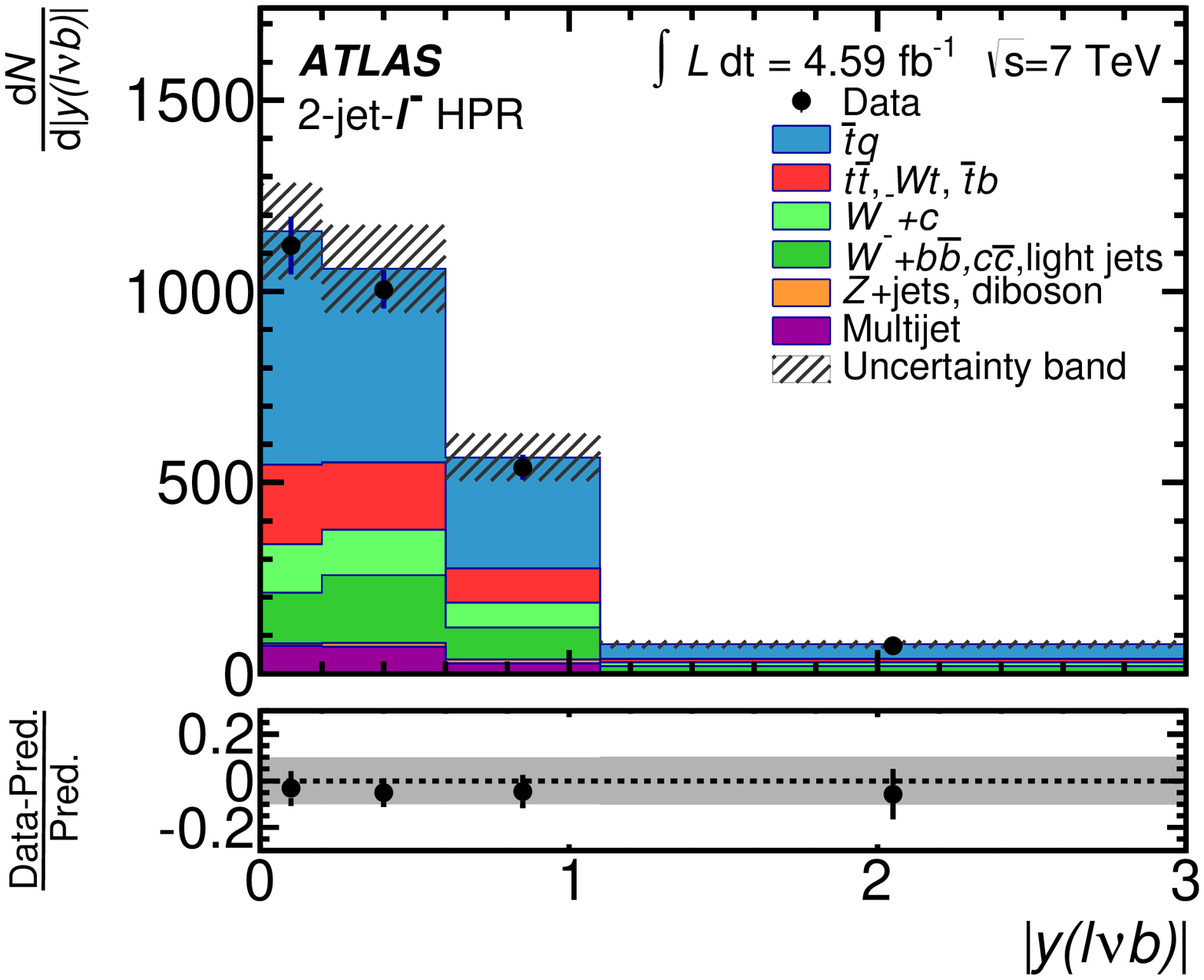}
     \label{subfig:2j_hpr_minus_toprap}  
  }
  \caption{\label{fig:measured_dist} Measured distributions of \subref{subfig:2j_hpr_plus_toppt} the top-quark $\pt$, \subref{subfig:2j_hpr_minus_toppt}
  top-antiquark $\pt$, \subref{subfig:2j_hpr_plus_toprap} top-quark $|y|$,
  and \subref{subfig:2j_hpr_minus_toprap} top-antiquark $|y|$ shown on reconstruction level
  in the HPR normalized to the result of the binned maximum-likelihood fit. 
    The uncertainty band represents the normalization uncertainty 
    of all processes after the fit and the Monte Carlo statistical uncertainty, added in quadrature. The relative difference between the observed and
    expected number of events in each bin is shown in the lower panels. 
  }
\end{figure*}

The signal and background composition in the 
2-jet-$\ell^+$ and the 2-jet-$\ell^-$ HPR channels can be found in Table~\ref{tab:evtyield_hp}. 
Figure~\ref{fig:measured_dist} shows the measured distributions of the reconstructed top-quark $\pT$ and the reconstructed top-quark 
$|y|$ normalized to the result of the binned maximum-likelihood fit performed to measure 
$\sigma(tq)$ and $\sigma(\bar{t}q)$.

The binning of the differential cross sections is chosen based on the experimental resolution of the $\pT$ and $|y|$ distributions
as well as the data statistical uncertainty.
Typical values for the resolution of the top-quark $\pt$ are 10~$\gev$, increasing to 25~$\gev$ in the tail of the
distribution. The resolution of the rapidity varies from 0.2 to 0.4 from central to forward rapidities.

\begin{table}[htbp]
  \centering
  \caption{Event yields for the 2-jet-$\ell^{+}$ and 2-jet-$\ell^{-}$ HPR channels.  
  The expectation for the signal and background yields correspond to the result of the binned maximum-likelihood fit
  described in Sec.~\ref{sec:fitresult}. The uncertainty of the expectations is the normalization uncertainty  
    of each processes after the fit, as described in Sec.~\ref{sec:unc_est}.
\label{tab:evtyield_hp}}
\begin{ruledtabular} 
    \begin{tabular}{lr@{\extracolsep{0pt}$\:\pm\:$}l@{\hspace{0.4cm}}@{\extracolsep{\fill}}
                     r@{\extracolsep{0pt}$\:\pm\:$}l@{\hspace{0.4cm}}@{\extracolsep{\fill}}}
     & \multicolumn{2}{c}{2-jet-$\ell^{+}$ HPR} & \multicolumn{2}{c}{2-jet-$\ell^{-}$ HPR} \\
    \hline 
    $tq$                                & 1210 & 150  & 1.3  & 0.2 \\
    $\bar{t}q$                          & 0.29 & 0.05 & 549  & 87\\
    $\ttbar$,$Wt$,$t\bbar$,$\tbar b$    & 161  & 18   & 175  & 19 \\
    $W^+$+$b\bbar$,$c\cbar$,light jets  & 250  & 48   & 0.35 & 0.07 \\
    $W^-$+$b\bbar$,$c\cbar$,light jets  & 0.7  & 0.2  & 166  & 40 \\
    $W$+$c$                             & 110  & 26   & 125  & 30 \\
    $Z$+jets, diboson                   & 15   & 10   & 11.4 & 6.8 \\
    Multijet                            & 59   & 30   & 62   & 31 \\
    \multicolumn{5}{c}{ } \\
    Total expectation                   & 1810 & 160  & 1090 & 110 \\
    Data  & \multicolumn{2}{c}{1813$\;\;\;\;\;$}  & \multicolumn{2}{c}{1034$\;\;\;\;\;\;$} \\
    \end{tabular}%
\end{ruledtabular}
\end{table}%

\subsection{Method}

\begin{figure*}[htb]
  \centering
\subfigure[]{
\includegraphics[width=0.46\textwidth]{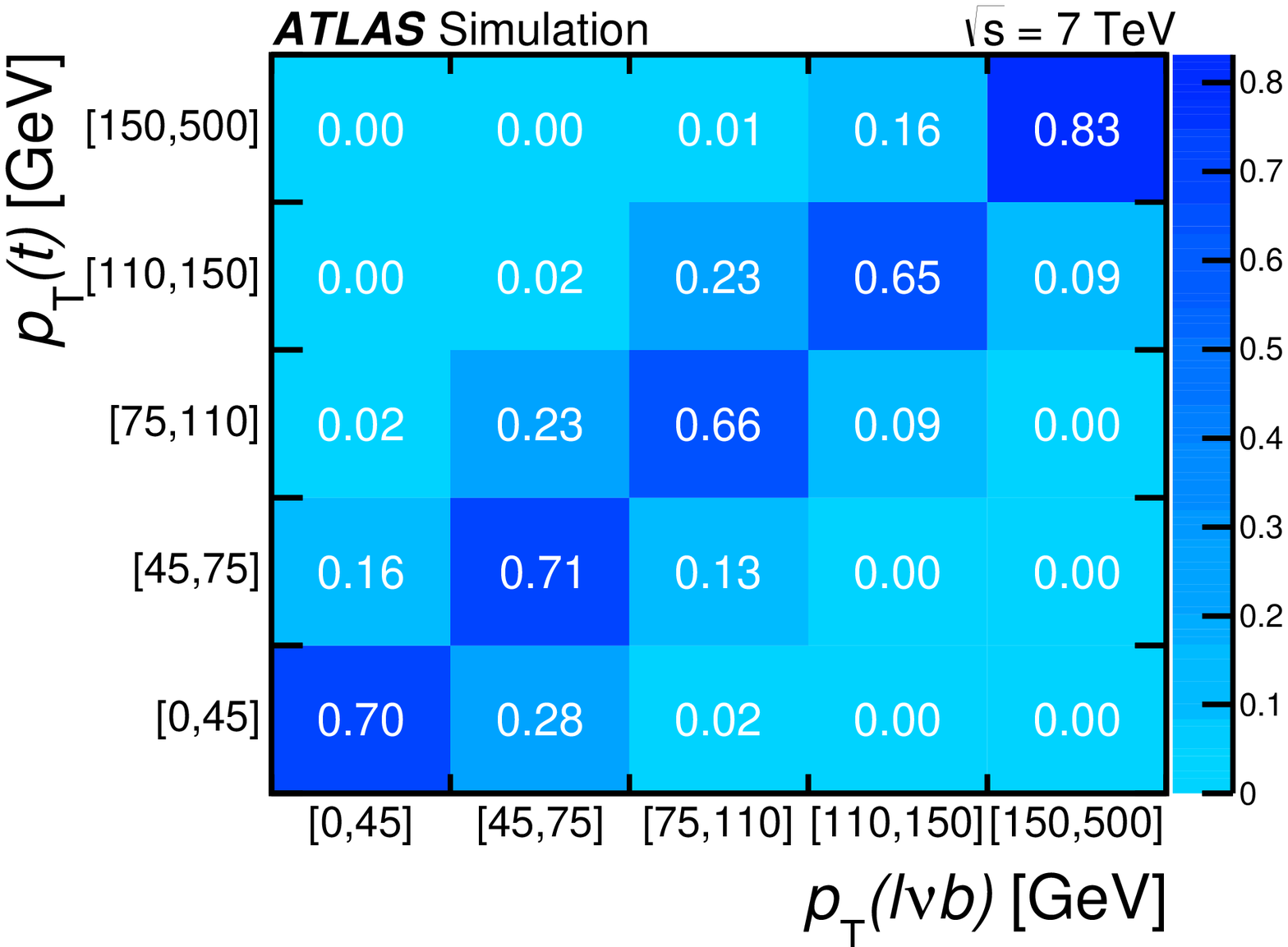}
     \label{subfig:mm_toppt}  
}
\subfigure[]{
\includegraphics[width=0.46\textwidth]{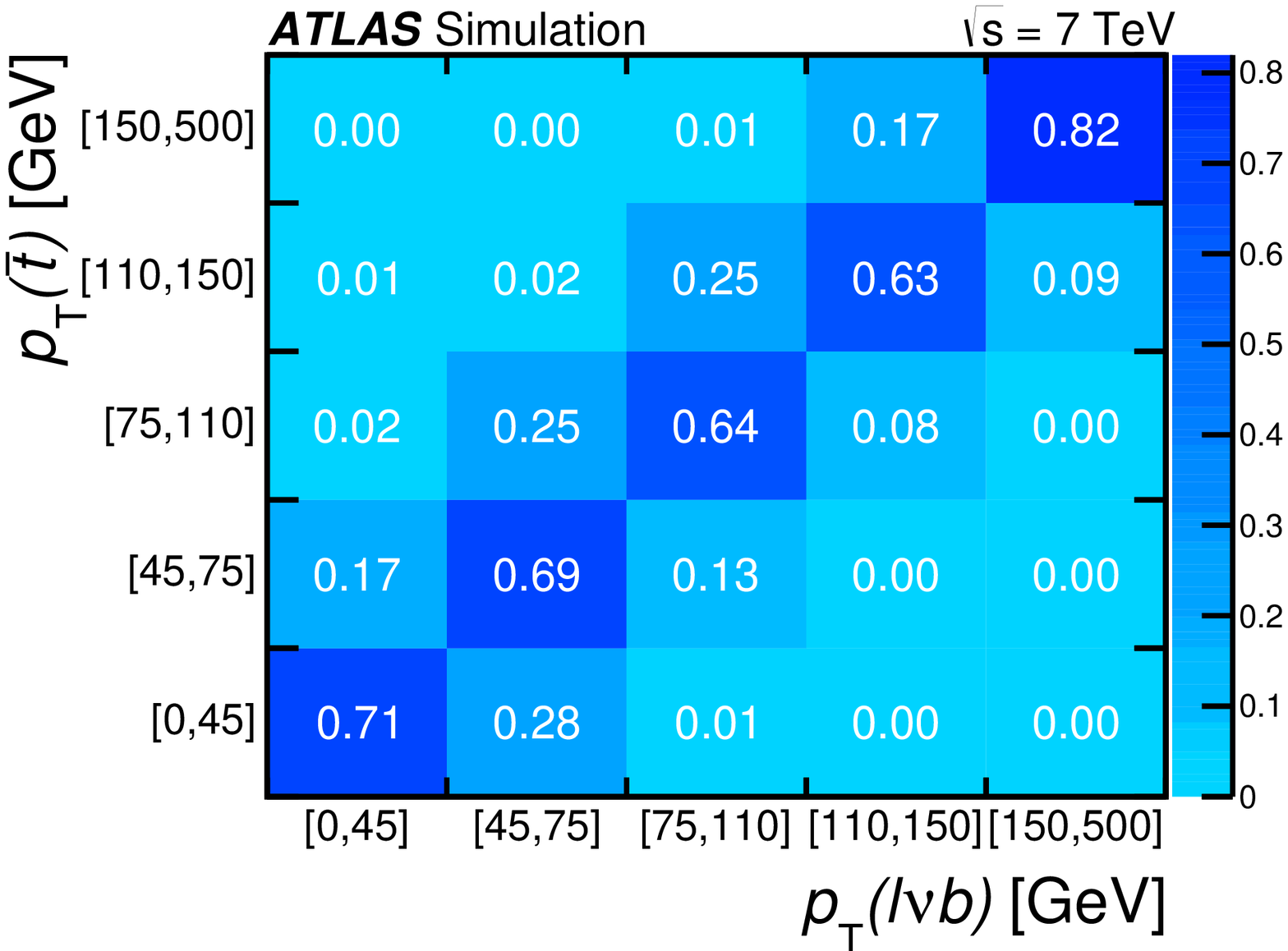}
     \label{subfig:mm_antitoppt}  
}
\subfigure[]{
\includegraphics[width=0.46\textwidth]{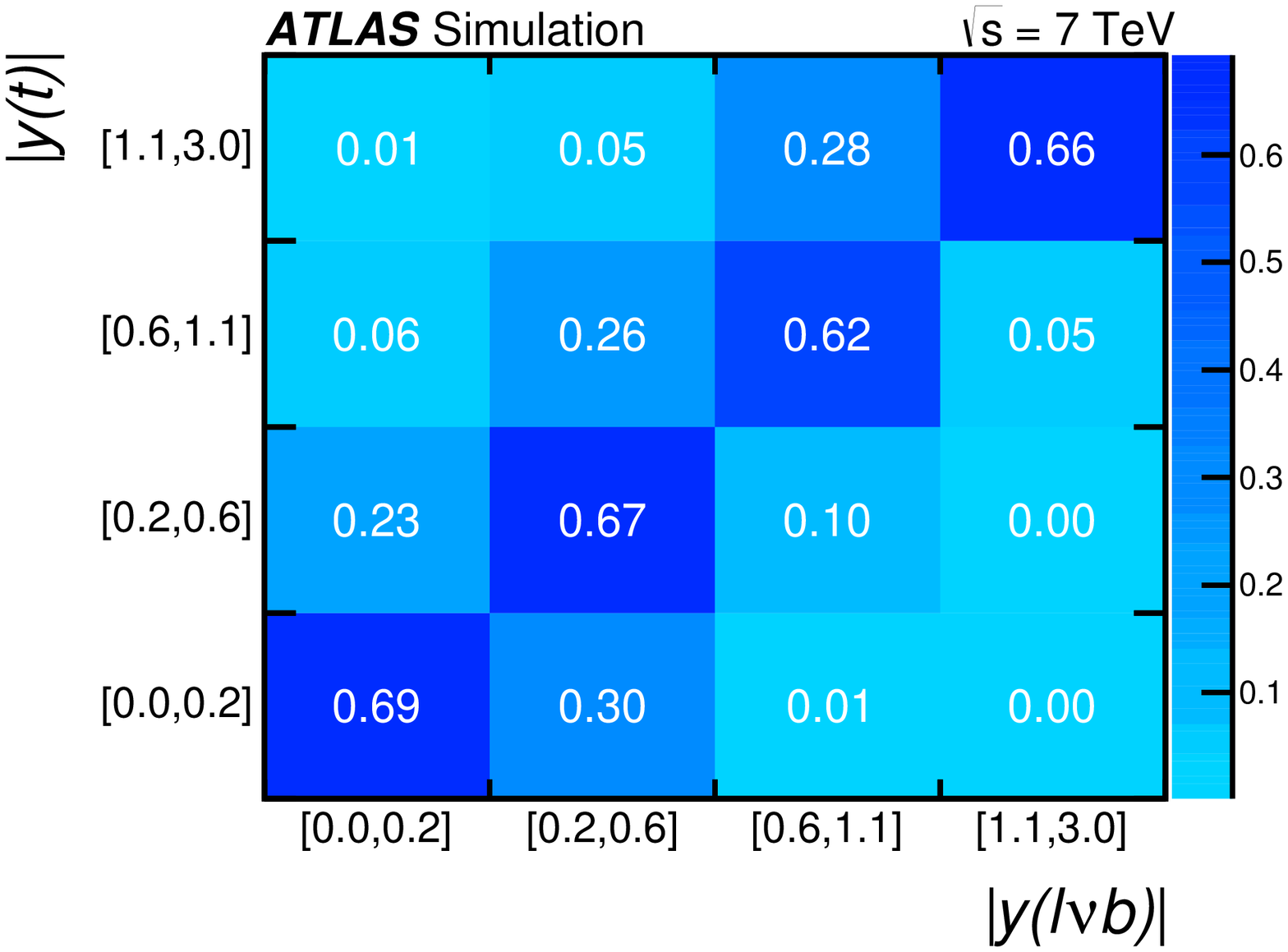}
     \label{subfig:mm_toprap}  
}
\subfigure[]{
\includegraphics[width=0.46\textwidth]{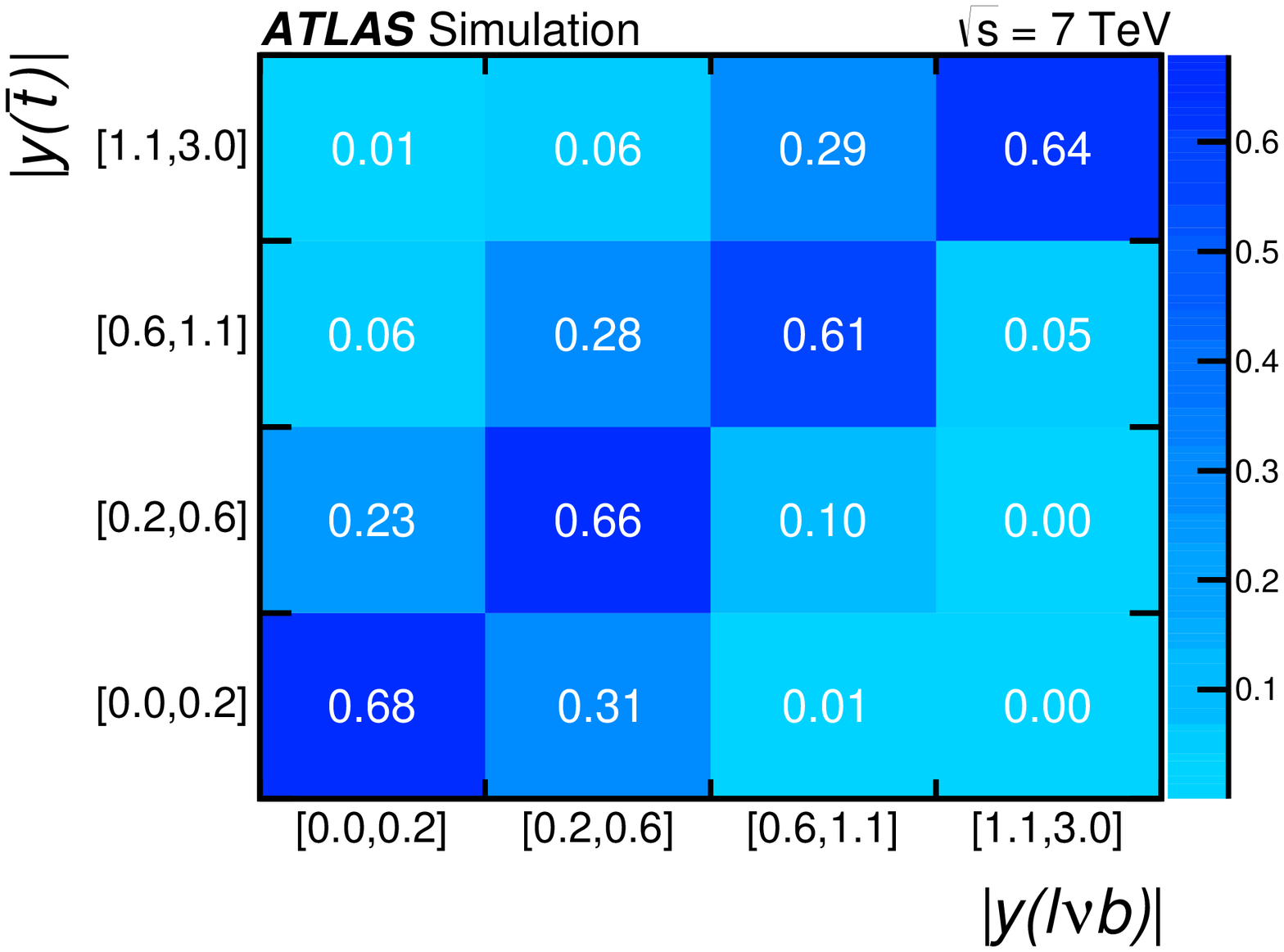}
     \label{subfig:mm_antitoprap}  
}
\caption{Migration matrices relating the parton level shown on the $y$ axis and reconstruction level shown on the $x$ axis for the \subref{subfig:mm_toppt} top-quark $\pt$, \subref{subfig:mm_antitoppt} top-antiquark $\pt$, \subref{subfig:mm_toprap} top-quark $|y|$, and \subref{subfig:mm_antitoprap} top-antiquark $|y|$ distribution.}
\label{fig:migrationm}
\end{figure*}

The measured distributions are distorted by detector effects and acceptance effects.
The observed distributions are unfolded to the (parton level) four-momenta of
the top quarks before the decay and after QCD radiation to correct for these distortions.
In the following, each bin of the measured distribution is referred to by the index $i$, while
each bin of the parton-level distribution is referred to by the index $j$. 
The relation between the measured distribution and the differential 
cross section in each bin $j$ of the parton-level distribution can be written as:
\begin{equation}
\frac{d\sigma}{dX_j} = \frac{1}{\Delta X_j}\cdot \frac{\sum\limits_{i} M_{ij}^{-1} \cdot (N_i -B_i)}{\mathcal{L} \cdot \epsilon_j \cdot \it{B}(t\rightarrow \ell\nu b)}
\end{equation}
where $\Delta X_j$ is the bin width of the parton-level distribution, $N_i$ ($B_i$) are the data (expected background) yields in 
each bin of the measured distribution, $\mathcal{L}$ is the integrated luminosity of the data sample,
$\epsilon_j$ is the event selection efficiency and $M_{ij}^{-1}$ is the inverse of the migration matrix.
The migration matrix accounts for the detector response and is defined as the probability to observe an 
event in bin $i$ when it is generated in bin $j$. The 
migration matrix is built by relating the variables at the reconstruction and at the parton level using the signal simulation.
Figure~\ref{fig:migrationm} shows the migration matrices for the $\pT$ and $|y|$ distributions of the top quark and top antiquark.
The inverse of the matrix is determined by applying Bayes' theorem iteratively~\cite{D'Agostini1995487} 
in order to perform the unfolding. The number of iterations is chosen such that 
the absolute change in the unfolded distributions 
is on average smaller than 1\% of the content in each bin. This procedure results in a total of five iterations for all distributions.
The selection efficiency $\epsilon_j$ in bin $j$ of each variable is defined 
as the ratio of the parton-level yield before and after selection and is evaluated using simulation.
The efficiencies are typically in the 0.5--2.2\% range.

The unfolding is applied to the reconstructed $\pT(\ell\nu b)$ and $|y(\ell\nu b)|$ distributions after subtraction 
of the background contributions. When subtracting the background, all backgrounds are normalized according
to Table~\ref{tab:evtyield_hp}.

Closure tests are performed in order to check the validity of the unfolding procedure. 
The shape of the parton-level distributions in the Monte Carlo simulation are altered to verify that the 
simulation does not bias the results. It is checked that the altered parton-level distributions are recovered by unfolding the reconstructed 
distributions with the nominal migration matrix.

\subsection{Treatment of uncertainties}

The statistical uncertainty of the unfolded results is estimated using pseudo-experiments, propagating the 
uncertainties from the measured distribution and from the size of the Monte Carlo signal and 
background samples through the unfolding process. All sources of systematic uncertainty described in 
Sec.~\ref{sec:systematics} are included for the unfolded distributions. 
In the case of the background normalization, the uncertainties quoted in Table~\ref{tab:evtyield_hp} are taken into account.
The impact of the systematic uncertainties is evaluated by modifying the subtracted background before unfolding 
in the case of uncertainties on the backgrounds. 
To assign uncertainties on the signal modeling, systematic shifts are applied to the simulated signal sample. 
The shifted reconstructed distribution is unfolded and then compared to the nominal distribution at parton level. 

\subsection{Results}

To reduce the impact of systematic uncertainties normalized differential cross sections $1/\sigma \cdot(d\sigma/dX_j)$ are 
calculated by dividing the differential cross section by the total cross section evaluated by integrating over 
all bins.

\begin{figure*}[p]
  \centering
  \subfigure[]{
     \includegraphics[width=0.45\textwidth]{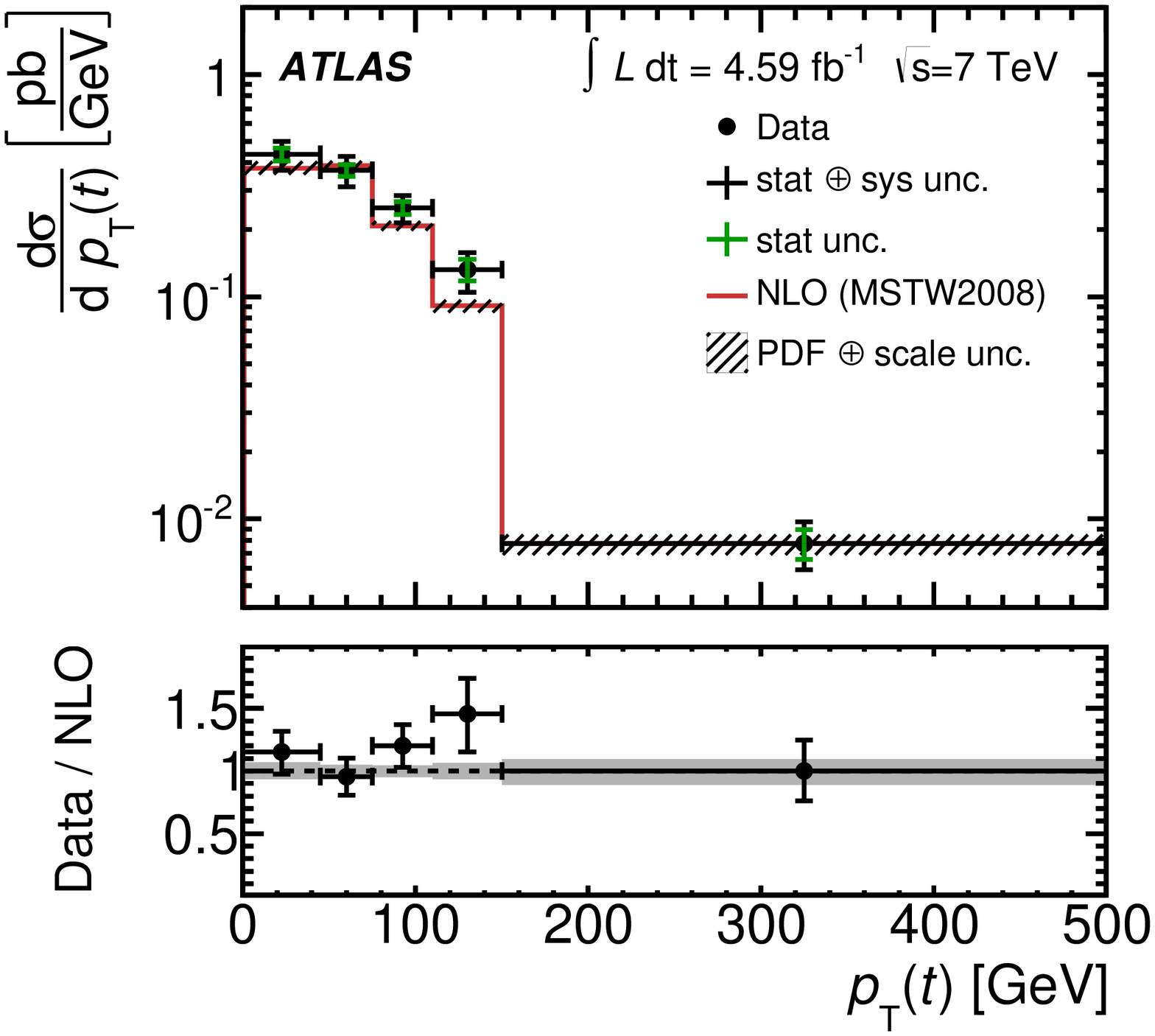}
     \label{subfig:diffdist_toppt}  
  }
  \subfigure[]{
     \includegraphics[width=0.45\textwidth]{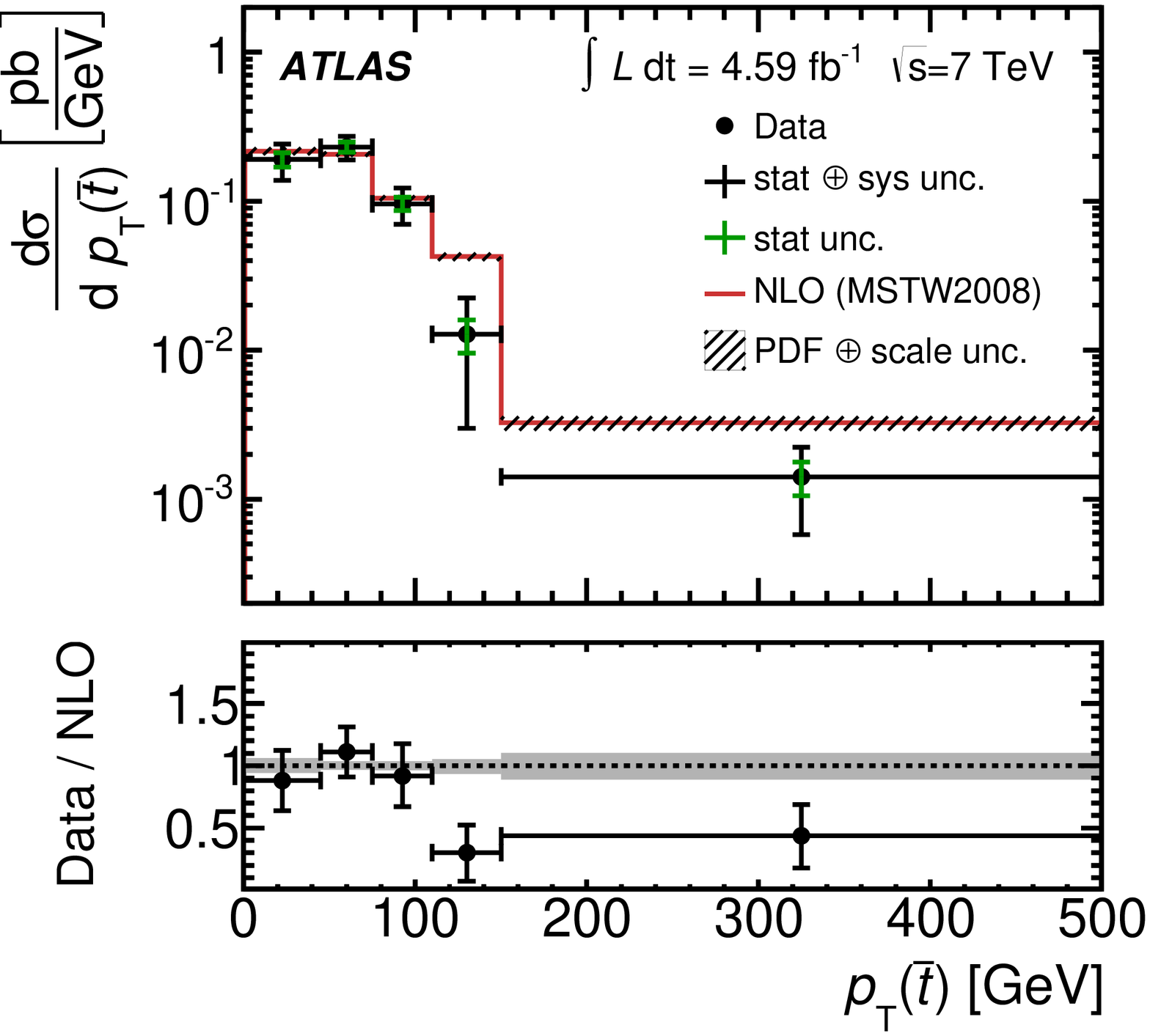}
     \label{subfig:diffdist_antitoppt}  
  }
  \subfigure[]{
     \includegraphics[width=0.45\textwidth]{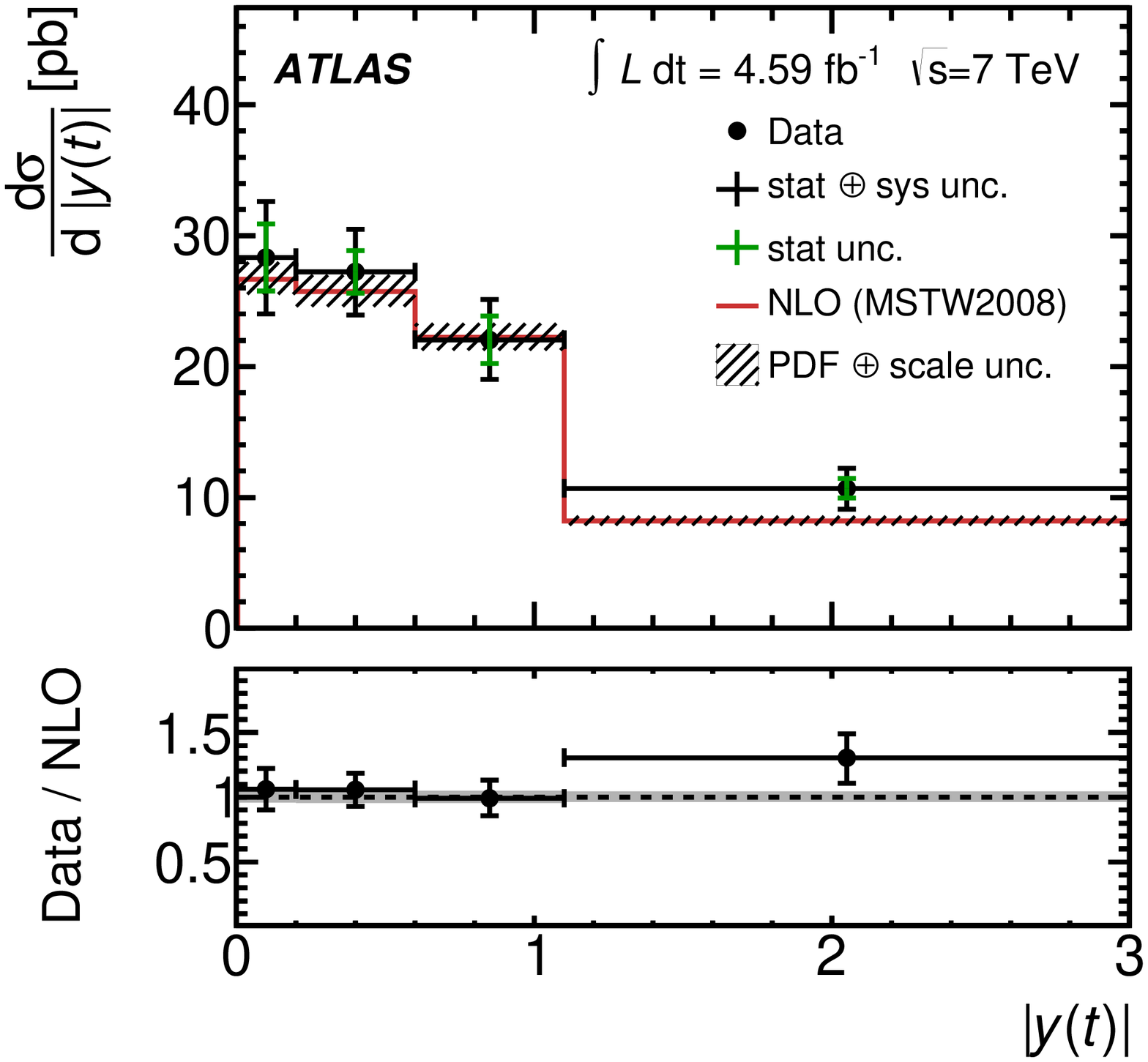}
     \label{subfig:diffdist_toprap}  
  }
  \subfigure[]{
     \includegraphics[width=0.45\textwidth]{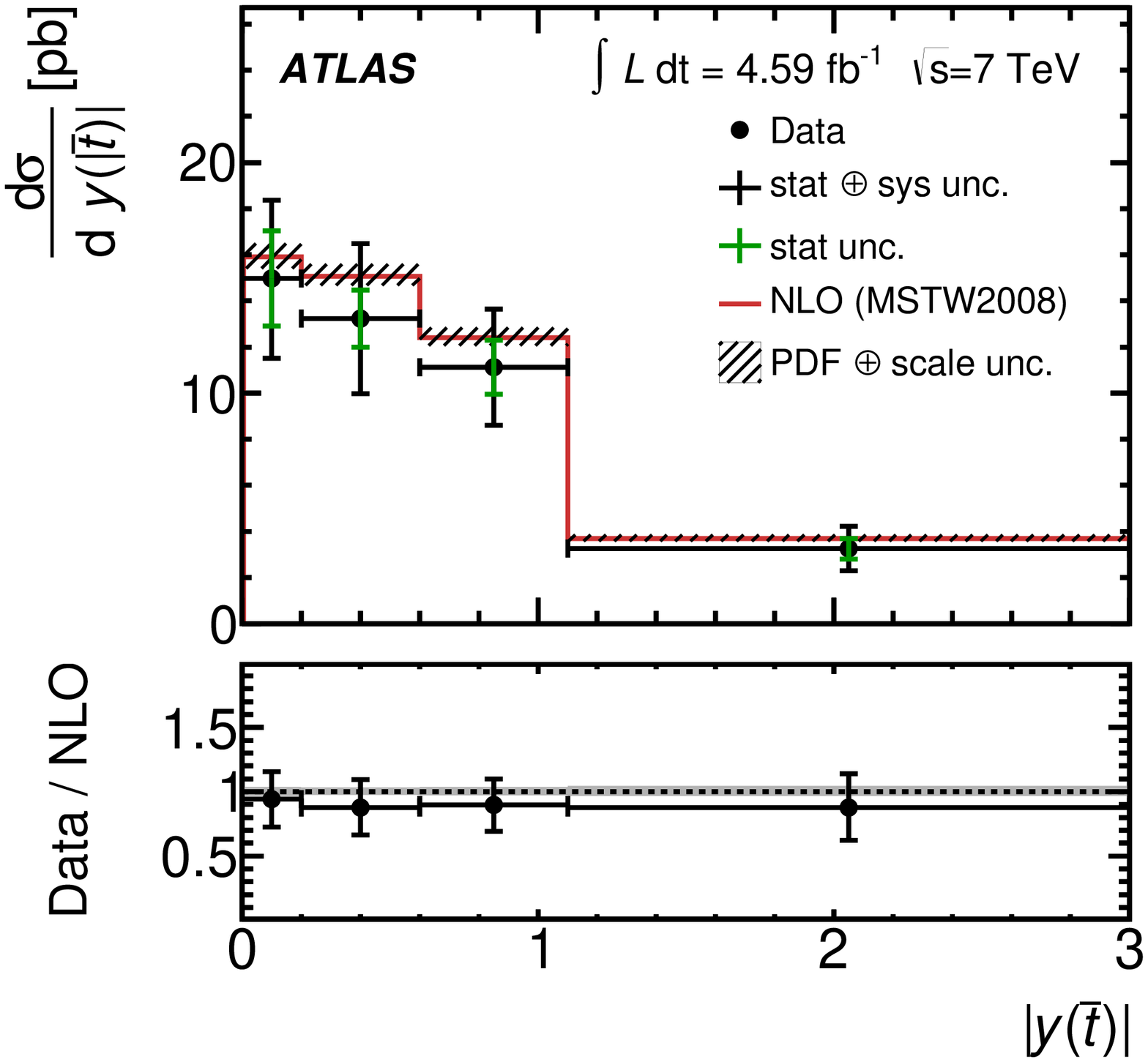}
     \label{subfig:diffdist_antitoprap}  
  }
  \caption{\label{fig:unfolded_dist} Differential cross section as a function of \subref{subfig:diffdist_toppt} $\pt(t)$, \subref{subfig:diffdist_antitoppt} $\pt(\tbar)$, \subref{subfig:diffdist_toprap} $|y(t)|$ and \subref{subfig:diffdist_antitoprap} $|y(\bar{t})|$. The differential distributions are compared to the QCD NLO calculation. The black vertical error bars on the data points denote the total combined uncertainty, the green error bars denote the statistical uncertainty, while the red band denotes the theory predictions calculated at NLO using {\sc MCFM}~\cite{Campbell:2010ff}. Uncertainties on the predicted values include the PDF and scale uncertainties. The horizontal error bars indicate the bin width.
 } 
\end{figure*}

\begin{figure*}[p]
  \centering
  \subfigure[]{
     \includegraphics[width=0.45\textwidth]{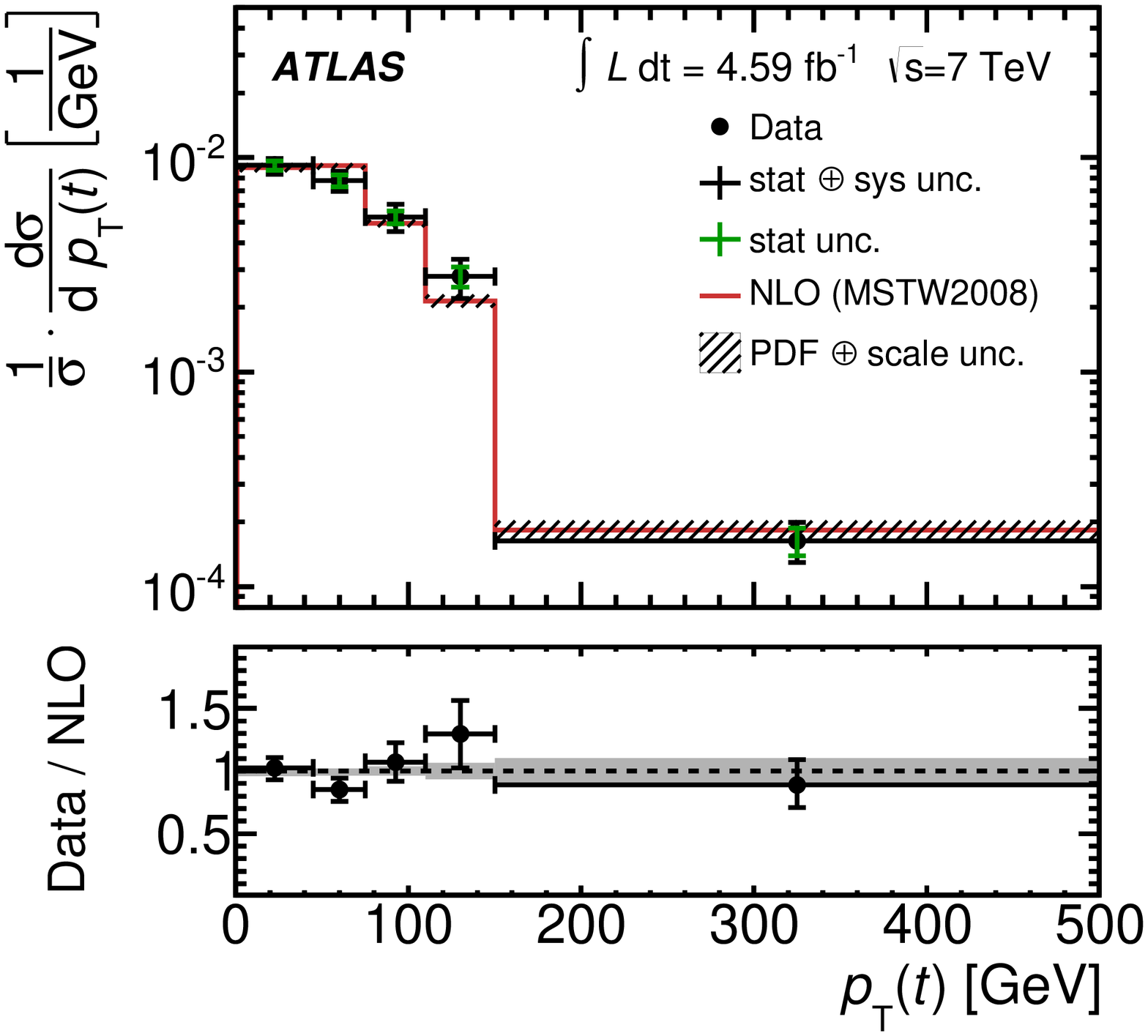}
     \label{subfig:normdiffdist_toppt}
  }
  \subfigure[]{
     \includegraphics[width=0.45\textwidth]{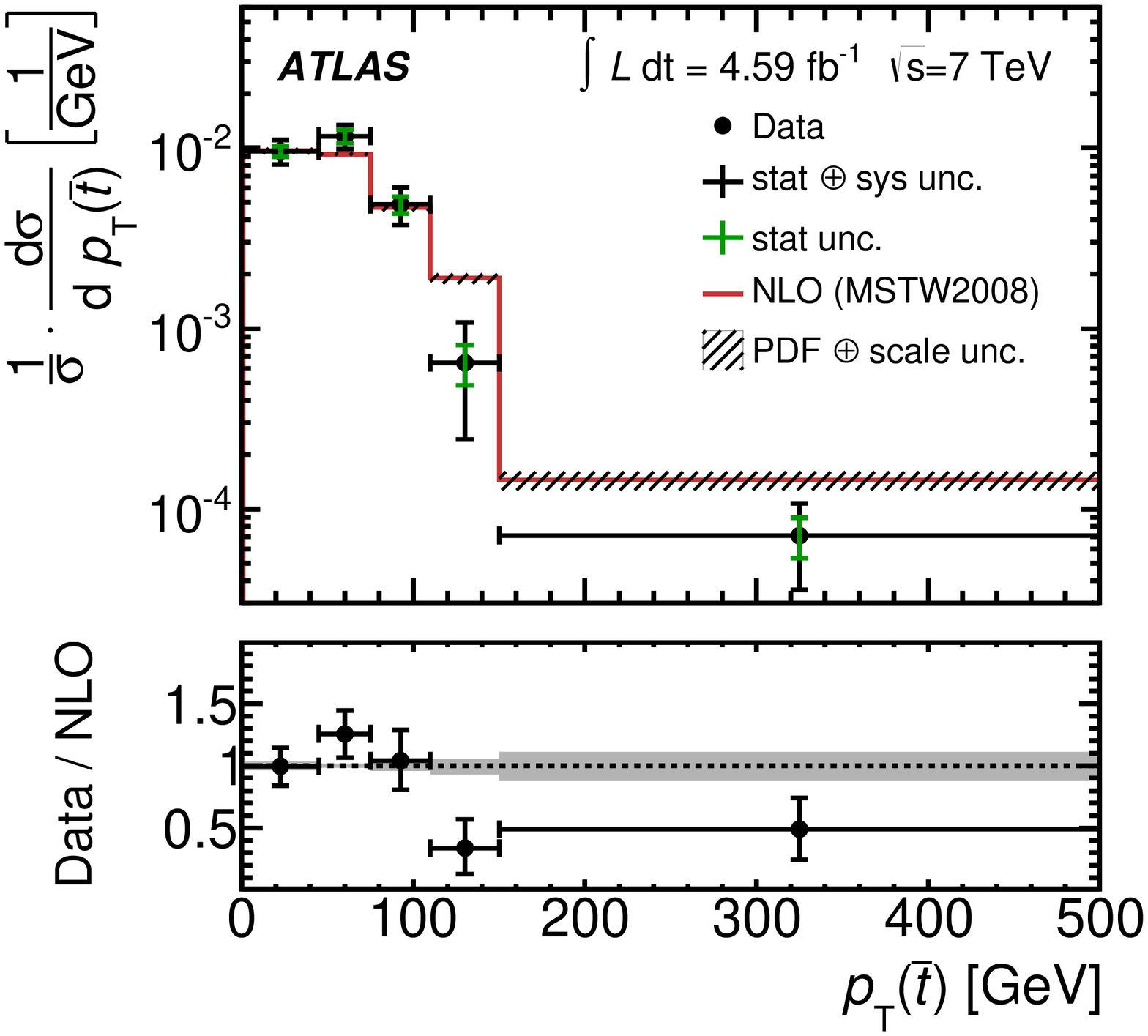}
     \label{subfig:normdiffdist_antitoppt}
  }
  \subfigure[]{
     \includegraphics[width=0.45\textwidth]{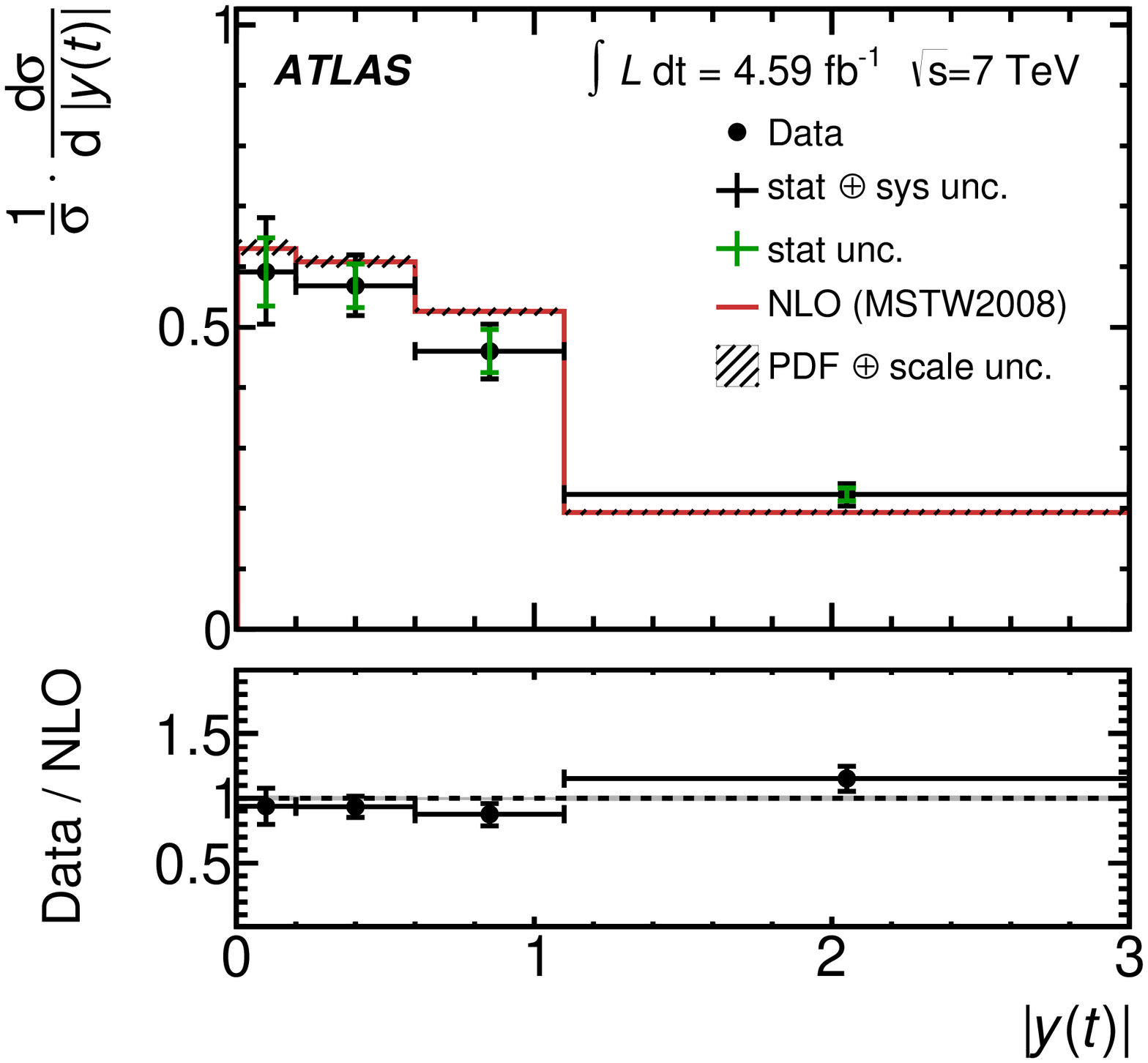}
     \label{subfig:normdiffdist_toprap}
  }
  \subfigure[]{
     \includegraphics[width=0.45\textwidth]{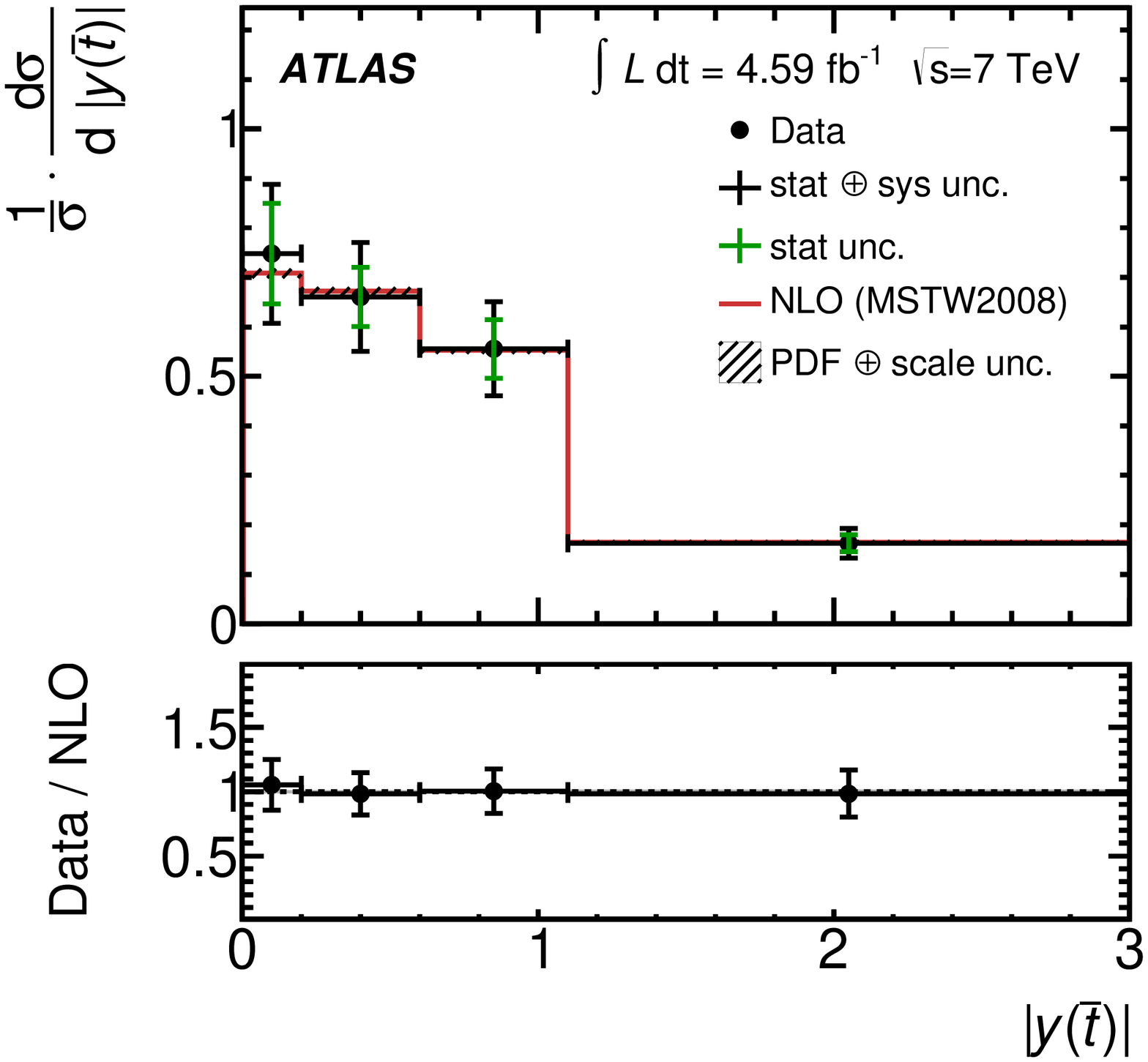}
     \label{subfig:normdiffdist_antitoprap}
  }
  \caption{\label{fig:unfolded_dist_norm}  Normalized differential cross section as a function of \subref{subfig:normdiffdist_toppt} $\pt(t)$, \subref{subfig:normdiffdist_antitoppt} $\pt(\tbar)$, \subref{subfig:normdiffdist_toprap} $|y(t)|$ and \subref{subfig:normdiffdist_antitoprap} $|y(\bar{t})|$. The normalized differential distributions are compared to the QCD NLO calculation. The black vertical error bars on the data points denote the total combined uncertainty, the green error bars denote the statistical uncertainty, while the red band denotes the theory predictions calculated at NLO using {\sc MCFM}~\cite{Campbell:2010ff}. Uncertainties on the predicted values include the PDF and scale uncertainties. The horizontal error bars indicate the bin width.
 }
\end{figure*}

The absolute differential cross-section results are listed in Table~\ref{tab:results_diff} and the normalized resutls in Table~\ref{tab:results_diffnorm}
as a function of $\pT$ and $|y|$ of the top quark.
A graphical representation of the results is shown in Fig.~\ref{fig:unfolded_dist} for the absolute cross sections and in 
Fig.~\ref{fig:unfolded_dist_norm} for the normalized case. They are compared to NLO predictions from 
MCFM~\cite{Campbell:2010ff} using the MSTW2008 PDF set for all variables. Uncertainties on the predicted values include the uncertainty on the scale and the PDF. 
To compare the NLO prediction with the measurement, $\chi^2$ values are computed with HERAfitter~\cite{Aaron:2009aa,Aaron:2009kv} taking into account 
the full correlation of the systematic and statistical uncertainties.
The $\chi^2$ values for the differential cross sections are listed in Table~\ref{tab:chi2}.

\begin{figure*}[htb]
  \centering
\subfigure[]{
\includegraphics[width=0.46\textwidth]{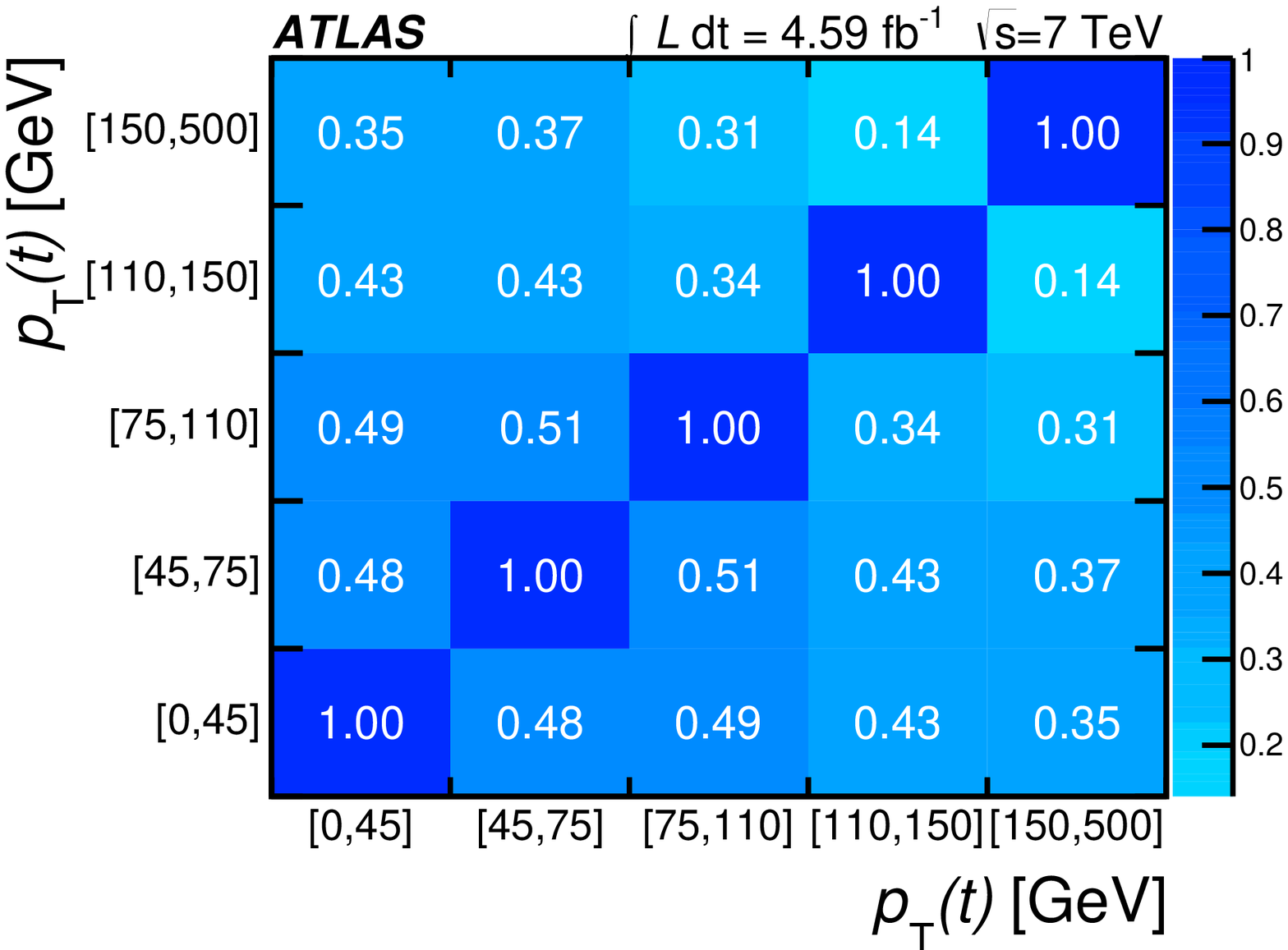}
     \label{subfig:corr_toppt}  
}
\subfigure[]{
\includegraphics[width=0.46\textwidth]{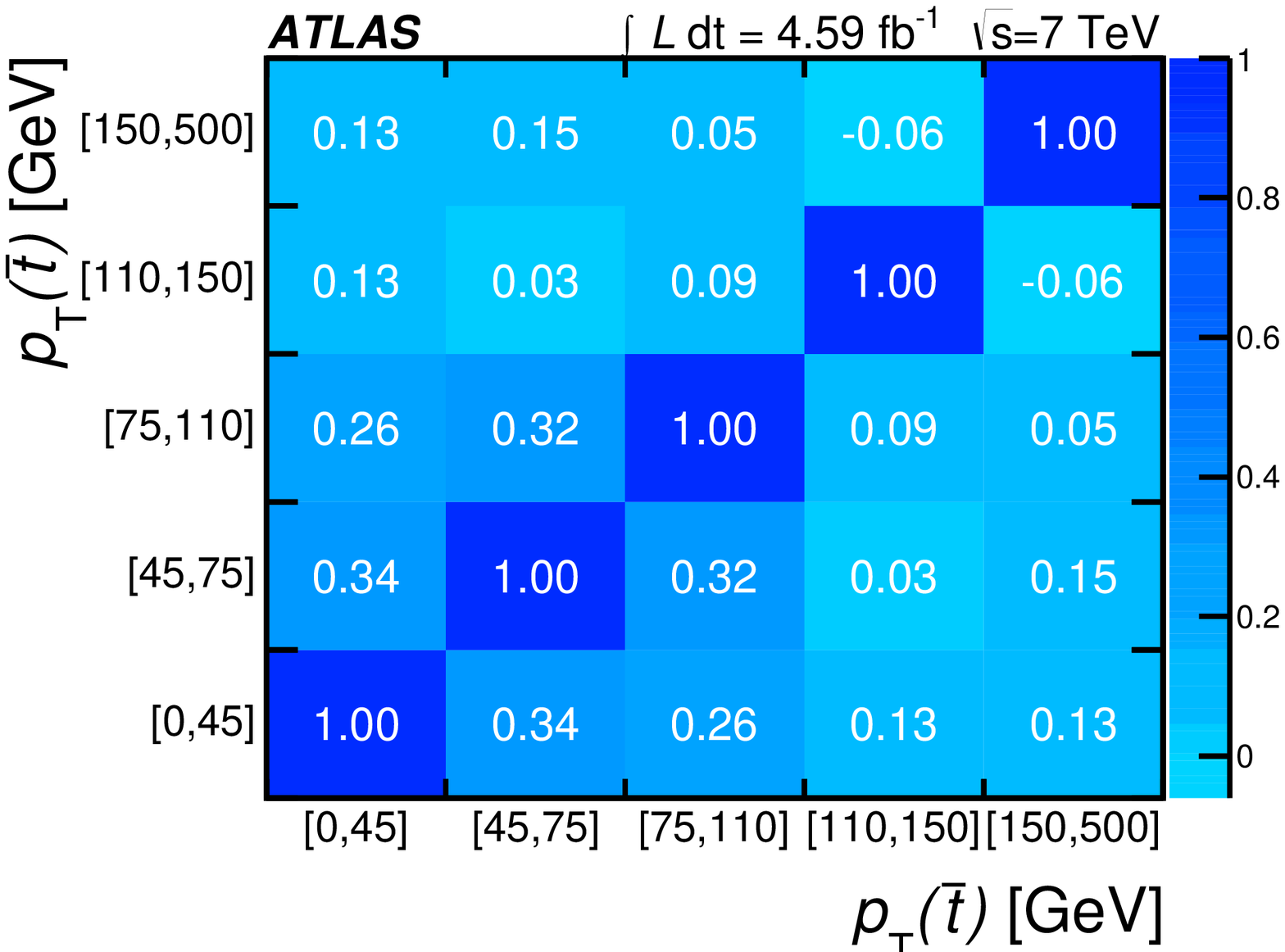}
     \label{subfig:corr_antitoppt}  
}
\subfigure[]{
\includegraphics[width=0.46\textwidth]{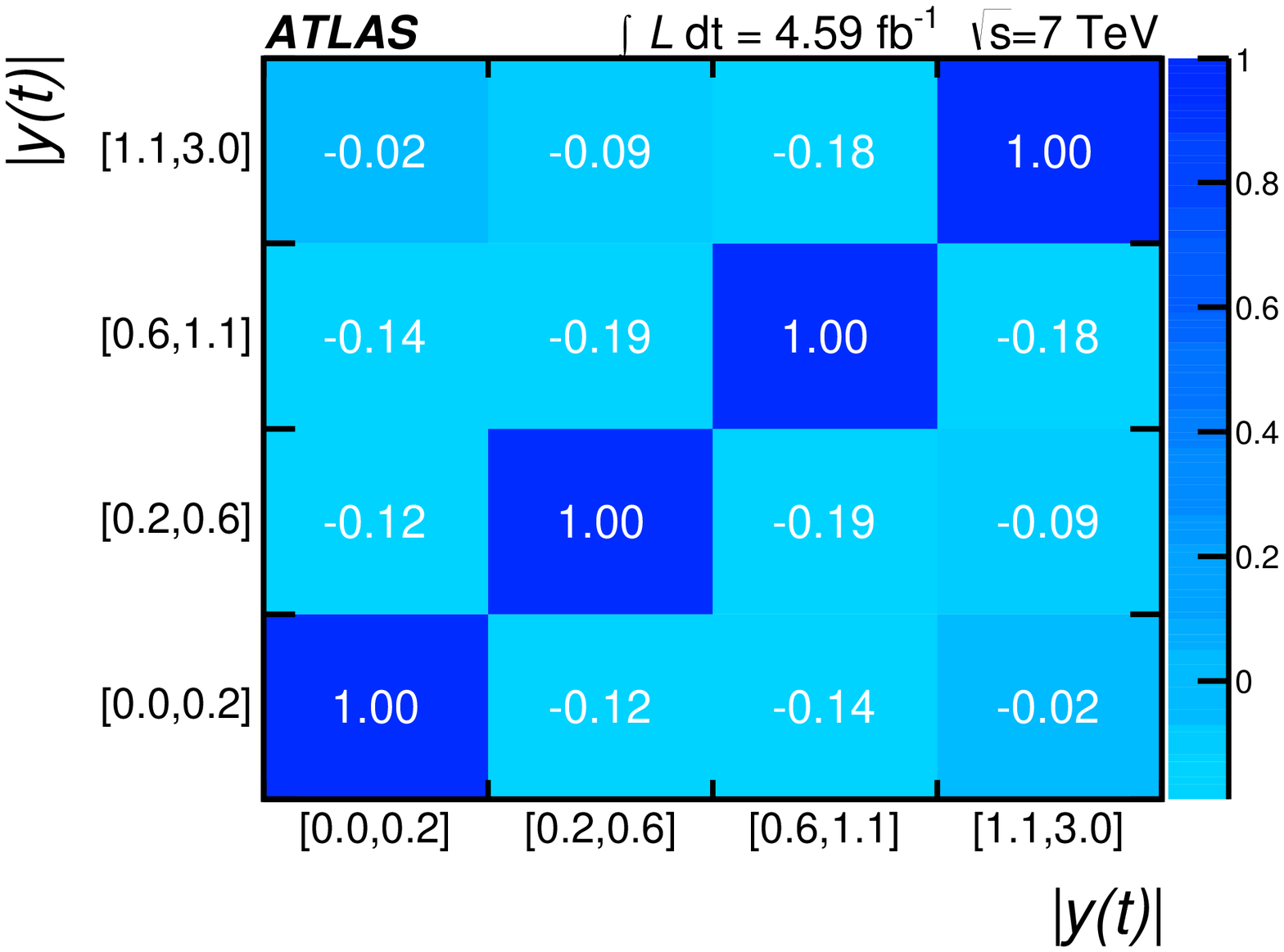}
     \label{subfig:corr_toprap}  
}
\subfigure[]{
\includegraphics[width=0.46\textwidth]{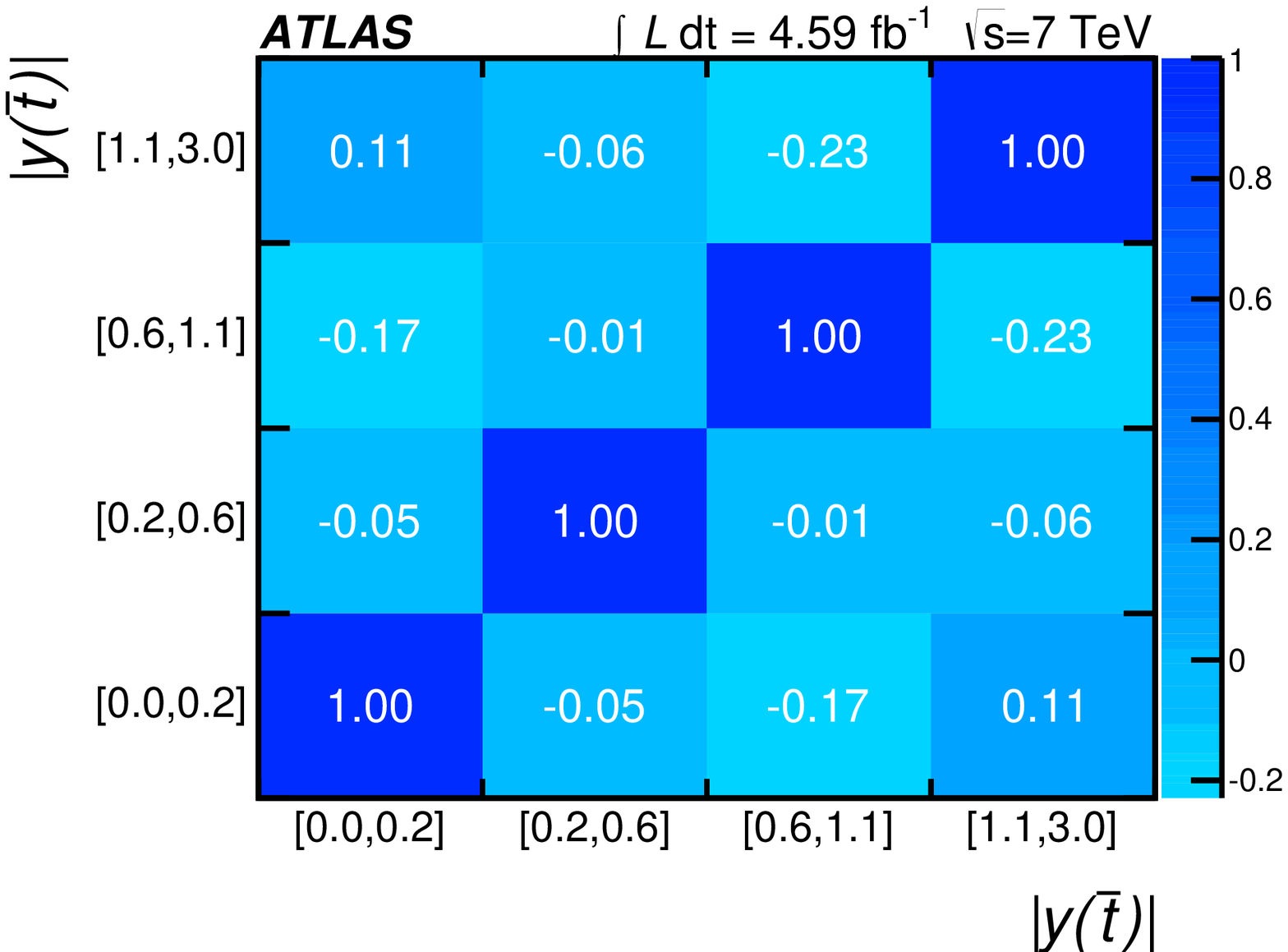}
     \label{subfig:corr_antitoprap}  
}
\caption{Statistical correlation matrices for the differential cross section as a function of \subref{subfig:corr_toppt} $\pt(t)$, \subref{subfig:corr_antitoppt} $\pt(\tbar)$, \subref{subfig:corr_toprap} $|y(t)|$ and \subref{subfig:corr_antitoprap} $|y(\bar{t})|$.
  The contents of this figure are provided in machine-readable format in the Supplemental Material~\cite{suppurl}.
}
\label{fig:corrAbs}
\end{figure*}

\begin{figure*}[htb]
  \centering
\subfigure[]{
\includegraphics[width=0.46\textwidth]{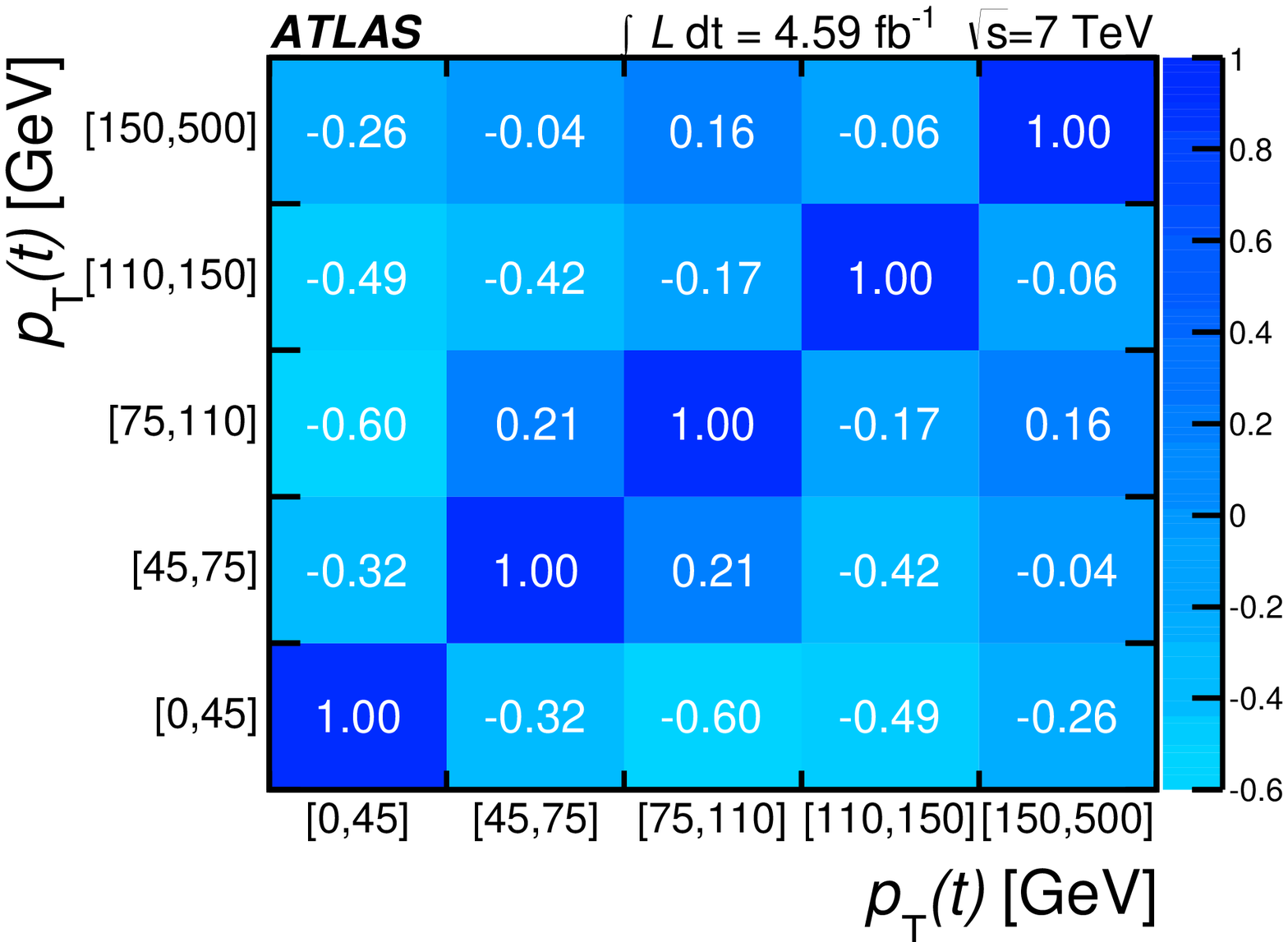}
     \label{subfig:corr2_toppt}  
}
\subfigure[]{
\includegraphics[width=0.46\textwidth]{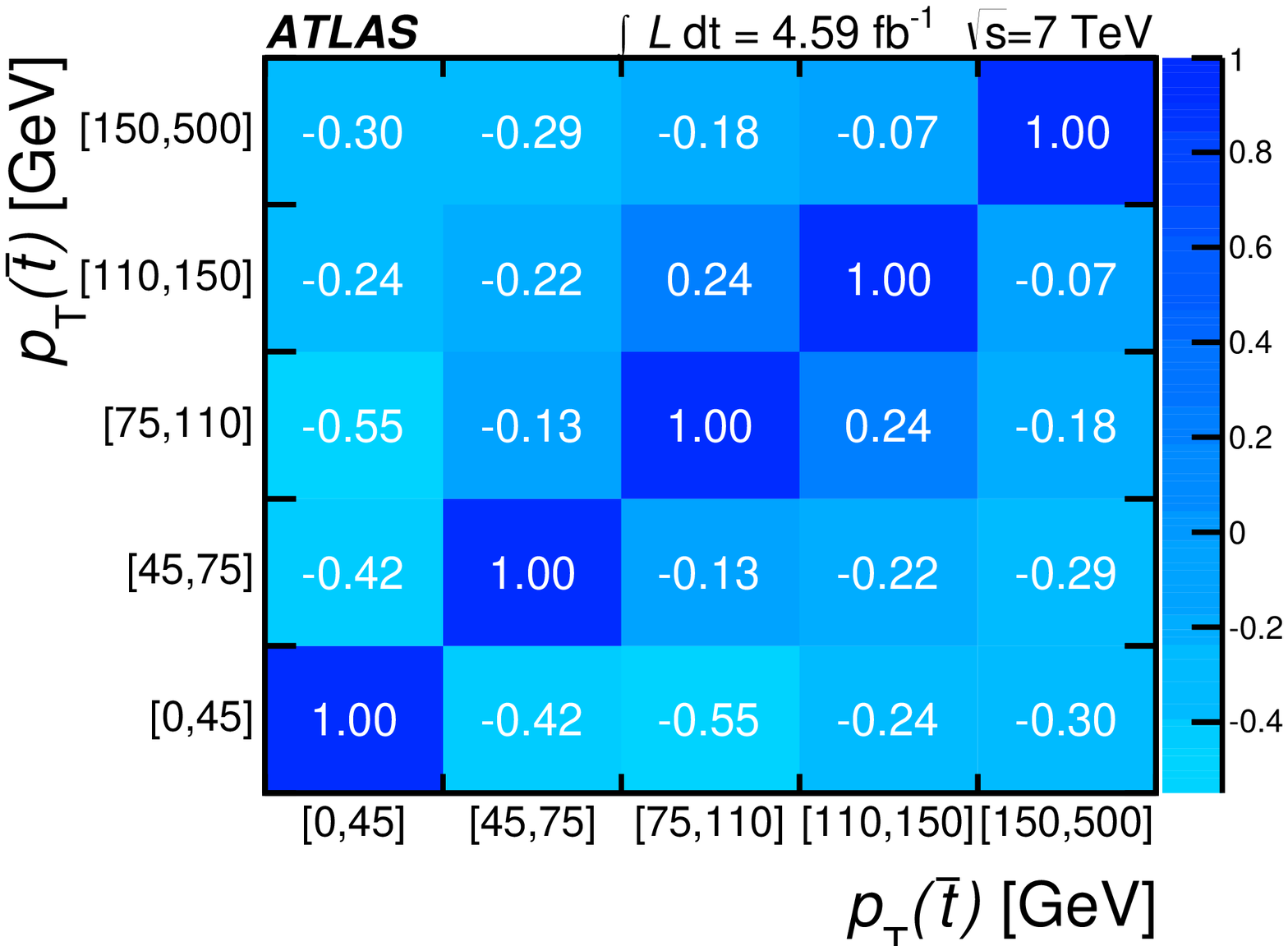}
     \label{subfig:corr2_antitoppt}  
}
\subfigure[]{
\includegraphics[width=0.46\textwidth]{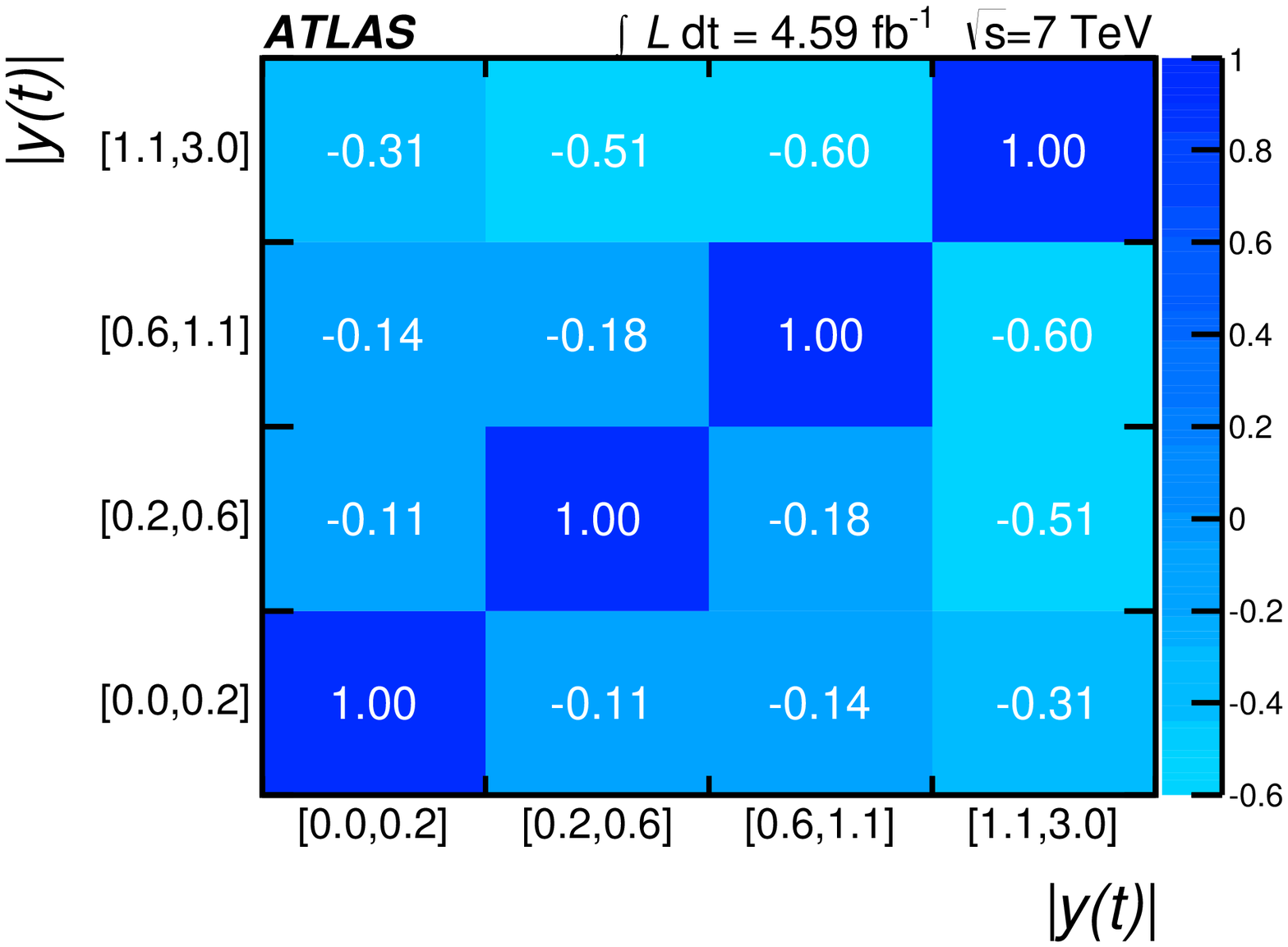}
     \label{subfig:corr2_toprap}  
}
\subfigure[]{
\includegraphics[width=0.46\textwidth]{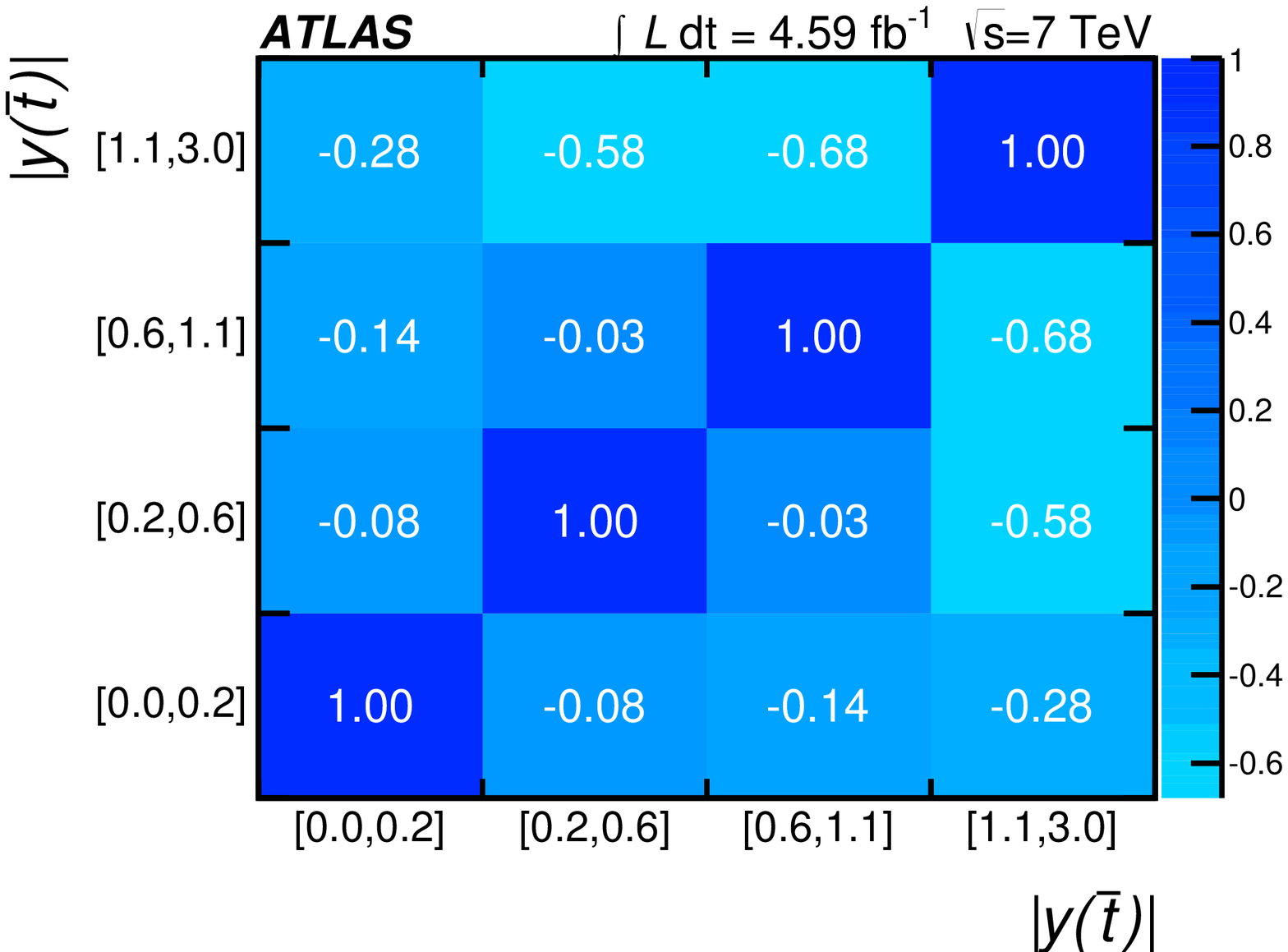}
     \label{subfig:corr2_antitoprap}  
}
\caption{Statistical correlation matrices for the normalized differential cross section as a function of \subref{subfig:corr2_toppt} $\pt(t)$, \subref{subfig:corr2_antitoppt} $\pt(\tbar)$, \subref{subfig:corr2_toprap} $|y(t)|$ and \subref{subfig:corr2_antitoprap} $|y(\bar{t})|$.
   The contents of this figure are provided in machine-readable format in the Supplemental Material~\cite{suppurl}.
}
\label{fig:corrNorm}
\end{figure*}

Systematic uncertainties dominate over the statistical uncertainty for the differential cross sections. Large uncertainties originate from the background
normalization, the $tq$ generator + parton shower uncertainty, the JES due to the uncertainty in the $\eta$ intercalibration as well as the 
PDF uncertainties mainly in the top-antiquark distributions.
A detailed list of the systematic contributions in each bin of each distributions is shown in Table~\ref{tab:topPtComb} for 
$d\sigma /d\pt (t)$, in Table~\ref{tab:antitopPtComb} for 
$d\sigma /d\pt (\tbar)$, in Table~\ref{tab:topRapComb} for 
$d\sigma /d|y(t)|$, and in Table~\ref{tab:antitopRapComb} for
$d\sigma /d|y(\tbar)|$.
In the case of the normalized differential cross sections many systematic uncertainties cancel and thus the measurement is dominated by
statistical uncertainties from the data distributions and the Monte Carlo sample size. 
The contribution of systematic uncertainties 
to the normalized distribution is again dominated by the background normalization, $tq$ generator + parton shower, and the JES $\eta$-intercalibration uncertainty. 
Details of the contribution of each systematic uncertainty in each bin of the normalized distributions are listed in Table~\ref{tab:topPtNormComb} 
for $1/\sigma \cdot d\sigma /d\pt (t)$, in Table~\ref{tab:antitopPtNormComb} for $1/\sigma \cdot d\sigma /d\pT(\tbar)$, in Table~\ref{tab:topRapNormComb} for $1/\sigma \cdot d\sigma /d|y(t)|$, and in Table~\ref{tab:antitopRapNormComb}
for $1/\sigma \cdot d\sigma /d|y(\tbar)|$.
Bin-wise correlation matrices for the statistical uncertainty are given in Fig.~\ref{fig:corrAbs} for the differential cross sections and in Fig.~\ref{fig:corrNorm} for
the normalized differential cross sections.

Overall, good agreement is observed between the NLO QCD predictions and the differential cross-section measurements. This is also supported by the $\chi^2$ values
listed in Table~\ref{tab:chi2}. 

The contents of Tables~\ref{tab:results_diff} to~\ref{tab:antitopRapNormComb} and the contents
of Fig.~\ref{fig:corrAbs} and Fig.~\ref{fig:corrNorm} are provided in machine-readable format
in the Supplemental Material~\cite{suppurl}.

\section{Conclusions}

In summary, measurements of the single top-quark production cross sections, $\sigma(tq)$, $\sigma(\bar{t}q)$, $R_t$, 
and $\sigma(tq + \bar{t}q)$, with the ATLAS detector at the LHC are presented using an integrated luminosity of 
4.59~fb$^{-1}$ $pp$ collision data at $\rts = 7 \tev$. 
All measurements are based on neural network (NN) discriminants separating signal events
from background events. Binned maximum-likelihood fits to the NN discriminants yield:
 $\sigma(tq)= 46\pm 6\, \mathrm{pb}$, $\sigma(\bar{t}q)= 23 \pm 4\, \mathrm{pb}$,
 and $\sigma(tq + \bar{t}q)= 68 \pm 8\, \mathrm{pb}$.
The measured cross-section ratio is $R_{t}=2.04\pm 0.18$.
The corresponding coupling at the $Wtb$ vertex is $|V_{tb}|=1.02 \pm 0.07$, and the
95\% CL lower limit on the CKM matrix element $\Vtb$ is 0.88.
A high-purity region is defined using the signal region of the NN discriminant for the differential cross-section measurements. Using an iterative Bayesian method, differential cross sections
are extracted as a function of $p_\mathrm{T}(t)$, $p_\mathrm{T}(\bar{t})$, $|y(t)|$, and $|y(\bar{t})|$.
Good agreement with the NLO QCD predictions is observed.

\section*{Acknowledgements}

We thank CERN for the very successful operation of the LHC, as well as the
support staff from our institutions without whom ATLAS could not be
operated efficiently.

We acknowledge the support of ANPCyT, Argentina; YerPhI, Armenia; ARC,
Australia; BMWF and FWF, Austria; ANAS, Azerbaijan; SSTC, Belarus; CNPq and FAPESP,
Brazil; NSERC, NRC and CFI, Canada; CERN; CONICYT, Chile; CAS, MOST and NSFC,
China; COLCIENCIAS, Colombia; MSMT CR, MPO CR and VSC CR, Czech Republic;
DNRF, DNSRC and Lundbeck Foundation, Denmark; EPLANET, ERC and NSRF, European Union;
IN2P3-CNRS, CEA-DSM/IRFU, France; GNSF, Georgia; BMBF, DFG, HGF, MPG and AvH
Foundation, Germany; GSRT and NSRF, Greece; ISF, MINERVA, GIF, I-CORE and Benoziyo Center,
Israel; INFN, Italy; MEXT and JSPS, Japan; CNRST, Morocco; FOM and NWO,
Netherlands; BRF and RCN, Norway; MNiSW and NCN, Poland; GRICES and FCT, Portugal; MNE/IFA, Romania; MES of Russia and ROSATOM, Russian Federation; JINR; MSTD,
Serbia; MSSR, Slovakia; ARRS and MIZ\v{S}, Slovenia; DST/NRF, South Africa;
MINECO, Spain; SRC and Wallenberg Foundation, Sweden; SER, SNSF and Cantons of
Bern and Geneva, Switzerland; NSC, Taiwan; TAEK, Turkey; STFC, the Royal
Society and Leverhulme Trust, United Kingdom; DOE and NSF, United States of
America.

The crucial computing support from all WLCG partners is acknowledged
gratefully, in particular from CERN and the ATLAS Tier-1 facilities at
TRIUMF (Canada), NDGF (Denmark, Norway, Sweden), CC-IN2P3 (France),
KIT/GridKA (Germany), INFN-CNAF (Italy), NL-T1 (Netherlands), PIC (Spain),
ASGC (Taiwan), RAL (UK) and BNL (USA) and in the Tier-2 facilities
worldwide.

\begin{table}[htbp]
  \centering
  \caption{Differential $t$-channel top-quark production cross section as a function of $\pT(t)$, $\pT(\bar{t})$, $|y(t)|$ and $|y(\bar{t})|$ 
  with the uncertainties for each bin given in percent.
  The contents of this table are provided in machine-readable format in the Supplemental Material~\cite{suppurl}.
  \label{tab:results_diff}}
\begin{ruledtabular} 
\begin{tabular}{ccccc}
             $\pT(t)$ [GeV]\  & $\frac{d\sigma}{d\pT(t)}$ $[\frac{\mathrm{fb}}{\gev}]$      & total [\%]\      & stat. [\%]\      & syst. [\%]\ \\
\hline
              $[0,45]$   &        $440 \pm 70$ & $\pm 15$   &   $\pm 7.4$ & $\pm 13$\\
             $[45,75]$   &        $370 \pm 60$ & $\pm 16$   &   $\pm 6.5$ & $\pm 14$\\
             $[75,110]$  &        $250 \pm 40$ & $\pm 15$   &   $\pm 7.7$ & $\pm 13$\\
            $[110,150]$  &        $133 \pm 27$ & $\pm 20$   &   $\pm 12$& $\pm 16$\\
            $[150,500]$  &         $7.8 \pm 1.9$ & $\pm 24$ &   $\pm 16$& $\pm 19$\\
    \multicolumn{5}{c}{ } \\
             $\pT(\bar{t})$ [GeV]\  & $\frac{d\sigma}{d\pT(\bar{t})}$ $[\frac{\mathrm{fb}}{\gev}]$    & total [\%]\      & stat. [\%]\      & syst. [\%]\ \\
\hline 
              $[0,45]$   &        $190 \pm 50$ & $\pm 28$   &   $\pm 12$  & $\pm 25$\\
             $[45,75]$   &        $230 \pm 40$ & $\pm 18$   &   $\pm 8.2$ & $\pm 17$\\
             $[75,110]$  &        $97  \pm 27$ & $\pm 27$   &   $\pm 13$  & $\pm 24$\\
            $[110,150]$  &        $13.0 \pm 9.7$ & $\pm 74$ &   $\pm 26$  & $\pm 70$\\
            $[150,500]$  &        $1.4 \pm 0.9$ & $\pm 59$  &   $\pm 26$ & $\pm 53$\\
    \multicolumn{5}{c}{ } \\

             $|y(t)|$\  & $\frac{d\sigma}{d|y(t)|}$[pb]\      & total [\%]\      & stat. [\%]\      & syst. [\%]\ \\
\hline 
              $[0,0.2]$  &        $28 \pm 4$      & $\pm 15$  &   $\pm 9.0$& $\pm 12$\\
            $[0.2,0.6]$  &        $27.3 \pm 3.3$  & $\pm 12$  &   $\pm 6.3$& $\pm 10$\\
            $[0.6,1.1]$  &        $22.1 \pm 3.0$  & $\pm 14$  &   $\pm 7.5$& $\pm 11$\\
            $[1.1,3.0]$  &        $10.7 \pm 1.6$  & $\pm 15$  &   $\pm 7.0$& $\pm 13$\\
    \multicolumn{5}{c}{ } \\

              $|y(\bar{t})|$  & $\frac{d\sigma}{d|y(\bar{t})|}$ [pb]\      & total [\%]\      & stat. [\%]\      & syst. [\%]\ \\
\hline
              $[0,0.2]$  &        $ 15.0 \pm 3.4$ & $\pm 23$&   $\pm 13$  & $\pm 18$\\
            $[0.2,0.6]$  &        $ 13.3 \pm 3.3$ & $\pm 25$&   $\pm 9.5$ & $\pm 23$\\
            $[0.6,1.1]$  &        $ 11.2 \pm 2.6$ & $\pm 23$&   $\pm 11$  & $\pm 20$\\
            $[1.1,3.0]$  &        $ 3.3 \pm 0.9$  & $\pm 29$&   $\pm 13$  & $\pm 25$\\
    \end{tabular}%
\end{ruledtabular}
\end{table}%

\begin{table}[htbp]
  \centering
  \caption{Normalized differential $t$-channel top-quark production cross section as a function of $\pT(t)$, $\pT(\bar{t})$, $|y(t)|$ and $|y(\bar{t})|$ 
  with the uncertainties for each bin given in percent.
  The contents of this table are provided in machine-readable format in the Supplemental Material~\cite{suppurl}.
  \label{tab:results_diffnorm}}
\begin{ruledtabular} 
\begin{tabular}{ccccc}
             $\pT(t)$ [GeV]\  & $\frac{1}{\sigma}\frac{d\sigma}{d\pT(t)} [\frac{10^{-3}}{\gev}]$\      & total [\%]\      & stat. [\%]\      & syst. [\%]\ \\
\hline 
              $[0,45]$   &        $9.2^{+0.8}_{-0.9}$ & $^{+8.4}_{-9.4}$ &   $\pm 5.3$ & $^{+6.5}_{-7.7}$\\
             $[45,75]$   &        $7.8 \pm 0.9$ & $\pm 11$      &   $\pm 6.9$ & $\pm 8.8$\\
             $[75,110]$  &        $5.3 \pm 0.8 $ & $\pm 15$     &   $\pm 8.0$ & $\pm 13$\\
            $[110,150]$  &        $2.8 \pm 0.6$  & $\pm 21$     &   $\pm 11$& $\pm 18$\\
            $[150,500]$  &        $0.16 \pm 0.04$ & $\pm 22$    &   $\pm 15$& $\pm 16$\\
    \multicolumn{5}{c}{ } \\
             $\pT(\bar{t})$ [GeV]\  & $\frac{1}{\sigma}\frac{d\sigma}{d\pT(\bar{t})} [\frac{10^{-3}}{\gev}]$     & total [\%]\      & stat. [\%]\      & syst. [\%]\ \\
\hline
              $[0,45]$   &        $9.6 \pm 1.6$  & $\pm 17$   &   $\pm 8.2$ & $\pm 15$\\
             $[45,75]$   &        $11.6 \pm 1.8$ & $\pm 15$   &   $\pm 8.8$ & $\pm 12$\\
             $[75,110]$  &        $4.9 \pm 1.2$  & $\pm 25$   &   $\pm 13$  & $\pm 21$\\
            $[110,150]$  &       $0.7 \pm 0.4$   & $^{+67}_{-61}$ &   $\pm 25.8$& $^{+62}_{-56}$\\
            $[150,500]$  &      $0.07 \pm 0.04$  & $\pm 51$   &   $\pm 26$  & $\pm 45$\\
    \multicolumn{5}{c}{ } \\

             $|y(t)|$\  & $\frac{1}{\sigma} \frac{d\sigma}{d|y(t)|}$\      & total [\%]\      & stat. [\%]\      & syst. [\%]\ \\
\hline
              $[0,0.2]$  &        $ 0.59 \pm 0.09$ & $\pm 15$&  $\pm 9.0$& $\pm 11$\\
            $[0.2,0.6]$  &        $ 0.57 \pm 0.05$ & $\pm 9.0$&   $\pm 6.4$& $\pm 6.3$\\
            $[0.6,1.1]$  &        $ 0.46 \pm 0.05$ & $\pm 9.7$&   $\pm 7.5$& $\pm 6.2$\\
            $[1.1,3.0]$  &        $ 0.223 \pm 0.019$ & $\pm 8.5$&   $\pm 4.9$& $\pm 6.9$\\
    \multicolumn{5}{c}{ } \\

              $|y(\bar{t})|$  & $\frac{1}{\sigma} \frac{d\sigma}{d|y(\bar{t})|}$\      & total [\%]\      & stat. [\%]\      & syst. [\%]\ \\
\hline
              $[0,0.2]$  &        $ 0.75 \pm 0.14$ & $\pm 19$&   $\pm 13$& $\pm 13$\\
            $[0.2,0.6]$  &        $ 0.66 \pm 0.11$ & $\pm 17$&   $\pm 9.1$& $\pm 14$\\
            $[0.6,1.1]$  &        $ 0.555 \pm 0.095$ & $\pm 17$&   $\pm 11$& $\pm 13$\\
            $[1.1,3.0]$  &        $ 0.163 \pm 0.030$ & $\pm 18$&   $\pm 11$& $\pm 15$\\
    \end{tabular}%
\end{ruledtabular}
\end{table}%


\begin{table}[htbp]
  \centering
  \caption{Comparison between the measured differential cross sections and the predictions from the NLO calculation using the MSTW2008 PDF set. For each variable and prediction a $\chi^2$ value is calculated with HERAfitter using the covariance matrix of each measured spectrum. The theory uncertainties of the predictions are treated as uncorrelated. The number of degrees of freedom (NDF) is equal to the number of bins in the measured spectrum. The contents of this table are provided in machine-readable format in the Supplemental Material~\cite{suppurl}.\label{tab:chi2}}
\begin{ruledtabular} 
\begin{tabular}{lcccc}
 & $\frac{d\sigma}{d\pT(t)}$ & $\frac{d\sigma}{d\pT(\tbar)}$ & $\frac{d\sigma}{d|y(t)|}$ & $\frac{d\sigma}{d|y(\tbar)|}$\\ 
\hline 
$\chi^2/\mathrm{NDF}$ & $7.55/5$ & $4.68/5$ & $6.30/4$ & $0.32/4$ \\ 
    \end{tabular}%
\end{ruledtabular}
\end{table}%

\begin{table*}[!thbp]
\begin{center}
\begin{ruledtabular} 
\caption{Detailed list of the contribution of each source of uncertainty to the  
   total relative uncertainty on the measured $\frac{d\sigma}{d\pT(t)}$ distribution given in percent for each bin.
   The list includes only those uncertainties that contribute with more than 1\%.
   The following uncertainties contribute to the total uncertainty with less than 1\% to each bin content: 
   JES detector, JES statistical, JES physics modeling, JES mixed detector and modeling, JES close-by jets, JES pile-up, JES flavor composition, JES flavor response, 
   jet vertex fraction, $b/\bbar$ acceptance, $\MET$ modeling, $W$+jets shape variation, and $\ttbar$ generator. 
The contents of this table are provided in machine-readable format in the Supplemental Material~\cite{suppurl}.
}
\label{tab:topPtComb}
\begin{tabular}{lccccc}
    $\frac{d\sigma}{d\pT(t)}$ & \multicolumn{5}{c}{$\pT(t)$ bins [\gev]} \\
    \multicolumn{6}{c}{ } \\
    Source & $[0,45]$ & 
             $[45,75]$ & 
             $[75,110]$ &
             $[110,150]$ &
             $[150,500]$ \\
    \hline
               Data statistical & $\pm$7.4 & $\pm$6.5 & $\pm$7.7 & $\pm$12 & $\pm$16\\
        Monte Carlo statistical & $\pm$5.5 & $\pm$5.3 & $\pm$4.8 & $\pm$6.0 & $\pm$9.4\\
    \multicolumn{6}{c}{ } \\
      Background normalization & $\pm$6.1 & $\pm$7.5 & $\pm$5.2 & $\pm$3.0 & $\pm$5.2\\
        JES $\eta$ intercalibration     & $<$ 1 &         $+2.6/-1.3$ &         $+3.4/-1.9$     & $<$ 1 &         $+9.0/-4.2$\\
                       $b$-JES     & $<$ 1 &         $+1.2/-2.3$     & $<$ 1 &$\pm$1.6     & $<$ 1\\
         Jet energy resolution &  $\pm$1.0 &  $\pm$2.4 &         $\pm$2.3 &        $\pm$3.0     & $<$ 1\\
         $b$-tagging efficiency  & $\pm$3.0 &  $\pm$3.1 &        $\pm$3.3 &         $\pm$3.6 &    $\pm$6.2\\
         $c$-tagging efficiency       &  $\pm$1.3 &  $\pm$1.5     & $<$ 1     & $<$ 1     & $<$ 1\\
         Mistag efficiency&  $\pm$2.0 &  $\pm$1.9     & $<$ 1     & $<$ 1 &   $\pm$1.2\\
         Lepton uncertainties &   $\pm$2.6 & $\pm$2.6 & $\pm$2.6 & $\pm$2.6 &  $\pm$2.6\\
    \multicolumn{6}{c}{ } \\
                         PDF   &         $\pm$3.0&        $\pm$1.8&         $\pm$2.3&         $\pm$2.8&         $\pm$2.4\\
     $tq$ generator + parton shower &         $\pm$6.8 &  $\pm$8.2 &  $\mp$7.9 & $\mp$12 &         $+9.2/-9.7$\\
   $tq$ scale variation &         $\pm$2.8 &  $<$ 1 &     $\pm$3.7 & $<$ 1  &         $+6.0/-6.4$\\
    \multicolumn{6}{c}{ } \\
                     Unfolding & $\pm$1.3 & $\pm$1.4  & $<$ 1     & $<$ 1     & $<$ 1\\
                    Luminosity & $\pm$1.8 & $\pm$1.8 & $\pm$1.8 & $\pm$1.8 & $\pm$1.8\\
    \multicolumn{6}{c}{ } \\
              Total systematic & $\pm$13  & $\pm$14   & $\pm$13   & $\pm$16   & $\pm$19\\
                        Total  & $\pm$15  & $\pm$16   & $\pm$15   & $\pm$20   & $\pm$25\\

   \end{tabular}
\end{ruledtabular}
\end{center}
\end{table*}

\begin{table*}[!thbp]
\begin{center}
\begin{ruledtabular} 
\caption{Detailed list of the contribution of each source of uncertainty to the  
   total relative uncertainty on the measured $\frac{d\sigma}{d\pT(\tbar)}$ distribution given in percent for each bin.
   The list includes only those uncertainties that contribute with more than 1\%.
   The following uncertainties contribute to the total uncertainty with less than 1\% to each bin content: 
   JES detector, JES statistical, JES physics modeling, JES mixed detector and modeling, JES close-by jets, JES pile-up, JES flavor composition, JES flavor response, 
   $b$-JES, jet vertex fraction, mistag efficiency, $b/\bbar$ acceptance, $\MET$ modeling, $W$+jets shape variation, and $\ttbar$ generator. 
 The contents of this table are provided in machine-readable format in the Supplemental Material~\cite{suppurl}.
}
\label{tab:antitopPtComb}
\begin{tabular}{lccccc}
    $\frac{d\sigma}{d\pT(\tbar)}$ & \multicolumn{5}{c}{$\pT(\tbar)$ bins [\gev]} \\
    \multicolumn{6}{c}{ } \\
    Source & $[0,45]$ & 
             $[45,75]$ & 
             $[75,110]$ &
             $[110,150]$ &
             $[150,500]$ \\
    \hline
               Data statistical & $\pm$12   & $\pm$8.2 & $\pm$13    & $\pm$26   & $\pm$26  \\
        Monte Carlo statistical & $\pm$12   & $\pm$9.1 & $\pm$14    & $\pm$28   & $\pm$28  \\
    \multicolumn{6}{c}{ } \\
      Background normalization & $\pm$14  & $\pm$11   & $\pm$16   & $\pm$48     & $\pm$33\\
JES $\eta$ intercalibration &         $-9.0/+8.7$ &         $+1.9/-3.7$ &         $+4.9/-1.3$ &        $+15/-13$     & $<$ 1\\
          Jet energy resolution &   $\pm$1.0 & $\pm$2.2 & $\pm$3.4     & $<$ 1 & $\pm$3.0\\
           $b$-tagging efficiency &   $\pm$3.0 & $\pm$3.1 & $\pm$3.2 & $\pm$3.6 & $\pm$5.9\\
          $c$-tagging efficiency &   $\pm$5.6 & $\pm$2.0 & $\pm$2.2 & $\pm$10  & $\pm$5.9\\
         Lepton uncertainties &   $\pm$2.6 & $\pm$2.6 & $\pm$2.6 & $\pm$2.6 & $\pm$2.7\\
    \multicolumn{6}{c}{ } \\
                          PDF  &       $\pm$3.8&      $\pm$4.3&         $\pm$5.3&         $\pm$7.2&         $\pm$8.2\\
     $tq$ generator + parton shower &        $\pm$12.2 & $<$ 1    &   $\mp$9.6 &   $\pm$11      & $<$ 1\\ 
   $tq$ scale variation &         $\pm$3.1 &         $<$ 1 &         $\pm$3.2 &         $\pm$1.9 &        $\pm$5.9\\
    \multicolumn{6}{c}{ } \\
                     Unfolding & $<$ 1     & $<$ 1     & $<$ 1    & $\pm$6.9 & $\pm$2.6\\
                    Luminosity & $\pm$1.8 & $\pm$1.8 & $\pm$1.8 & $\pm$1.8 & $\pm$1.8\\
    \multicolumn{6}{c}{ } \\
              Total systematic & $\pm$25    & $\pm$17   & $\pm$24  & $\pm$70   & $\pm$53 \\
                        Total  & $\pm$27    & $\pm$18   & $\pm$27  & $\pm$74   & $\pm$59 \\
   \end{tabular}
\end{ruledtabular}
\end{center}
\end{table*}

\begin{table*}[!thbp]
\begin{center}
\begin{ruledtabular} 
\caption{Detailed list of the contribution of each source of uncertainty to the  
   total relative uncertainty on the measured $\frac{d\sigma}{d|y(t)|}$ distribution given in percent for each bin.
   The list includes only those uncertainties that contribute with more than 1\%.
   The following uncertainties contribute to the total uncertainty with less than 1\% to each bin content: 
   JES detector, JES statistical, JES physics modeling, JES mixed detector and modeling, JES close-by jets, JES pile-up, JES flavor composition, JES flavor response, 
   jet vertex fraction, $b/\bbar$ acceptance, $\MET$ modeling, $W$+jets shape variation, $\ttbar$ generator, $\ttbar$ ISR/FSR, and unfolding. The contents of this table are provided in machine-readable format in the Supplemental Material~\cite{suppurl}.}
\label{tab:topRapComb}
\begin{tabular}{lcccc}
    $\frac{d\sigma}{d|y(t)|}$ & \multicolumn{4}{c}{$|y(t)|$ bins } \\
    \multicolumn{5}{c}{ } \\
    Source & $[0,0.2]$ & 
             $[0.2,0.6]$ & 
             $[0.6,1.1]$ &
             $[1.1,3.0]$ \\
    \hline
              Data statistical & $\pm$9.0 & $\pm$6.3 & $\pm$7.5 & $\pm$7.1\\
       Monte Carlo statistical & $\pm$5.9 & $\pm$4.8 & $\pm$5.0 & $\pm$4.4 \\
    \multicolumn{5}{c}{ } \\
      Background normalization & $\pm$5.3 & $\pm$6.5 & $\pm$6.7 & $\pm$4.7\\
   JES $\eta$ intercalibration     &     $+1.7/-0.6$     & $<$ 1 &         $+1.7/-0.4$     & $<$ 1\\
                         b-JES &         $+1.1/-1.7$     & $<$ 1 &         $+1.1/+0.2$     & $<$ 1\\
         Jet energy resolution &   $\pm$3.2 & $\pm$1.7     & $<$ 1 & $\pm$3.1\\
        $b$-tagging efficiency &   $\pm$3.3 & $\pm$3.4 & $\pm$3.4 & $\pm$3.2\\
        $c$-tagging efficiency &   $\pm$1.3 & $\pm$1.2 & $\pm$1.2 & $\pm$1.0\\
             Mistag efficiency     & $<$ 1 &   $\pm$1.3 & $\pm$2.0 & $\pm$1.4\\
          Lepton uncertainties &   $\pm$2.6 & $\pm$2.7 & $\pm$2.6 & $\pm$2.5\\
    \multicolumn{5}{c}{ } \\
                          PDF  &     $\pm$3.6&     $\pm$3.6&    $\pm$2.8&     $\pm$2.8\\
     $tq$ generator + parton shower &         $\mp$5.7 &  $\pm$0.8 & $\pm$4.0 & $\pm$8.7\\
   $tq$ scale variation &         $\pm$3.5 &  $<$ 1 &  $\pm$2.6 &  $\pm$4.7\\
    \multicolumn{5}{c}{ } \\
                    Luminosity & $\pm$1.8 & $\pm$1.8 & $\pm$1.8 & $\pm$1.8\\
    \multicolumn{5}{c}{ } \\
              Total systematic & $\pm$12    & $\pm$10    & $\pm$11  & $\pm$14 \\
                        Total  & $\pm$15    & $\pm$12    & $\pm$14  & $\pm$15 \\
   \end{tabular}
\end{ruledtabular}
\end{center}
\end{table*}

\begin{table*}[!thbp]
\begin{center}
\begin{ruledtabular} 
\caption{Detailed list of the contribution of each source of uncertainty to the  
   total relative uncertainty on the measured $\frac{d\sigma}{d|y(\tbar)|}$ distribution given in percent for each bin.
   The list includes only those uncertainties that contribute with more than 1\%.
   The following uncertainties contribute to the total uncertainty with less than 1\% to each bin content: 
   JES detector, JES statistical, JES physics modeling, JES mixed detector and modeling, JES close-by jets, JES pile-up, JES flavor composition, JES flavor response, 
   $b$-JES, jet vertex fraction, $b/\bbar$ acceptance, mistag efficiency, $\MET$ modeling, $W$+jets shape variation, $\ttbar$ generator, $\ttbar$ ISR/FSR, and unfolding.
   The contents of this table are provided in machine-readable format in the Supplemental Material~\cite{suppurl}.
}
\label{tab:antitopRapComb}
\begin{tabular}{lcccc}
    $\frac{d\sigma}{d|y(\tbar)|}$ & \multicolumn{4}{c}{$|y(\tbar)|$ bins } \\
    \multicolumn{5}{c}{ } \\
    Source & $[0,0.2]$ & 
             $[0.2,0.6]$ & 
             $[0.6,1.1]$ &
             $[1.1,3.0]$ \\
    \hline
               Data statistical & $\pm$13 & $\pm$9.5 & $\pm$11 & $\pm$13\\
        Monte Carlo statistical & $\pm$11 & $\pm$12 & $\pm$11 & $\pm$17 \\
    \multicolumn{5}{c}{ } \\
      Background normalization & $\pm$11 & $\pm$16 & $\pm$13 & $\pm$15\\
        JES $\eta$ intercalibration     & $<$ 1 &         $+1.0/-1.8$     & $<$ 1 &         $+2.3/-0.9$\\
         Jet energy resolution &   $\pm$2.3 & $\pm$2.2 & $\pm$1.0 & $\pm$3.2\\
        $b$-tagging efficiency &   $\pm$3.4 & $\pm$3.3 & $\pm$3.2 & $\pm$3.2\\
        $c$-tagging efficiency &   $\pm$2.5 & $\pm$3.6 & $\pm$2.9 & $\pm$4.0\\
          Lepton uncertainties &   $\pm$2.7 & $\pm$2.7 & $\pm$2.6 & $\pm$2.4\\
    \multicolumn{5}{c}{ } \\
                         PDF   &    $\pm$6.0&    $\pm$5.3&     $\pm$4.4&      $\pm$4.1\\
     $tq$ generator + parton shower &         $\pm$1.0 &  $\mp$5.6 & $\pm$6.6 & $\pm$6.2\\
   $tq$ scale variation &        $\pm$2.1 &  $\pm$2.6 &   $\pm$1.6 &  $\pm$4.3\\
    \multicolumn{5}{c}{ } \\
                    Luminosity & $\pm$1.8 & $\pm$1.8 & $\pm$1.8 & $\pm$1.8\\
    \multicolumn{5}{c}{ } \\
             Total systematic & $\pm$18 & $\pm$23 & $\pm$20 & $\pm$25 \\
                        Total  & $\pm$23 & $\pm$25 & $\pm$23 & $\pm$29 \\
   \end{tabular}
\end{ruledtabular}
\end{center}
\end{table*}

\begin{table*}[!thbp]
\begin{center}
\begin{ruledtabular} 
\caption{Detailed list of the contribution of each source of uncertainty to the  
   total relative uncertainty on the measured $\frac{1}{\sigma}\frac{d\sigma}{d\pT(t)}$ distribution given in percent for each bin.
   The list includes only those uncertainties that contribute with more than 1\%.
   The JES $\eta$ intercalibration uncertainty has a sign switch from the first to the second bin. For the $tq$ generator + parton shower uncertainty a sign switch is denoted with $\mp$.
   The following uncertainties contribute to the total uncertainty with less than 1\% to each bin content: 
   JES detector, JES statistical, JES physics modeling, JES mixed detector and modeling, JES close-by jets, JES pile-up, JES flavor composition, JES flavor response, 
   $b$-JES, jet vertex fraction, $b/\bbar$ acceptance, $c$-tagging efficiency, $\MET$ modeling, lepton uncertainties, $W$+jets shape variation, and $\ttbar$ generator.
   The contents of this table are provided in machine-readable format in the Supplemental Material~\cite{suppurl}.
}
\label{tab:topPtNormComb}
\begin{tabular}{lccccc}
    $\frac{1}{\sigma}\frac{d\sigma}{d\pT(t)}$ & \multicolumn{5}{c}{$\pT(t)$ bins [\gev]} \\
    \multicolumn{6}{c}{ } \\
    Source & $[0,45]$ & 
             $[45,75]$ & 
             $[75,110]$ &
             $[110,150]$ &
             $[150,500]$ \\
    \hline
               Data statistical & $\pm$5.3 & $\pm$6.9 & $\pm$8.0 & $\pm$11 & $\pm$15\\
        Monte Carlo statistical & $\pm$4.2 & $\pm$5.5 & $\pm$5.2 & $\pm$6.2 & $\pm$9.3\\
    \multicolumn{6}{c}{ } \\
          Background normalization & $<$ 1      & $\pm$1.7 & $<$ 1     & $\pm$3.0 & $<$ 1\\ 
   JES $\eta$ intercalibration &         $-4.7/+1.5$ &         $+3.5/-2.3$ &         $+4.1/-0.8$     & $<$ 1 &         $+9.6/-3.1$\\
         Jet energy resolution     & $<$ 1     & $<$ 1     & $<$ 1 &   $\mp$1.4 &  $\pm$2.7\\
         $b$-tagging efficiency    & $<$ 1     & $<$ 1     & $<$ 1     & $<$ 1 &         $\pm$2.8\\
             Mistag efficiency     & $<$ 1     & $<$ 1     & $<$ 1 &   $\pm$1.0     & $<$ 1\\
    \multicolumn{6}{c}{ } \\
    $tq$ generator + parton shower &        $\pm$3.9 &     $\pm$5.4 &   $\mp$11 &     $\mp$14 &    $\pm$6.9\\
   $tq$ scale variation & $<$ 1 &        $\mp$1.8 &    $\pm$1.3 &  $\mp$2.7 &   $+4.4/-5.1$\\
    \multicolumn{6}{c}{ } \\
                     Unfolding & $<$ 1     & $\pm$1.7 & $<$ 1     & $<$ 1     & $\pm$1.1\\
    \multicolumn{6}{c}{ } \\
              Total systematic &          +6.5/-7.7 & $\pm$8.8 & $\pm$13 & $\pm$18 & $\pm$16\\
                        Total  &          +8.4/-9.4 & $\pm$11 & $\pm$15 & $\pm$21 & $\pm$22\\

   \end{tabular}
\end{ruledtabular}
\end{center}
\end{table*}

\begin{table*}[!thbp]
\begin{center}
\begin{ruledtabular} 
\caption{Detailed list of the contribution of each source of uncertainty to the  
   total relative uncertainty on the measured $\frac{1}{\sigma}\frac{d\sigma}{d\pT(\tbar)}$ distribution given in percent for each bin.
   The list includes only those uncertainties that contribute with more than 1\%.
   Sign switches within one uncertainty are denoted with $\mp$ and $\pm$.
   The following uncertainties contribute to the total uncertainty with less than 1\% to each bin content: 
   JES detector, JES statistical, JES physics modeling, JES mixed detector and modeling, JES close-by jets, JES pile-up, JES flavor composition, JES flavor response, 
   $b$-JES, jet vertex fraction, $b/\bbar$ acceptance, mistag efficiency, $\MET$ modeling, lepton uncertainties, $W$+jets shape variation, and $\ttbar$ generator.
  The contents of this table are provided in machine-readable format in the Supplemental Material~\cite{suppurl}.
}
\label{tab:antitopPtNormComb}
\begin{tabular}{lccccc}
    $\frac{1}{\sigma}\frac{d\sigma}{d\pT(\tbar)}$ & \multicolumn{5}{c}{$\pT(\tbar)$ bins [\gev]} \\
    \multicolumn{6}{c}{ } \\
    Source & $[0,45]$ & 
             $[45,75]$ & 
             $[75,110]$ &
             $[110,150]$ &
             $[150,500]$ \\
    \hline
               Data statistical & $\pm$8.2 & $\pm$8.8 & $\pm$13 & $\pm$26 & $\pm$26\\
        Monte Carlo statistical & $\pm$8.7 & $\pm$9.6 & $\pm$14 & $\pm$28 & $\pm$27\\
    \multicolumn{6}{c}{ } \\
          Background normalization & $<$ 1     & $\pm$4.5 & $\pm$1.8 & $\pm$39 & $\pm$22\\
   JES $\eta$ intercalibration     &         $-7.5/+6.7$ &         $+3.8/-5.3$ &         $+6.9/-3.1$ &        $+17/-9.9$     & $<$ 1\\
         Jet energy resolution     & $<$ 1     & $<$ 1 &   $\mp$1.6 & $\pm$1.8 &  $\mp$1.2\\
        $b$-tagging efficiency     & $<$ 1     & $<$ 1     & $<$ 1     & $<$ 1 &         $+2.4/-2.8$\\
        $c$-tagging efficiency &   $\mp$1.8 & $\pm$2.0 & $\pm$1.7 &         $-6.2/+5.9$ &   $\mp$2.0\\
    \multicolumn{6}{c}{ } \\
                          PDF  & $<$ 1 & $<$ 1 & $<$ 1 &         $\pm$2.5&         $\pm$3.6\\
     $tq$ generator + parton shower &         $+7.7/-8.2$ &  $-3.6/+3.7$ &        $-13/+14$ &         $+6.4/-7.0$ &         $-4.2/+4.5$\\
     $tq$ scale variation &        $\pm$ 1.3 &   $\mp$ 3.0 &   $\pm$ 1.4 &  $\mp$ 1.8 &  $\pm$ 5.1\\
    \multicolumn{6}{c}{ } \\
                     Unfolding & $<$ 1     & $<$ 1     & $<$ 1      & $\pm$6.7 & $\pm$2.8\\
    \multicolumn{6}{c}{ } \\
              Total systematic & $\pm$15 & $\pm$13 & $\pm$21 &      $+62/-56$ & $\pm$45\\
                        Total  & $\pm$17 & $\pm$15 & $\pm$25 &      $+67/-61$ & $\pm$52\\

   \end{tabular}
\end{ruledtabular}
\end{center}
\end{table*}

\begin{table*}[!thbp]
\begin{center}
\begin{ruledtabular} 
\caption{Detailed list of the contribution of each source of uncertainty to the  
   total relative uncertainty on the measured $\frac{1}{\sigma} \frac{d\sigma}{d|y(t)|}$ distribution given in percent for each bin.
   The list includes only those uncertainties that contribute with more than 1\%.
   Sign switches within one uncertainty are denoted with $\mp$ and $\pm$.
   The following uncertainties contribute to the total uncertainty with less than 1\% to each bin content: 
   JES detector, JES statistical, JES physics modeling, JES mixed detector and modeling, JES close-by jets, JES pile-up, JES flavor composition, JES flavor response, 
   $b$-JES, jet vertex fraction, $b/\bbar$ acceptance, $b$-tagging efficiency, $c$-tagging efficiency, mistag efficiency, $\MET$ modeling, lepton uncertainties, $W$+jets shape variation, $\ttbar$ generator, $\ttbar$ ISR/FSR, and unfolding.
  The contents of this table are provided in machine-readable format in the Supplemental Material~\cite{suppurl}.
}
\label{tab:topRapNormComb}
\begin{tabular}{lcccc}
    $\frac{1}{\sigma} \frac{d\sigma}{d|y(t)|}$ & \multicolumn{4}{c}{$|y(t)|$ bins } \\
    \multicolumn{5}{c}{ } \\
    Source & $[0,0.2]$ & 
             $[0.2,0.6]$ & 
             $[0.6,1.1]$ &
             $[1.1,3.0]$ \\
    \hline
               Data statistical & $\pm$9.0 & $\pm$6.4 & $\pm$7.5 & $\pm$5.0\\
        Monte Carlo statistical & $\pm$5.9 & $\pm$4.8 & $\pm$4.9 & $\pm$3.2 \\
    \multicolumn{5}{c}{ } \\
      Background normalization & $<$ 1     & $<$ 1     & $\pm$1.1 & $\pm$1.0\\
        JES $\eta$ intercalibration &         $+1.6/-1.5$ &         $-0.5/+2.3$ &         $+1.4/-1.5$     & $<$ 1\\
         Jet energy resolution     &   $\pm$1.2     & $<$ 1 & $\mp$1.6 & $\pm$1.0\\
    \multicolumn{5}{c}{ } \\
                           PDF &     $\pm$1.7&    $\pm$1.8& $<$ 1  &     $\pm$2.3\\
     $tq$ generator + parton shower &         $-9.0/+9.8$ &         $-2.8/+3.0$ &    $<$ 1 &         $+4.8/-5.2$\\
   $tq$ scale variation &   $<$ 1     & $<$ 1     & $<$ 1 &  $\pm$1.5\\
    \multicolumn{5}{c}{ } \\
             Total systematic & $\pm$11 & $\pm$6.3 & $\pm$6.2 & $\pm$6.9 \\
                        Total  & $\pm$15 & $\pm$9.0 & $\pm$9.7 & $\pm$8.5 \\
   \end{tabular}
\end{ruledtabular}
\end{center}
\end{table*}

\begin{table*}[!thbp]
\begin{center}
\begin{ruledtabular} 
\caption{Detailed list of the contribution of each source of uncertainty to the  
   total relative uncertainty on the measured $\frac{1}{\sigma} \frac{d\sigma}{d|y(\tbar)|}$ distribution given in percent for each bin.
   The list includes only those uncertainties that contribute with more than 1\%.
   Sign switches within one uncertainty are denoted with $\mp$ and $\pm$.
   The following uncertainties contribute to the total uncertainty with less than 1\% to each bin content: 
   JES detector, JES statistical, JES physics modeling, JES mixed detector and modeling, JES close-by jets, JES pile-up, JES flavor composition, JES flavor response, 
   $b$-JES, jet energy resolution, jet vertex fraction, $b/\bbar$ acceptance, $b$-tagging efficiency, $c$-tagging efficiency, mistag efficiency, $\MET$ modeling, lepton uncertainties, $W$+jets shape variation, $\ttbar$ generator, $\ttbar$ ISR/FSR, and unfolding.
    The contents of this table are provided in machine-readable format in the Supplemental Material~\cite{suppurl}.
}
\label{tab:antitopRapNormComb}
\begin{tabular}{lcccc}
    $\frac{1}{\sigma} \frac{d\sigma}{d|y(\tbar)|}$ & \multicolumn{4}{c}{$|y(\tbar)|$ bins } \\
    \multicolumn{5}{c}{ } \\
    Source & $[0,0.2]$ & 
             $[0.2,0.6]$ & 
             $[0.6,1.1]$ &
             $[1.1,3.0]$ \\
    \hline
               Data statistical & $\pm$13 & $\pm$9.1 & $\pm$11 & $\pm$11\\
        Monte Carlo statistical & $\pm$12 & $\pm$11 & $\pm$12 & $\pm$14\\
    \multicolumn{5}{c}{ } \\
          Background normalization & $\pm$3.4 & $\pm$2.4 & $\pm$1.1 & $<$ 1\\ 
        JES $\eta$ intercalibration & $<$ 1 &         $+0.5/-1.9$     & $<$ 1 &         $+1.5/-0.8$\\
    \multicolumn{5}{c}{ } \\
                           PDF &     $\pm$1.6&    $\pm$1.0&   $<$ 1&     $\pm$1.8\\
     $tq$ generator + parton shower &         $\mp$1.4 &         $-7.8/+8.2$ &         $+4.0/-4.3$ &         $+3.8/-3.9$\\
   $tq$ scale variation &         $\pm$1.9 &   $<$ 1     & $<$ 1     & $<$ 1\\
    \multicolumn{5}{c}{ } \\
             Total systematic & $\pm$13 & $\pm$14 & $\pm$13 & $\pm$15 \\
                        Total  & $\pm$19 & $\pm$17 & $\pm$17 & $\pm$18 \\
   \end{tabular}
\end{ruledtabular}
\end{center}
\end{table*}

\clearpage

\bibliographystyle{apsrev4-1}
\bibliography{ST_tchannel}

\clearpage 
\onecolumngrid
\clearpage 

\input{atlas_authlist}

\end{document}

%% file: atlas_authlist.tex
\begin{flushleft}
{\Large The ATLAS Collaboration}

\bigskip

G.~Aad$^{\rm 84}$,
B.~Abbott$^{\rm 112}$,
J.~Abdallah$^{\rm 152}$,
S.~Abdel~Khalek$^{\rm 116}$,
O.~Abdinov$^{\rm 11}$,
R.~Aben$^{\rm 106}$,
B.~Abi$^{\rm 113}$,
M.~Abolins$^{\rm 89}$,
O.S.~AbouZeid$^{\rm 159}$,
H.~Abramowicz$^{\rm 154}$
% ...
\bigskip
\\
% ...
$^{11}$ Institute of Physics, Azerbaijan Academy of Sciences, Baku, Azerbaijan\\
% ...
$^{84}$ CPPM, Aix-Marseille Universit{\'e} and CNRS/IN2P3, Marseille, France\\
% ...
$^{89}$ Department of Physics and Astronomy, Michigan State University, East Lansing MI, United States of America\\
% ...
$^{106}$ Nikhef National Institute for Subatomic Physics and University of Amsterdam, Amsterdam, Netherlands\\
% ...
$^{112}$ Homer L. Dodge Department of Physics and Astronomy, University of Oklahoma, Norman OK, United States of America\\
$^{113}$ Department of Physics, Oklahoma State University, Stillwater OK, United States of America\\
% ...
$^{116}$ LAL, Universit{\'e} Paris-Sud and CNRS/IN2P3, Orsay, France\\
% ...
$^{152}$ Institute of Physics, Academia Sinica, Taipei, Taiwan\\
% ...
$^{154}$ Raymond and Beverly Sackler School of Physics and Astronomy, Tel Aviv University, Tel Aviv, Israel\\
% ...
$^{159}$ Department of Physics, University of Toronto, Toronto ON, Canada\\
% ...
\end{flushleft}
